\documentclass[11pt]{article}
\usepackage{geometry}
\usepackage[utf8]{inputenc}
\usepackage{mathrsfs}
\usepackage{bbm}
\usepackage{bm}
\usepackage{graphicx}          % Include figure files
\usepackage{epstopdf}
\usepackage[export]{adjustbox}
\usepackage{dcolumn}           % Align table columns on decimal point
\usepackage{framed}
\usepackage{setspace}
\usepackage[toc,page]{appendix}
\usepackage{amsmath}
\usepackage{amssymb}
\usepackage{amsthm}
\usepackage{authblk}
\usepackage{color}
\usepackage{tikz}
\usepackage{tikz-cd}
\usepackage{comment}
\usepackage[
    colorlinks=true,
    linkcolor=blue,  % internal references
    citecolor=blue,  % citations
    urlcolor=blue    % URLs
]{hyperref}

\usepackage{xcolor}

\newtheorem{theorem}{Theorem}

\newtheorem{lemma}{Lemma}
\newtheorem{claim}{Claim}

\newtheorem{cor}{Corollary}
\newtheorem{definition}{Definition}

\newcommand{\cC}{\mathcal{C}}

\newcommand{\cE}{\mathcal{E}}
\newcommand{\cF}{\mathcal{F}}
\newcommand{\cG}{\mathcal{G}}
\newcommand{\cH}{\mathcal{H}}

\newcommand{\cL}{\mathcal{L}}

\newcommand{\cN}{\mathcal{N}}
\newcommand{\cO}{\mathcal{O}}

\newcommand{\cR}{\mathcal{R}}

\newcommand{\cU}{\mathcal{U}}

\newcommand{\cW}{\mathcal{W}}
\newcommand{\cX}{\mathcal{X}}

\newcommand{\cZ}{\mathcal{Z}}

\newcommand{\cc}{\mathbb{C}}

\newcommand{\bE}{\mathbb{E}}
\newcommand{\ff}{\mathbb{F}}

\newcommand{\zz}{\mathbb{Z}}

\newcommand{\beq}{\begin{equation}}
\newcommand{\eeq}{\end{equation}}
\newcommand{\ba}{\begin{array}{ccc}}
\newcommand{\ea}{\end{array}}
\newcommand{\eea}{\end{eqnarray}}

\makeatletter

\renewcommand{\section}{%
  \@startsection{section}{1}{0pt}%
  {3.5ex plus 1ex minus .2ex}%
  {2.3ex plus .2ex}%
  {\normalfont\large\bfseries}%
}

\renewcommand{\subsection}{%
  \@startsection{subsection}{2}{0pt}%
  {3.25ex plus 1ex minus .2ex}%
  {1.5ex plus .2ex}%
  {\normalfont\normalsize\bfseries}%
}

\renewcommand{\subsubsection}{%
  \@startsection{subsubsection}{3}{0pt}%
  {3.25ex plus 1ex minus .2ex}%
  {1.0ex plus .2ex}%
  {\normalfont\normalsize\itshape}%
}

\def\@hangfrom@section#1#2#3{\@hangfrom{#1#2}#3}
\def\@hangfroms@section#1#2{#1#2}

\makeatother

\newcommand{\unstable}{\texttt{Unstable}}

\newcommand{\RR}{\mathbb{R}}

\numberwithin{equation}{section}

\AddToHook{cmd/appendix/after}{%
}

\newif\ifCLshow \CLshowtrue   % \CLshowfalse + \excludecomment{CLdraft} hides all

\newcommand{\CLgap}[2]{\ifCLshow\textcolor{\csname CLcol#1\endcsname}{[\textbf{CLAUDE/#1}: #2]}\fi}

\begin{document}

\title{A thermally and perturbatively stable classical spin model in two dimensions}

\author[1,$\ast$]{Shankar Balasubramanian}

\author[1,$\ast$]{Margarita Davydova}

\author[2,$\ast$]{Ting-Chun Lin}

\author[3,$\ast$]{Ryan Thorngren}

\affil[1]{\small \small Walter Burke Institute for Theoretical Physics and Institute for Quantum Information and Matter,
  California Institute of Technology, Pasadena, CA 91125, USA}

\affil[2]{\small Department of Physics, Stanford University, Stanford, CA 94305, USA}

\affil[3]{\small Mani L. Bhaumik Institute for Theoretical Physics, Department of Physics and Astronomy,
University of California, Los Angeles, CA 90095, USA}

\affil[$\ast$]{\small Alphabetical order}

\date{}
\maketitle

\begin{abstract}
We show that the 2D classical spin model constructed in Ref.~\cite{balasubramanian2026passive} hosts many metastable states at non-zero temperature which are stable to arbitrary Hamiltonian perturbations, providing the first such example.  Our proof relies on a recursive application of the cluster expansion.
\end{abstract}

\noindent\rule{\textwidth}{0.2pt}
\vspace{-25 pt}
\setcounter{tocdepth}{2}
{
  \hypersetup{linkcolor=black}
  \tableofcontents
}

\noindent\rule{\textwidth}{0.2pt}

\vspace{5 pt}

\section{Introduction}
A prominent goal of statistical physics is to understand the stability of matter at non-zero temperature.  Beyond its theoretical value, this has relevance to information storage technologies such as magnetic hard drives. For this paper, we will specifically consider low-dimensional (2D) spin models that can be used to store information, and will refer to such systems as ``classical memories''.

Classical memories are generally associated with phenomena like spontaneous symmetry breaking or spin glass order. However, memories relying on spontaneous symmetry breaking, such as the Ising model, are fragile under Hamiltonian perturbations that explicitly break the symmetry.  Low-dimensional spin glasses are unstable to an external field. Furthermore, in two dimensions, there are no examples of spin glasses that are thermally stable.
Thus, it is interesting to ask whether there is a classical memory that is simultaneously thermally stable and stable to Hamiltonian perturbations.

A well-known example of such a system is a discrete classical gauge theory in three dimensions, the Wegner gauge theory~\cite{wegner1971duality}, which unlike the Ising model, does not admit a local order parameter.  It is known to have a thermally and perturbatively stable deconfined phase, which was first studied in~\cite{wegner1971duality, PhysRevD.19.3682} and later revisited in~\cite{PhysRevB.99.094103,stahl2026slow,ryan-wegner}. However, in two dimensions, the situation is much less clear.  All two-dimensional classical gauge theories that we know are in a confined phase at all non-zero temperatures and thus cannot store classical information.  In fact, a widely-held belief is that a perturbatively and thermally stable classical memory is \emph{not possible} in two dimensions.

In this paper, we show an unexpected example of a thermally and perturbatively stable spin model in two dimensions.  Recently, in Ref.~\cite{balasubramanian2026passive}, Balasubramanian, Davydova, and Lin constructed a self-correcting quantum memory in three dimensions.  They also constructed a 2D classical code using similar ideas and conjectured that it was stable.  The Hamiltonian associated to this code has $\exp(N^\gamma)$ metastable states for some $\gamma > 0$, where $N$ is the number of spins.  In this paper, we resolve this conjecture by showing that all of these states are thermally and perturbatively stable.

Informally, the properties of this model come from its recursive construction: at odd recursion levels, the energy cost of excitations is enhanced, which improves thermal stability, while at even recursion levels, gauge generators are added, which improves perturbative stability.  The resulting code can be viewed as a classical gauge theory with a gauge group generated by a hierarchy of generators whose support sizes vary from $O(1)$ to the system size. At the same time, the excitations of this system are ``confined'' energetically, i.e. the energy of an excitation increases with the size of the support of the operator that created it. The code is also geometrically local in two dimensions; its fractal dimension can be as small as $1 + \varepsilon$ where $\varepsilon$ can be tuned to $0$ by tuning a geometric parameter.

\subsection{Results}

\subsubsection{Thermal dynamics and main theorem statement}
We will first explain how to show thermal and perturbative stability for a pair of metastable states associated to the largest length scale of the model, which we may use to reliably store a classical bit. We will then explain how the result can be straightforwardly generalized to all $O(\exp(N^\gamma))$ metastable states.

Consider a system of spins occupying a subset of a two-dimensional lattice $\cL$ described by a Hamiltonian $\cH$.  The system evolves under general detailed-balance/reversible Markov chain dynamics with bounded transition rates whose stationary distribution is the Gibbs distribution at inverse temperature $\beta$.  A standard example is Glauber dynamics with single spin updates.  More precisely, calling $P(\vec{s}, t)$ the probability of spin configuration $\vec{s}$ at time $t$, the evolution of this probability distribution is given by the differential equation
\begin{equation}
\frac{d}{dt} P(\vec{s},t) = \sum_{i \in \cL} w_i(-s_i, \vec{s}_{-i}) P(-s_i, \vec{s}_{-i}, t) - w_i(s_i, \vec{s}_{-i}) P(s_i, \vec{s}_{-i}, t)
\end{equation}
where $\vec{s} = (s_i, \vec{s}_{-i})$ separates the spin configuration into the value at site $i$ and the vector of values at all other sites.  The transition probability $w_i$ is given by
\begin{equation}
w_i(s_i, \vec{s}_{-i}) = \frac{1}{1 + \exp(\beta \cH(-s_i, \vec{s}_{-i}) - \beta \cH(s_i, \vec{s}_{-i} ))}.
\end{equation}
When $\cH$ is a local Hamiltonian in the lattice, $w_i$ depends only on spins in a $O(1)$ neighborhood of $s_i$.  The dynamics obeys detailed balance/reversibility, and the unique stationary distribution $\lim_{t \to \infty} P(\vec{s},t) = \pi_\cH(\vec{s})$, assuming ergodicity,  is given by
\begin{equation}
\pi_\cH(\vec{s}) = \frac{1}{\mathcal Z} \exp(-\beta \cH(\vec{s})).
\end{equation}
We say that the Hamiltonian $\cH$ describes a classical memory storing a bit if there exists an encoding and decoding scheme such that after encoding the bit and allowing it to thermalize for a time $t \leq t_{\rm{mem}}$, the bit can be retrieved by decoding, with high probability.  We refer to $t_{\rm{mem}}$ as the memory lifetime for this reason.  More formally:
\begin{definition}
Consider a subset $\cL$ of a two-dimensional lattice with $N$ classical spins.  A Hamiltonian $\cH$ undergoing Glauber dynamics\footnote{This result can be extended to any continuous-time single-spin-flip Markov dynamics with uniformly bounded transition rates and stationary Gibbs distribution $\pi_{\mathcal H}$. In this case, detailed balance/reversibility is not required~\cite{dyer2006markov}.} with stationary distribution $\pi_{\cH}$  is a classical memory if there exists an encoding map $\rm{Enc}: \{0,1\} \to \mathscr{P}_\cL$ and a decoding map $\rm{Dec}: \mathscr{P}_\cL \to \{0,1\}$, where $\mathscr{P}_\cL$ is the space of probability distributions on spin configurations in $\{-1,1\}^{|\cL|}$, such that
\begin{equation}
\mathbb{P}(\mathrm{Dec} \circ \Phi_{\beta, t} \circ \mathrm{Enc}(b) \neq b) \lesssim t \cdot \exp(-N^\alpha)
\end{equation}
for $\beta > \beta_c = O(1)$, where $\Phi_{\beta, t}$ is the Glauber dynamics evolution operator for time $t$ at inverse temperature $\beta$, and $\alpha > 0$.  We define the memory lifetime to be
\begin{equation}
t_{\rm{mem}} = \max\{t: \max_{b \in \{0,1\}}\mathbb{P}(\mathrm{Dec} \circ \Phi_{\beta, t} \circ \mathrm{Enc}(b) \neq b) \leq 1/100\}
\end{equation}
which, combined with the previous definition, implies that $t_{\rm{mem}} \gtrsim \exp(N^\alpha)$.
\end{definition}
Our main theorem is that such a classical memory exists in two dimensions and is perturbatively stable.
\begin{theorem} []
There exists an infinite family of Hamiltonians $\{\cH(N): N \in \mathbb{Z}_{\geq 0}\}$ defined on a growing family of subsets of $\mathbb{Z}^2$, along with a family of encoding and decoding maps $\rm{Enc}$ and $\rm{Dec}$, such that $(\cH(N), \rm{Enc}, \rm{Dec})$ corresponds to a classical memory with $t_{\rm{mem}} \gtrsim \exp(N^{\alpha})$ for some $\alpha > 0$ at inverse temperatures $\beta > \beta_c = O(1)$.  Furthermore, there exist constants $\epsilon>0$ and $\beta_c'>0$, independent of $N$ and $\beta$, such that for every $\beta>\beta_c'$ and every local classical perturbation $V(N)$ whose strength satisfies
\begin{equation} \label{eq:condition-1}
    \beta |V(N)|_{\eta, \mathrm{loc}}\leq\epsilon
\end{equation}
for some $\eta > 0$, the perturbed model has memory lifetime $t_{\mathrm{mem}}\gtrsim\exp(N^{\alpha'})$ for some $\alpha'>0$.
\end{theorem}
Informally, $ |V(N)|_{\eta, \mathrm{loc}}$ is an exponentially weighted interaction norm
commonly used in perturbative stability analysis.
When this norm is finite,
it forces the strength of long range interactions or high weight terms to decay exponentially.
As a special example, a finite-range perturbation whose terms have norm $\le \epsilon$ has $|V (N)|_{\eta, \rm loc} = O(\epsilon)$.
More formally, write $V(N) = \sum_{X \subseteq \zz^2} V_X(N)$ where $V_X(N)$ acts only on sping in $X$, and define $l_{\mathrm{loc}}(X)=|H_{\zz^2}(X)|$ where $H_{\zz^2}(X)$ is the smallest connected hull of $X$ in $\zz^2$ (this makes localization volumetric). Then, for $\eta > 0$, we define 
\begin{equation}
    |V (N)|_{\eta, \rm loc}
    =\sup_{x \in \zz^2}
      \sum_{\substack{X: x\in X}}
      \|V_X(N)\|_\infty e^{\eta l_{\mathrm{loc}}(X)},
    \label{eq:def-geometric-local-norm}
\end{equation}
where $ \|V_X(N)\|_\infty$ is the maximum over spin configurations. 

Furthermore, the Hamiltonian family $\{\cH(N): N \in \mathbb{Z}_{\geq 0}\}$ is local in $\mathbb{R}^2$: all interactions involve spins an $O(1)$ distance away and each $O(1)$-radius ball contains an $O(1)$ number of spins.

\subsubsection{Overview of the construction}
The Hamiltonian family $\{\cH (N): N \in \mathbb{Z}_{\geq 0}\}$ is associated with the classical code constructed in Sec. 6 of Ref.~\cite{balasubramanian2026passive}.   The construction is illustrated in Fig.~\ref{fig:code-iterations} and is informally defined as follows. The starting Hamiltonian $\cH_0$ is a 2-bit repetition code (Ising model) with periodic boundary conditions, which is then iteratively \emph{refined}.  Each round of refinement increases the system size but preserves geometric locality in $\mathbb{R}^2$.  

Before explaining the construction, let us briefly explain how to read the figures in this paper. Fig.~\ref{fig:code-iterations} shows the interaction graph of the model at different levels of the interaction. Bits are shown as solid dots, checks/interactions are shown as hollow dots, and a check acts on all spins connected to it by edges. All the checks are  parity checks, i.e. they are satisfied if the parity of the state on adjacent bits is even. Alternatively, a check acting on bits $\{ i_1, ... i_k \}$ corresponds to a $-Z_{i_1} ... Z_{i_k}$ interaction in the Hamiltonian, where $Z$ is the Pauli-$Z$ operator. All checks throughout the construction have degree 2 or 3, and the vertices and edges of the interaction graph occupy a subset of a square lattice in two dimensions. Some locations contain two separate degree-2 vertices, in which case the vertices are shown with an offset.

Starting from the base model, which has Hamiltonian $\cH_{(0)}$ associated with a classical code $\cC_{(0)}$, we alternately apply ``repetition'' and ``parity''-type refinement procedures, which we denote $\cR_{\rm Rep}$ and $\cR_{\rm Par}$.
These procedures are defined in Fig.~\ref{fig:segment-replacements}. Each edge of the current code, which has a bit at one endpoint and a check at the other, is replaced with a ladder-like structure, which is different in $\cR_{\rm Rep}$ and $\cR_{\rm Par}$.  The ladder consists of two parallel rails subdivided into $\ell$ segments and offset from each other by a fixed shift.  The rails are joined by $(\ell-3)/2$ rungs, and we take $\ell = 7$ in the illustrations. We call a repetition step followed by a parity step one \emph{iteration}, $\cR \equiv \cR_{\rm Par} \circ \cR_{\rm Rep}$, and say that the code obtained after $i$ iterations,
\begin{equation}
    \cC_{(2i)} = \cR^{i}(\cC_{(0)}),
\end{equation}
is \emph{at level $i$}.  The first iteration is shown in Fig.~\ref{fig:code-iterations}. We fix the total number of iterations to be $n$, resulting in the physical code is $\cC_{(2n)}$ whose associated Hamiltonian is $\mathcal{H}$.  Each replacement multiplies the number of bits and interactions at the physical level by $\Theta(\ell)$, thus the system size for $\cH_n$ is $N=\Theta(\ell^n)$. 

We now explain the effect of the $\cR_{\rm Rep}$ and $\cR_{\rm Par}$ steps. Note that in the $\cR_{\rm Rep}$ step, an edge is replaced with a local ladder-like structure, which is a repetition code of width 2. One can check that a single $\cR_{\rm Rep}$ replacement increases the energy barrier by a factor of 2, thus increasing thermal protection of the code.  In the $\cR_{\rm Par}$ step, new local codewords are introduced, associated to the smallest cycles of the ladder, whose size is $O(1)$. These correspond to the gauge transformations of the model. If such operators are applied by the Glauber dynamics, they do not change the value of the encoded bit, which is instead stored in a larger-scale codeword.  However, they help protect the code under Hamiltonian perturbations: by including gauge codewords in our encoding, the codespace gains the property of local indistinguishability. Informally speaking, this prevents a perturbation from ``detecting'' the difference between the logical states.  Thus, by alternating between repetition and parity refinements, we ensure both thermal and perturbative stability.  In Sec.~\ref{sec:code}, we provide a rigorous characterization of the logical states of the code.

\subsubsection{Proof outline}
To compute the memory lifetime of the encoded bit in the code, we appeal to standard results from Markov chain theory.  Consider a probability distribution $\pi$ over a space $\Omega$ of outcomes, and a Markov chain whose stationary distribution is $\pi$.  Construct a graph whose vertices are labeled by $\omega \in \Omega$ and whose edges are $(\omega, \omega')$ if $\omega$ and $\omega'$ are related by a single transition in the Markov chain.  Consider a connected subset of the graph corresponding to $A \in \Omega$ and construct $\partial A$, which is the set of nodes not in $A$ but neighboring some node in $A$.  Call $\pi(A) = \sum_{\omega \in A} \pi(\omega)$.  Then, define $\phi(A) = \pi(\partial A)/\pi(A)$.  This quantity is related to the  conductance\footnote{Strictly speaking, the conductance is upper bounded by Eqn.~\ref{eq:conductance}.}
\begin{equation}\label{eq:conductance}
\phi = \min_{A: \pi(A) \leq 1/2} \phi(A),
\end{equation}
which is related to the mixing time of the Markov chain via Cheeger's inequality~\cite{levin2026markov}.  For the memory lifetime, we will need to compute $\phi(A)$ for specific choices of $A$ corresponding to the codespace.  Choose $A_0$ and $A_1$ to be subsets of $\Omega$ such that configurations in $A_0$ correspond to logical $0$ and configurations in $A_1$ correspond to logical $1$.  We define the encoding map
\begin{equation}
    0 \overset{\rm Enc}{\mapsto} P_0(\omega) = \frac{\pi(\omega)}{\sum_{\omega \in A_0} \pi(\omega)} \mathbbm{1}_{\omega \in A_0}, \qquad 1 \overset{\rm Enc}{\mapsto} P_1(\omega) = \frac{\pi(\omega)}{\sum_{\omega \in A_1} \pi(\omega)} \mathbbm{1}_{\omega \in A_1}
\end{equation}
also called the thermal encoding map in Refs.~\cite{alicki2010thermal,chesi2010thermodynamic,balasubramanian2026passive}.  The decoding map $\rm{Dec}: \Omega \to \{0,1\}$ which maps $\omega \in A_0$ to $0$ and $\omega \in A_1$ to $1$.  Our choice of ``$\rm{Dec}$'' thus is crucial to upper bounding $\phi(A_b)$.  With these choices of encoding and decoding maps, one can show that (see Lemma~\ref{lem:conductancetomemlife}) the memory lifetime starting from logical $b$ is roughly $t_{\rm{mem}}\gtrsim 1/\phi(A_b)$, where $\gtrsim$ hides polynomial factors in $N$.  Thus, if $\phi(A_b)$ is exponentially small in a power of $N$, then the memory lifetime is exponentially long in a power of $N$.

We apply this to the Hamiltonian corresponding to the classical code $\cH + V$, where $V$ is a perturbation.  The goal is to define the sets $A_{0,1}$ and estimate $\phi(A_{0,1}) = \frac{\pi_{\cH + V}(\partial A_{0,1})}{\pi_{\cH + V}(A_{0,1})}$.  To do so, we define a decoder ``$\rm{Dec}$'' that operates in an iterative ``renormalization-group'' way.  The decoder only depends on error syndromes rather than the spin configurations.  
It first applies a correction at the finest scale (within the ladders created in the last iteration), then at the next scale, and so on, undoing the effect of each replacement step until no syndrome is left. Then, the logical value of the bit can be read out. This defines the sets $A_0$ and $A_1$.  Note that applying a gauge transformation to $A_0$ permutes its elements (since two configurations with the same syndrome are both in $A_0$ or both in $A_1$); the same is true for $A_1$.

The unperturbed Hamiltonian $\cH$ has a gauge group $G$ with a hierarchical structure: the generators created in the parity step of iteration $i$ form a level-$i$ subgroup, and their supports grow exponentially with $n-i$.
For a function $f$ of the spin configuration $s$, we introduce the gauge average
\begin{equation}
  (\bE_G f)(s) = \frac{1}{|G|} \sum_{g \in G} f(s+g),
\end{equation}
where $s+g$ denotes the configuration obtained from $s$ by flipping the spins in the support of $g$ (in operator language, $\bE_G f=\frac{1}{|G|}\sum_{g\in G}X^{\mathrm{supp}(g)}\,f\,X^{\mathrm{supp}(g)}$, with $X^{S}$ the product of Pauli-$X$ operators supported in $S$). We define the effective gauge-invariant interaction via
\begin{equation}
    e^{-\beta V_{\rm{eff}}} = \bE_G e^{- \beta V}.
\end{equation}
This allows us to simplify the memory lifetime estimate, since for any set of configurations $B$ that are gauge invariant, we have
    \begin{align} \label{eq:pi}
\pi_{\cH + V}(B)
&=\frac{1}{\cZ} \sum_{s \in B} \exp(-\beta \left ( \cH(s) + V(s) \right )) \nonumber
\\
&=   \frac{1}{\cZ}  \sum_{s \in B} \bE_G \exp(-\beta \left ( \cH(s) + V(s) \right ))
\\
&= \frac{1}{\cZ} \sum_{s \in B} \exp(-\beta \left ( \cH(s) + V_{\rm eff}(s) \right ))  = \pi_{\cH + V_{\rm eff} }(B)
\end{align}
Since the decoder depends on the configuration only through its syndrome, which is gauge-invariant, the sets $A_0$, $A_1$, $\partial A_0$ and $\partial A_1$ are gauge-invariant. Thus, by \eqref{eq:pi}
\begin{equation} \label{eq:phi}
  \phi(A_b) \leq \frac{\pi_{\cH+V}(\partial A_b)}{\pi_{\cH+V}(A_b)} = \frac{\pi_{\cH+V_{\rm eff}}(\partial A_b)}{\pi_{\cH+V_{\rm eff}}(A_b)} .
\end{equation}

For small enough perturbation $V$, we want this to be exponentially small in a power of the system size $N$.

We will use a hierarchical averaging method to compute $V_{\rm eff}$ and bound its strength. We first write
\begin{equation}
e^{-\beta V_{\rm{eff}}} = \bE_G e^{- \beta V}   = \bE_{1}   \bE_{2}...  \bE_{n}   e^{- \beta V}
\end{equation}
where $\bE_i$ averages over the level-$i$ subgroup. We then perform the average $\mathbb{E}_n$, which gives us $V_{\rm{eff}}^{(n)}$.
\begin{equation}
    e^{-\beta V_{\rm{eff}}^{(n)}} = \bE_{n}   e^{- \beta V}.
\end{equation}
Every time we perform averaging, we decompose the effective potential into two contributions

\begin{equation}
    V_{\rm{eff}}^{(i)} = \cH_{\rm eff}^{(i)} + \cU^{(i)}, \qquad \cH_{\rm eff}^{(i)} = \bE_G  V_{\rm{eff}}^{(i)}, \qquad  \cU^{(i)} = (1- \bE_G)  V_{\rm{eff}}^{(i)},
\end{equation}
where the first part contains gauge-invariant terms with respect to the \emph{full} gauge group, and the second part contains partially gauge-invariant operators only.
We can define the next effective potential recursively via
\begin{equation} \label{eq:veff-i}
    e^{-\beta V_{\rm{eff}}^{(i)}} = \bE_{i}   e^{- \beta \cU^{(i+1)}}.
\end{equation}
Note that $\cU^{(i)}$ can only contain ``partial'' Wilson operators, whose support grows as the level $i$ decreases, i.e. as the scale coarsens. Generation of these operators from the local perturbation $V$ becomes increasingly unlikely with level, which contributes to suppression of the strength of $\cU^{(i)}$ with $n-i$. By Eqn.~\eqref{eq:veff-i}, $\cU^{(i+1)}$ bounds all the terms in the effective interaction at level $i$, and hence, if its norm appropriately decays with $n-i$, the norms of $\cH_{\rm eff}^{(i)}$ also decay.

After performing all steps in the hierarchical gauge-averaging procedure,  we are left with
\begin{equation} \label{eq:veff-sum}
    V_{\rm{eff}} = \sum_{i=1}^n \cH_{\rm{eff}}^{(i)}.
\end{equation}

One of the main results, proved in Sec.~\ref{sec:perturbation-strength}, bounds the norm of this effective interaction:
\begin{theorem}[Informal]
There exists a norm $|\cdot|_{i}$ depending on the level $i$ in the hierarchy, a geometrically-local norm $|\cdot |_{\eta, \rm loc }$ and an $O(1)$ constant $\eta$, as well as and $O(1)$ parameter $\epsilon$, such that if $\beta |V|_{\eta, \rm loc } < \epsilon$, then

\begin{equation}
|\cU^{(i)}|_{i} \lesssim \frac{|V|_{\eta, \rm loc }}{c^{n-i}}, \qquad |\cH_{\rm eff}^{(i)}|_{i} \lesssim A \frac{|V|_{\eta, \rm loc }}{c^{n-i}}
\end{equation}
\end{theorem}
The proof of this theorem relies on a substantial improvement to the gauge-averaging result of Ref.~\cite{ryan-wegner}, which we show in Subsec.~\ref{sec:gauge-averaging-thm}.  This establishes that the perturbation has a ``small'' effect on $V_{\rm{eff}}$, and potentially a small effect on $\phi(A_b)$.  

Having bounded the terms entering $V_{\rm{eff}}$, in Sec.~\ref{sec:peierls}, we apply a Peierls argument to upper-bound $\phi(A_b)$ and obtain a lower bound on the memory lifetime. Once we do this, we can establish thermal and perturbative stability of the memory associated with $\cH + V$ for small enough $V$. 

We now elaborate on the Peierls argument. We first define an unstable configuration $S$ as one where $\mathrm{Dec}(S) \neq \mathrm{Dec}(S \oplus e)$ for some error $e$ corresponding to a transition in the Markov process.  Choosing $\rm{Dec}$ to be a renormalization-group decoder (i.e. one that successively corrects errors at larger and larger scales), the unstable configurations are a subset of all configurations that still have an error syndrome at the last stage of the decoder.  Such configurations contain a witness syndrome $\tau$, which is a large-weight syndrome configuration.  The energy cost of such a witness under the unperturbed Hamiltonian is $|\tau|$, but under the perturbed Hamiltonian, we want to show that the energy cost is $\leq (1+\epsilon') |\tau|$ and $\geq (1-\epsilon') |\tau|$ for some $\epsilon'$.  Then, we show that the Boltzmann weight of configurations containing a witness $\tau$ is $\sim \exp(-\beta (1-\epsilon') |\tau|)$ and the number of witnesses of fixed size $m$ is $A^m$ for some $A > 1$. 
We then argue that the minimum size of a witness is $\geq 2^n$, where the maximum level in the construction is $n$. Finally, combined with the Peierls argument, this shows that the $\phi$ is exponentially small in $2^n$ and thus the memory lifetime is exponentially large in $2^n$.  The number of bits is $N = O(\ell^n)$, and thus, the memory lifetime is exponentially large in $N^{\alpha}$ for some $\alpha > 0$, establishing the main theorem for a single encoded bit.

Finally, classical code has $O(N^{\gamma
})$ bits that are thermally and perturbatively stable.  This is because the gauge transformations introduced at level $i$ of the hierarchy have support size $O(\ell^{n-i})$. Let $k' = \theta n$ with $0 < \theta <1$. If we only include gauge transformations introduced at some level $k' < n$ and higher in our definition of the gauge group (instead of all newly introduced local codewords as we did earlier), the codewords of the model will all be large, i.e. the minimal support size of their logical-$X$ operators are at least $O(\ell^{n-k'})$. All the earlier arguments carry over with appropriate redefinitions of encoding and decoding maps, which now will utilize these codewords for encoding additional bits. The resulting memory lifetime scales as $\exp (N^{\alpha (1-\theta)})$.

\subsection{Outlook}

Our 2D model exhibits some similarities to spin glasses, since it can store $O(N^{\gamma
})$ bits.  However, not only are the $2^{O(N^\gamma)}$ configurations thermally robust, but they are also robust to arbitrary Hamiltonian perturbations.  In contradistinction to standard spin glasses, our model is not disordered, leading to a complex but structured energy landscape.  However, we believe that it exhibits many properties of spin glasses, such as shattering and incongruence.  The finite-dimensional analog of the shattering property seems plausible given that there are many metastable states; the configurational entropy likely scales as $\sim N^{\alpha}$, but a proof is needed.  The finite-dimensional analog of incongruence is related to showing that a large class of low-energy states are metastable, which is the case in our model and would follow from the same proof technique. If these properties are carefully proved, then our 2D model could correspond to a ``topological'' version of a spin glass~\cite{placke2025expansion} in two dimensions.

Another interesting question concerns adding quantum perturbations to the code.  At non-zero temperature, quantum perturbations do not destroy the classical memory; a proof of this will be provided elsewhere.  At zero temperature, the effect of quantum perturbations seems quite unusual.  Consider adding a transverse field $V = \sum_i X_i$. Typically adding this perturbation to the Hamiltonian of a classical gauge theory results in quantum topological order. Working perturbatively in the field strength generates the gauge symmetries that enter the perturbed Hamiltonian. In this code, recall that the gauge symmetries are hierarchical: a generator created at level $i$ has support growing exponentially in $n-i$. The larger-support terms will be generated with increasingly smaller magnitude, and the model will have a dense gapless spectrum.  If the dual logical $Z$ operators are of large-weight, they might potentially be stable to perturbations.  Showing this formally would require strengthening the result in Ref.~\cite{Bravyi_2010}. If this holds, this model would correspond to an interesting gapless zero-$T$ quantum memory with large number of logical qubits.  Another model that would exhibit such a property would be a collection of decoupled toric codes at different scales.  However, our model is not in the same phase since it functions as a nontrivial classical memory at non-zero temperature, unlike the toric code.  We leave exploring this direction to future work.

Finally, one could ask more about the phase diagram of this model.  Consider the classical Hamiltonian $H_0 = -J \sum_S Z_S$ associated to the 2D classical code  ($S$ denotes sets of spins and $Z_S = \prod_{i \in S} Z_i$) and add a longitudinal field $V = -h \sum_i Z_i$.  Then, defining $K = e^{-2\beta J}$ and $\tau = \tanh \beta h$, the partition function $\mathcal{Z}(K, \tau)$ for the classical code $C$ admits a Kramers-Wannier duality where $C \leftrightarrow C^T$ and $K \leftrightarrow \tau$.  Furthermore, the transpose code $C^T$ is essentially equivalent to exchanging the repetition and parity iterations in the code construction.  Thus, $C^T$ has essentially the same kind of behavior as $C$ and this code almost possesses a kind of self-duality (which is expected to lead to an exact self-duality in the IR).  This situation is reminiscent to the self-duality in $\mathbb{Z}_2$ gauge theory with matter~\cite{PhysRevD.19.3682}.  It would be interesting to study the nature of the phase transitions and especially the multicritical point in our new gauge theory.  Preliminary numerical computations of the phase diagram of this model as a function of temperature and external field showed that it had a similar shape to that of the Wegner gauge theory~\cite{wegner1971duality}.

\section{The 2D classical code}\label{sec:code}

In this section,
  we construct the 2D classical code with the desired memory property.
This code was first constructed by Balasubramanian, Davydova, and Lin in Ref.~\cite{balasubramanian2026passive}.  Here, we prove some properties of it that will be helpful in the later analysis.  Throughout this section, we adopt the code-embedding formalism introduced in Ref.~\cite{yuan} and later refined in Ref.~\cite{balasubramanian2026passive}.

The code is constructed iteratively, starting from a base code $\cC_{(0)}$ with two bits and two checks arranged around a unit square.
We then alternately apply the two operations $\cR_{\mathrm{Rep}}$ and $\cR_{\mathrm{Par}}$:
\begin{align}
  \cC_{(0)},\quad
  \cC_{(1)} = \cR_{\mathrm{Rep}}\bigl(\cC_{(0)}\bigr),\quad
  \cC_{(2)} = \cR_{\mathrm{Par}}\bigl(\cC_{(1)}\bigr),\quad \ldots,\quad
  \cC_{(2k-1)} = \cR_{\mathrm{Rep}}\bigl(\cC_{(2k-2)}\bigr),\quad
  \cC_{(2k)} = \cR_{\mathrm{Par}}\bigl(\cC_{(2k-1)}\bigr).
\end{align}

Each of the operations, $\cR_{\mathrm{Rep}}$ and $\cR_{\mathrm{Par}}$ turn a 2D local code into another 2D local code.
$\cR_{\mathrm{Rep}}$ introduces redundant bits that improve the code's thermal stability,
  while $\cR_{\mathrm{Par}}$ introduces gauge generators that improve its perturbative stability.

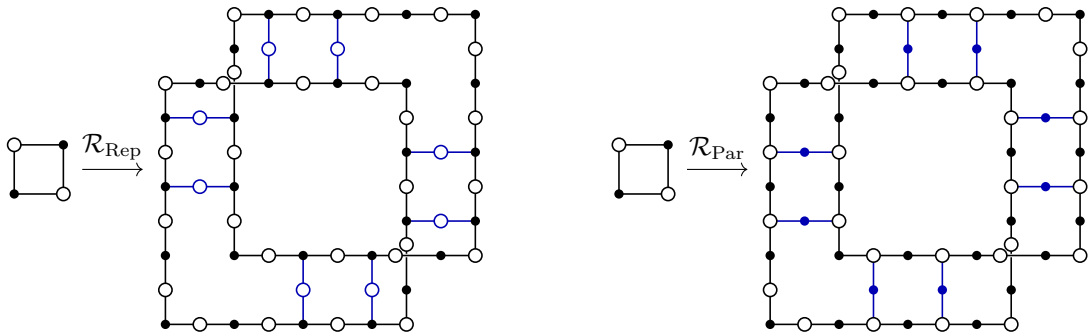
\begin{figure}[h]
  \centering
    \begin{tikzpicture}[x=1cm,y=1cm,
    iteration edge/.style={line width=0.6pt},
    iteration spin/.style={circle,fill=black,inner sep=0pt,minimum size=3.5pt},
    iteration check/.style={circle,draw=black,fill=white,line width=0.6pt,
      inner sep=0pt,minimum size=5pt}]
    % C_(0); crossing vertices remain distinct.
    \draw[iteration edge]
      (0.000,1.725) -- (0.000,2.375) (0.650,2.375) -- (0.650,1.725);
    \draw[iteration edge,preaction={draw=white,line width=1.3pt}]
      (0.000,2.375) -- (0.650,2.375) (0.650,1.725) -- (0.000,1.725);
    \foreach \point in {
      (0.000,1.725),(0.650,2.375)}
      \node[iteration spin] at \point {};
    \foreach \point in {
      (0.000,2.375),(0.650,1.725)}
      \node[iteration check] at \point {};
    % C_(1); crossing vertices remain distinct.
    \draw[iteration edge]
      (2.000,0.000) -- (2.000,0.456) (2.000,0.456) -- (2.000,0.911) (2.000,0.911) -- (2.000,1.367)
      (2.000,1.367) -- (2.000,1.822) (2.000,1.822) -- (2.000,2.278) (2.000,2.278) -- (2.000,2.733)
      (2.000,2.733) -- (2.000,3.189) (5.189,0.000) -- (5.189,0.456) (5.189,0.456) -- (5.189,1.057)
      (5.189,1.057) -- (5.189,1.367) (5.189,1.367) -- (5.189,1.822) (5.189,1.822) -- (5.189,2.278)
      (5.189,2.278) -- (5.189,2.733) (5.189,2.733) -- (5.189,3.189) (2.911,0.911) -- (2.911,1.367)
      (2.911,1.367) -- (2.911,1.822) (2.911,1.822) -- (2.911,2.278) (2.911,2.278) -- (2.911,2.733)
      (2.911,2.733) -- (2.911,3.335) (2.911,3.335) -- (2.911,3.644) (2.911,3.644) -- (2.911,4.100)
      (6.100,0.911) -- (6.100,1.367) (6.100,1.367) -- (6.100,1.822) (6.100,1.822) -- (6.100,2.278)
      (6.100,2.278) -- (6.100,2.733) (6.100,2.733) -- (6.100,3.189) (6.100,3.189) -- (6.100,3.644)
      (6.100,3.644) -- (6.100,4.100);
    \draw[iteration edge,blue!70!black]
      (3.367,3.189) -- (3.367,3.644) (3.367,3.644) -- (3.367,4.100) (4.278,3.189) -- (4.278,3.644)
      (4.278,3.644) -- (4.278,4.100) (3.822,0.000) -- (3.822,0.456) (3.822,0.456) -- (3.822,0.911)
      (4.733,0.000) -- (4.733,0.456) (4.733,0.456) -- (4.733,0.911);
    \draw[iteration edge,preaction={draw=white,line width=1.3pt}]
      (2.000,3.189) -- (2.456,3.189) (2.456,3.189) -- (2.765,3.189) (2.765,3.189) -- (3.367,3.189)
      (3.367,3.189) -- (3.822,3.189) (3.822,3.189) -- (4.278,3.189) (4.278,3.189) -- (4.733,3.189)
      (4.733,3.189) -- (5.189,3.189) (2.000,0.000) -- (2.456,0.000) (2.456,0.000) -- (2.911,0.000)
      (2.911,0.000) -- (3.367,0.000) (3.367,0.000) -- (3.822,0.000) (3.822,0.000) -- (4.278,0.000)
      (4.278,0.000) -- (4.733,0.000) (4.733,0.000) -- (5.189,0.000) (2.911,4.100) -- (3.367,4.100)
      (3.367,4.100) -- (3.822,4.100) (3.822,4.100) -- (4.278,4.100) (4.278,4.100) -- (4.733,4.100)
      (4.733,4.100) -- (5.189,4.100) (5.189,4.100) -- (5.644,4.100) (5.644,4.100) -- (6.100,4.100)
      (2.911,0.911) -- (3.367,0.911) (3.367,0.911) -- (3.822,0.911) (3.822,0.911) -- (4.278,0.911)
      (4.278,0.911) -- (4.733,0.911) (4.733,0.911) -- (5.043,0.911) (5.043,0.911) -- (5.644,0.911)
      (5.644,0.911) -- (6.100,0.911);
    \draw[iteration edge,blue!70!black,preaction={draw=white,line width=1.3pt}]
      (2.000,1.822) -- (2.456,1.822) (2.456,1.822) -- (2.911,1.822) (2.000,2.733) -- (2.456,2.733)
      (2.456,2.733) -- (2.911,2.733) (5.189,1.367) -- (5.644,1.367) (5.644,1.367) -- (6.100,1.367)
      (5.189,2.278) -- (5.644,2.278) (5.644,2.278) -- (6.100,2.278);
    \foreach \point in {
      (2.000,0.000),(5.189,3.189),(2.000,0.911),(2.000,1.822),(2.000,2.733),(2.456,3.189),
      (3.367,3.189),(4.278,3.189),(5.189,0.456),(5.189,1.367),(5.189,2.278),(2.911,0.000),
      (3.822,0.000),(4.733,0.000),(2.911,0.911),(6.100,4.100),(2.911,1.822),(2.911,2.733),
      (2.911,3.644),(3.367,4.100),(4.278,4.100),(5.189,4.100),(6.100,1.367),(6.100,2.278),
      (6.100,3.189),(3.822,0.911),(4.733,0.911),(5.644,0.911)}
      \node[iteration spin] at \point {};
    \foreach \point in {
      (2.000,3.189),(5.189,0.000),(2.000,0.456),(2.000,1.367),(2.000,2.278),(2.765,3.189),
      (3.822,3.189),(4.733,3.189),(5.189,1.057),(5.189,1.822),(5.189,2.733),(2.456,0.000),
      (3.367,0.000),(4.278,0.000),(2.911,4.100),(6.100,0.911),(2.911,1.367),(2.911,2.278),
      (2.911,3.335),(3.822,4.100),(4.733,4.100),(5.644,4.100),(6.100,1.822),(6.100,2.733),
      (6.100,3.644),(3.367,0.911),(4.278,0.911),(5.043,0.911)}
      \node[iteration check] at \point {};
    \foreach \point in {
      (2.456,1.822),(2.456,2.733),(3.367,3.644),(4.278,3.644),(5.644,1.367),(5.644,2.278),
      (3.822,0.456),(4.733,0.456)}
      \node[iteration check,draw=blue!70!black] at \point {};
    % C_(0); crossing vertices remain distinct.
    \draw[iteration edge]
      (8.000,1.725) -- (8.000,2.375) (8.650,2.375) -- (8.650,1.725);
    \draw[iteration edge,preaction={draw=white,line width=1.3pt}]
      (8.000,2.375) -- (8.650,2.375) (8.650,1.725) -- (8.000,1.725);
    \foreach \point in {
      (8.000,1.725),(8.650,2.375)}
      \node[iteration spin] at \point {};
    \foreach \point in {
      (8.000,2.375),(8.650,1.725)}
      \node[iteration check] at \point {};
    % C_(1); crossing vertices remain distinct.
    \draw[iteration edge]
      (10.000,0.000) -- (10.000,0.456) (10.000,0.456) -- (10.000,0.911) (10.000,0.911) -- (10.000,1.367)
      (10.000,1.367) -- (10.000,1.822) (10.000,1.822) -- (10.000,2.278) (10.000,2.278) -- (10.000,2.733)
      (10.000,2.733) -- (10.000,3.189) (13.189,0.000) -- (13.189,0.456) (13.189,0.456) -- (13.189,1.057)
      (13.189,1.057) -- (13.189,1.367) (13.189,1.367) -- (13.189,1.822) (13.189,1.822) -- (13.189,2.278)
      (13.189,2.278) -- (13.189,2.733) (13.189,2.733) -- (13.189,3.189) (10.911,0.911) -- (10.911,1.367)
      (10.911,1.367) -- (10.911,1.822) (10.911,1.822) -- (10.911,2.278) (10.911,2.278) -- (10.911,2.733)
      (10.911,2.733) -- (10.911,3.335) (10.911,3.335) -- (10.911,3.644) (10.911,3.644) -- (10.911,4.100)
      (14.100,0.911) -- (14.100,1.367) (14.100,1.367) -- (14.100,1.822) (14.100,1.822) -- (14.100,2.278)
      (14.100,2.278) -- (14.100,2.733) (14.100,2.733) -- (14.100,3.189) (14.100,3.189) -- (14.100,3.644)
      (14.100,3.644) -- (14.100,4.100);
    \draw[iteration edge,blue!70!black]
      (11.822,3.189) -- (11.822,3.644) (11.822,3.644) -- (11.822,4.100) (12.733,3.189) -- (12.733,3.644)
      (12.733,3.644) -- (12.733,4.100) (11.367,0.000) -- (11.367,0.456) (11.367,0.456) -- (11.367,0.911)
      (12.278,0.000) -- (12.278,0.456) (12.278,0.456) -- (12.278,0.911);
    \draw[iteration edge,preaction={draw=white,line width=1.3pt}]
      (10.000,3.189) -- (10.456,3.189) (10.456,3.189) -- (10.765,3.189) (10.765,3.189) -- (11.367,3.189)
      (11.367,3.189) -- (11.822,3.189) (11.822,3.189) -- (12.278,3.189) (12.278,3.189) -- (12.733,3.189)
      (12.733,3.189) -- (13.189,3.189) (10.000,0.000) -- (10.456,0.000) (10.456,0.000) -- (10.911,0.000)
      (10.911,0.000) -- (11.367,0.000) (11.367,0.000) -- (11.822,0.000) (11.822,0.000) -- (12.278,0.000)
      (12.278,0.000) -- (12.733,0.000) (12.733,0.000) -- (13.189,0.000) (10.911,4.100) -- (11.367,4.100)
      (11.367,4.100) -- (11.822,4.100) (11.822,4.100) -- (12.278,4.100) (12.278,4.100) -- (12.733,4.100)
      (12.733,4.100) -- (13.189,4.100) (13.189,4.100) -- (13.644,4.100) (13.644,4.100) -- (14.100,4.100)
      (10.911,0.911) -- (11.367,0.911) (11.367,0.911) -- (11.822,0.911) (11.822,0.911) -- (12.278,0.911)
      (12.278,0.911) -- (12.733,0.911) (12.733,0.911) -- (13.043,0.911) (13.043,0.911) -- (13.644,0.911)
      (13.644,0.911) -- (14.100,0.911);
    \draw[iteration edge,blue!70!black,preaction={draw=white,line width=1.3pt}]
      (10.000,1.367) -- (10.456,1.367) (10.456,1.367) -- (10.911,1.367) (10.000,2.278) -- (10.456,2.278)
      (10.456,2.278) -- (10.911,2.278) (13.189,1.822) -- (13.644,1.822) (13.644,1.822) -- (14.100,1.822)
      (13.189,2.733) -- (13.644,2.733) (13.644,2.733) -- (14.100,2.733);
    \foreach \point in {
      (10.000,0.000),(13.189,3.189),(10.000,0.911),(10.000,1.822),(10.000,2.733),(10.456,3.189),
      (11.367,3.189),(12.278,3.189),(13.189,0.456),(13.189,1.367),(13.189,2.278),(10.911,0.000),
      (11.822,0.000),(12.733,0.000),(10.911,0.911),(14.100,4.100),(10.911,1.822),(10.911,2.733),
      (10.911,3.644),(11.367,4.100),(12.278,4.100),(13.189,4.100),(14.100,1.367),(14.100,2.278),
      (14.100,3.189),(11.822,0.911),(12.733,0.911),(13.644,0.911)}
      \node[iteration spin] at \point {};
    \foreach \point in {
      (10.456,1.367),(10.456,2.278),(11.822,3.644),(12.733,3.644),(13.644,1.822),(13.644,2.733),
      (11.367,0.456),(12.278,0.456)}
      \node[iteration spin,fill=blue!70!black] at \point {};
    \foreach \point in {
      (10.000,3.189),(13.189,0.000),(10.000,0.456),(10.000,1.367),(10.000,2.278),(10.765,3.189),
      (11.822,3.189),(12.733,3.189),(13.189,1.057),(13.189,1.822),(13.189,2.733),(10.456,0.000),
      (11.367,0.000),(12.278,0.000),(10.911,4.100),(14.100,0.911),(10.911,1.367),(10.911,2.278),
      (10.911,3.335),(11.822,4.100),(12.733,4.100),(13.644,4.100),(14.100,1.822),(14.100,2.733),
      (14.100,3.644),(11.367,0.911),(12.278,0.911),(13.043,0.911)}
      \node[iteration check] at \point {};
    \draw[->] (0.9,2.05) -- node[above,font=\small]
      {$\cR_{\mathrm{Rep}}$} (1.7,2.05);
    \draw[->] (8.9,2.05) -- node[above,font=\small]
      {$\cR_{\mathrm{Par}}$} (9.7,2.05);
  \end{tikzpicture}
  \caption{The replacement rules $\cR_{\mathrm{Rep}}$ (left) and
    $\cR_{\mathrm{Par}}$ (right), shown separately for the basic square.
    Filled circles denote spins and open circles denote the parity checks. If the check is connected to spins $\{i_1,..., i_s\}$, the corresponding interaction term is proportional to $-Z_{i_1}...Z_{i_s}$. 
    Here $\ell=7$: each incidence edge is subdivided into
    $7$ segments and is turned into a ladder with 2 rungs.
    Rungs are shown in blue; each contains a check in the Rep rule and a spin
    in the Par rule. 
    The two refined copies differ by the integer
    shift $(2,2)$; the overlapping vertices are displaced slightly for clarity. Crossing wires are disjoint.}
  \label{fig:replacement-rules}
\end{figure}

\begin{figure}[h]
  \centering
  \input{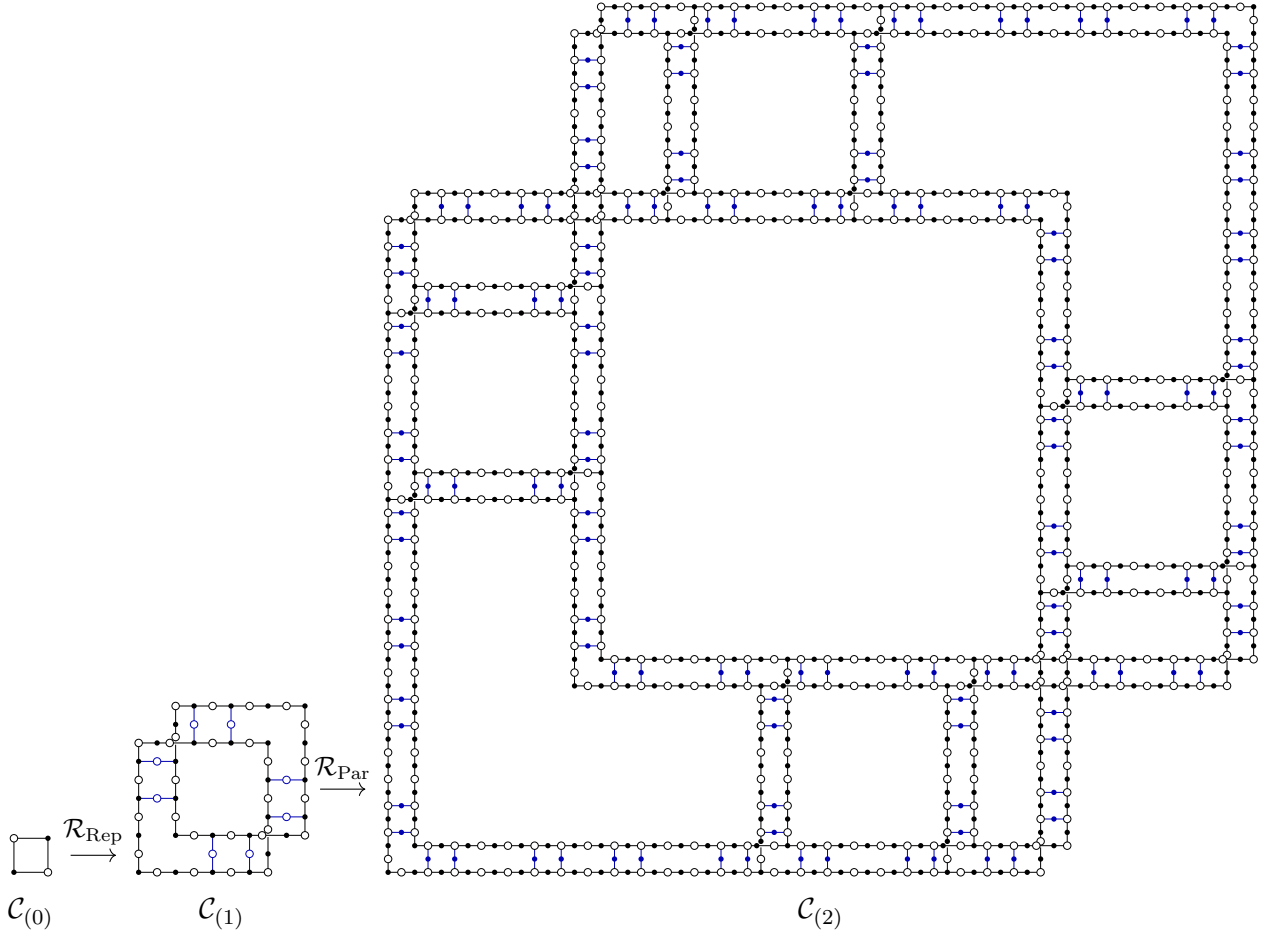}
  \caption{The first two iterations, $\cC_{(0)}$,
    $\cC_{(1)}=\cR_{\mathrm{Rep}}(\cC_{(0)})$, and
    $\cC_{(2)}=\cR_{\mathrm{Par}}(\cC_{(1)})$, using the spin/check
    convention of Figure~\ref{fig:replacement-rules}.
    At each step, $\ell=7$ alternating segments replace every
    incidence edge before doubling, and $2$ rungs
    are added per ladder. Blue highlights the latest rungs.
    The three examples are shown at different scales.}
  \label{fig:code-iterations}
\end{figure}

Figure~\ref{fig:replacement-rules} illustrates the two replacement rules, and
  Figure~\ref{fig:code-iterations} shows the first two iterations.
We will discuss them in more detail in the following subsections.
% Generated by draw_code_iterations.py --figure rules --ell 7.

% Generated by draw_code_iterations.py --figure iterations --ell 7.

\subsection{Classical gauge theory as a cochain complex}

At level $i$, the code is described by a 3-term cochain complex over $\ff_2$:
\begin{equation}
  \cC_{(i)}:\ C_{(i)}^0
  \overset{\delta_{(i)}^0}{\longrightarrow} C_{(i)}^1
  \overset{\delta_{(i)}^1}{\longrightarrow} C_{(i)}^2,
\end{equation}
interpreted as
\begin{equation}
  \text{symmetry generators}
  \longrightarrow
  \text{bits}
  \longrightarrow
  \text{checks}.
\end{equation}
In particular, let $\Lambda_{G,(i)}$, $\Lambda_{B,(i)}$, and $\Lambda_{C,(i)}$ label
  the chosen symmetry generators, bits (spins), and checks, respectively:
\begin{equation}
  C_{(i)}^0=\ff_2^{\Lambda_{G,(i)}},\qquad
  C_{(i)}^1=\ff_2^{\Lambda_{B,(i)}},\qquad
  C_{(i)}^2=\ff_2^{\Lambda_{C,(i)}}.
\end{equation}
We use $g\in \Lambda_{G,(i)}$, $b\in \Lambda_{B,(i)}$, and $c\in \Lambda_{C,(i)}$ for labels in these sets and $g\in C_{(i)}^0$, $s\in C_{(i)}^1$, $\sigma\in C_{(i)}^2$ for the
  corresponding coefficient vectors.
For a spin configuration $s$, the syndrome is $\sigma=\delta_{(i)}^1s$.
We drop the subscript $(i)$ when the level is clear from context.

The Hamiltonian of the $i$-th code is defined through the number of violated checks:
\begin{equation}
  \cH_{0,(i)}(s)=|\delta_{(i)}^1s|=|\sigma|,
\end{equation}
where $|\sigma|$ denotes the Hamming weight of the syndrome.

A coefficient vector $g\in C_{(i)}^0$ defines the symmetry transformation
  $s\mapsto s+\delta^0g$, flipping the spins in
  $\operatorname{supp}(\delta^0g)$.
The identity $\delta^1\delta^0=0$ gives
\begin{equation}
  \delta^1(s+\delta^0g)=\delta^1s,
  \qquad
  \cH_0(s+\delta^0g)=\cH_0(s),
\end{equation}
so symmetry transformations preserve the syndrome and energy.

\subsubsection{The two logical states}
\label{sec:logical-states}

At every level, the zero-energy configurations modulo symmetry transformations correspond to non-trivial classes of the first cohomology group
\begin{equation}
  H^1(\cC_{(i)})=\ker\delta_{(i)}^1/\operatorname{im}\delta_{(i)}^0
    \cong\ff_2,
\end{equation}
whose two classes encode one classical bit.
We use $\alpha_{(i)}\in\ker\delta_{(i)}^1\setminus\operatorname{im}\delta_{(i)}^0$
  for a representative of the nontrivial class, defined recursively in Eq.~\eqref{eq:iterated-logical-class}.
The logical spin-flip transformation $s\mapsto s+\alpha_{(i)}$ exchanges
  these classes while preserving the syndrome and energy (under the unperturbed Hamiltonian).

\subsection{Tanner graph and embedding}
\label{sec:tanner-embedding}

The code $\cC_{(i)}$ has a bipartite Tanner graph $T_{(i)}$.
Its vertex and edge sets are
\begin{equation}
  \begin{gathered}
  T_{(i)}(0)=\Lambda_{B,(i)}\sqcup\Lambda_{C,(i)},\\
  T_{(i)}(1)=\{(b,c)\in\Lambda_{B,(i)}\times\Lambda_{C,(i)}:
    (\delta_{(i)}^1)_{cb}=1\}.
  \end{gathered}
  \label{eq:ladder-tanner-graph}
\end{equation}
Thus $\delta_i^1$ is its biadjacency matrix.
Note that the Tanner graph only contains the information about the spins and checks and not the symmetry generators.

An \emph{embedding} $I_{(i)}:T_{(i)}\to\RR^2$ assigns vertex positions
  and extends linearly along edges.
We call the embeddings local if
  there exist constants $L,M$, independent of $i$, satisfy
\begin{equation}
  \begin{gathered}
  |I_{(i)}(b)-I_{(i)}(c)|\le L\quad \forall\, (b,c)\in T_{(i)}(1),\\
  |\{v\in T_{(i)}(0):I_{(i)}(v)\in \textnormal{Ball}_{1/2}(x)\}|\le M\quad \forall x\in\RR^2,
  \end{gathered}
  \label{eq:ladder-locality-definition}
\end{equation}
where $\textnormal{Ball}_{1/2}(x)$ is the open ball with radius $1/2$ centered at $x$.
The two bounds control interaction range and spatial density.
Our embeddings satisfies $L = 1$ and $M = 2$,
  which are proved in Sec.~\ref{sec:ladder-geometry}.

\subsection{Construction: Base code $\cC_{(0)}$}
\label{sec:ladder-base-code}

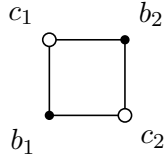
\begin{figure}[htbp]
  \centering
  \begin{tikzpicture}[x=1cm,y=1cm,
  base spin/.style={circle,fill=black,inner sep=0pt,minimum size=3.5pt},
  base check/.style={circle,draw=black,fill=white,line width=0.6pt,
    inner sep=0pt,minimum size=5pt}]
  \draw[line width=0.6pt] (0,0) -- (0,1) -- (1,1) -- (1,0) -- cycle;
  \node[base spin,label=below left:{$b_1$}] at (0,0) {};
  \node[base spin,label=above right:{$b_2$}] at (1,1) {};
  \node[base check,label=above left:{$c_1$}] at (0,1) {};
  \node[base check,label=below right:{$c_2$}] at (1,0) {};
\end{tikzpicture}
  \caption{The base code $\cC_{(0)}$: two spins (filled circles) and
    two identical parity checks (open circles).  Note that the two parity checks are completely redundant, but we include both for symmetry and to aid further construction. }
  \label{fig:base-code}
\end{figure}

There are two spins $b_1,b_2$ and two checks $c_1,c_2$, with
\begin{equation}
  \cC_{(0)}:\quad 0\xrightarrow{\ 0 \ }\ff_2^2
    \xrightarrow{\delta_{(0)}^1}\ff_2^2,
  \qquad
  \delta_{(0)}^1=\begin{pmatrix}1&1\\1&1\end{pmatrix}.
  \label{eq:ladder-base-code}
\end{equation}
Embedding this code in $\RR^2$ is trivial; place $b_1,b_2$ at $(0,0),(1,1)$ and $c_1,c_2$ at $(0,1),(1,0)$,
  giving the Tanner graph in Figure~\ref{fig:base-code}.

 There are no symmetry generators at this level, so the zero-energy configurations
  $(0,0)$ and $(1,1)$ give $H^1(\cC_{(0)})\cong\ff_2$, with
  $\alpha_{(0)}=(1,1)$.

\subsection{Construction: Replacement process: $\cR_{\mathrm{Rep}}$ and $\cR_{\mathrm{Par}}$}
\label{sec:ladder-replacements}

The goal of this section is to define the two replacement operations $\cR_{\mathrm{Rep}}$ and $\cR_{\mathrm{Par}}$,
  which recursively transform a code $\cC_{(i)}$ into a new code $\cC_{(i+1)}$.
We provide the graph replacement followed by its cochain formulation,
  following the organization of Sec.~3 of Ref.~\cite{balasubramanian2026passive}.
The geometry follows the ladder structure in Sec.~6.1 of Ref.~\cite{balasubramanian2026passive}, but our
  vertices label spins and checks, not repetition and parity junctions.

\subsubsection{Replacement in the graph framework}
\label{sec:ladder-graph-rule}

The elementary replacement acts on a single bit--check segment
  $e=(b,c)$, turning it into a ladder with unequal ends.
Figure~\ref{fig:segment-replacements} below shows the two local rules and
  Figure~\ref{fig:replacement-rules} applies them to all each edge
  of the base code.
The replacement depends on the chosen embedding
  $I_{(i)}:T_{(i)}\to\RR^2$.
It acts on each embedded incidence edge
  and thereby determines $I_{(i+1)}$.

% Generated by draw_code_iterations.py --figure segments --ell 7.
\begin{figure}[htbp]
  \centering
  \input{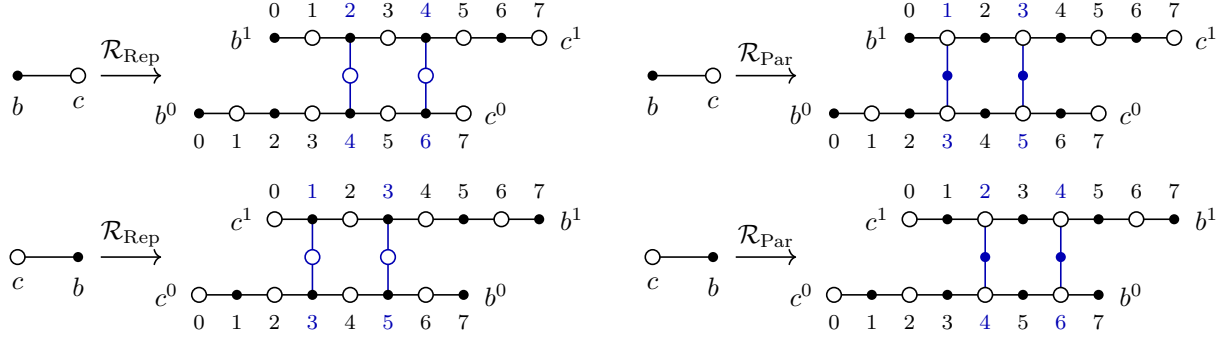}
  \caption{Replacement of a single bit--check segment by
    $\cR_{\mathrm{Rep}}$ (left) or $\cR_{\mathrm{Par}}$ (right),
    with bit--check inputs above and check--bit inputs below.
    Here $\ell=7$. Each rail has $7$ alternating incidence segments;
    the two copies differ by the shift $(2,2)$.
    The $2$ blue rungs have check midpoints for $\cR_{\mathrm{Rep}}$ and bit midpoints for $\cR_{\mathrm{Par}}$.
    Filled circles denote bits and open circles checks.
    Input segments are drawn at a different scale.}
  \label{fig:segment-replacements}
\end{figure}

The above Figure gives a well-defined replacement rule. Below, we provide an alternative definition of how to obtain this replacement constructively; the reader should be free to skip it if they find the Figure-based definition satisfactory.

Fix an odd integer $\ell\ge7$. The edge to be replaced is either horizontal or vertical. 
Scale it by
  $\ell$ and subdivide each segment into $\ell$ alternating incidence edges (by this, we mean edges where one end is a bit and the other is a check). Oddness of $\ell$ preserves the endpoint types.
Then duplicate the full subdivided structure, and shift one copy by $(2,2)$ in the lattice.
The two copies form the rails of the ladder.

To construct the rungs, number the vertices along each rail by $0,1,\ldots,\ell$ rightward for
  horizontal rails and upward for vertical rails, regardless of endpoint
  type.
Vertex types alternate along the rails.
For selected indices $j$ with $3\le j\le\ell-1$, a rung joins vertex $j$ of the stationary rail
  to vertex $j-2$ of the shifted rail.
These vertices have the same type
  and are separated by two lattice units perpendicular to the rails.
For $\cR_{\mathrm{Rep}}$, select the indices $j$ whose vertices are bits
  and insert a check at each rung midpoint, giving bit--check--bit rungs.
For $\cR_{\mathrm{Par}}$, select those whose vertices are checks
  and insert a bit at each midpoint, giving check--bit--check rungs.

In either rule there are $(\ell-3)/2$ rungs per ladder.
Thus, $\ell=7$ gives the two rungs shown in the figures.

  The replacement description gives a straightforward definition of the code and its embedding.
However, to complete the code definition, we also need to specify the symmetry generators and their action, which are encoded by $C^0$ and $\delta^0$. These data are essential to the proof of perturbative stability and are supplied by the cochain formulation in the following subsections.

\subsubsection{Review: spectral sequences and code embedding}
\label{sec:spectral-sequence-review}

We summarize the comparison argument of
Ref.~\cite[Secs.~2.3]{balasubramanian2026passive}.
Our labels $(G,B,C)$ correspond to $(X,Q,Z)$ in that reference.
Over $\ff_2$, take finite-dimensional local cochain complexes
\begin{equation}
  \cC_g:C_g^0\longrightarrow C_g^1\longrightarrow C_g^2,
  \qquad
  \cC_b:C_b^0\longrightarrow C_b^1\longrightarrow C_b^2,
  \qquad
  \cC_c:C_c^0\longrightarrow C_c^1\longrightarrow C_c^2.
  \label{eq:review-individual-complexes}
\end{equation}
Lowercase labels $g,b,c$ denote individual pieces; uppercase labels
$G,B,C$ denote their direct sums:
\begin{equation}
  C_G^\bullet=\bigoplus_{g\in\Lambda_G} C_g^\bullet,
  \qquad C_B^\bullet=\bigoplus_{b\in\Lambda_B} C_B^\bullet,
  \qquad C_C^\bullet=\bigoplus_{c\in\Lambda_C} C_c^\bullet.
  \label{eq:review-aggregate-pieces}
\end{equation}
The global complex $\cC=(C,\delta)$ has the form
\begin{equation}
  \cC:\ C_G^0\oplus C_B^0\oplus C_C^0
    \xrightarrow{\delta^0}C_G^1\oplus C_B^1\oplus C_C^1
    \xrightarrow{\delta^1}C_G^2\oplus C_B^2\oplus C_C^2.
  \label{eq:review-global-complex}
\end{equation}
For $m=0,1$, the local coboundary maps and connecting maps are
\begin{equation}
  \begin{gathered}
  \delta_g^m:C_g^m\to C_g^{m+1},\qquad
  \delta_b^m:C_B^m\to C_B^{m+1},\qquad
  \delta_c^m:C_c^m\to C_c^{m+1},\\
  \kappa_{gb}^m:C_g^m\to C_B^{m+1},\qquad
  \kappa_{bc}^m:C_B^m\to C_c^{m+1},\qquad
  \kappa_{gc}^m:C_g^m\to C_c^{m+1}.
  \end{gathered}
  \label{eq:review-local-connecting-maps}
\end{equation}
Each local differential squares to zero. Assembling these maps gives
$\delta=\delta_{\mathrm{loc}}+\kappa$, with $\delta^2=0$ required; the
connecting part $\kappa$ strictly increases the order $G<B<C$.
Connecting maps are denoted by $\kappa$ to distinguish them from generator labels $g$.
% Adapted from the height-two diagram in refs/memory-paper/paper.tex.
% The memory paper's (X,Q,Z) pieces are relabelled (G,B,C).
\begin{equation}
 \begin{tikzpicture}[baseline]
 \matrix (a) [
   matrix of math nodes, nodes in empty cells,
   nodes={minimum size=25pt},
   row sep=2em, column sep=2em,
   text height=1.25ex, text depth=0.25ex
 ] {
   {} & {} & \textcolor{blue}{C_G^0}
     & \textcolor{blue}{C_G^1} & \textcolor{blue}{C_G^2} \\
   {} & C_B^0 & C_B^1 & C_B^2 & {} \\
   \textcolor{red}{C_C^0} & \textcolor{red}{C_C^1}
     & \textcolor{red}{C_C^2} & {} & {} \\
 };
 \path[->,blue,font=\scriptsize]
   (a-1-3) edge node[above]{$\delta_G^0$} (a-1-4)
   (a-1-4) edge node[above]{$\delta_G^1$} (a-1-5);
 \path[->,font=\scriptsize]
   (a-2-2) edge node[above]{$\delta_B^0$} (a-2-3)
   (a-2-3) edge node[above]{$\delta_B^1$} (a-2-4);
 \path[->,red,font=\scriptsize]
   (a-3-1) edge node[above]{$\delta_C^0$} (a-3-2)
   (a-3-2) edge node[above]{$\delta_C^1$} (a-3-3);
 \path[->,font=\scriptsize]
   (a-1-3) edge node[right]{$\kappa_{GB}^0$} (a-2-3)
   (a-1-4) edge node[right]{$\kappa_{GB}^1$} (a-2-4)
   (a-2-2) edge node[right]{$\kappa_{BC}^0$} (a-3-2)
   (a-2-3) edge node[right]{$\kappa_{BC}^1$} (a-3-3);
 \path[->,dashed,font=\scriptsize]
   (a-1-3) edge[bend right=70]
     node[left]{$\kappa_{GC}^0$} (a-3-2)
   (a-1-4) edge[bend left=70]
     node[right]{$\kappa_{GC}^1$} (a-3-3);
 \end{tikzpicture}
 \label{eq:code-embedding-diagram}
\end{equation}
With $(A_0,A_1,A_2)=(G,B,C)$, the
first two pages are
\begin{equation}
  E_0^{p,q}=C_{A_q}^{p+q},
  \qquad E_1^{p,q}=H^{p+q}(C_{A_q},\delta_{A_q}),
  \qquad \operatorname{Tot}(E_1)^n=\bigoplus_{p+q=n}E_1^{p,q}.
\end{equation}
Successive pages take cohomology with respect to the leading
differential; the remaining terms are transferred by the homological
perturbation lemma.

The homological perturbation lemma gives a transferred differential
$d$ and a cochain-homotopy equivalence
\begin{equation}
  J:(\operatorname{Tot}(E_1),d)
    \xrightarrow{\simeq}(C,\delta).
\end{equation}
Here $d$ includes all higher transferred terms.

\subsubsection{Reduced complexes and comparison maps}
\label{sec:code-embedding-review}

For the code ${\cC}_{(i)}$, define the following two auxiliary modifications:
\begin{equation}
  \widehat{\cC}_{(i)}=
  \left[0\longrightarrow C_{(i)}^1
    \xrightarrow{\delta_{(i)}^1}C_{(i)}^2\right],
  \qquad
  \widetilde{\cC}_{(i+1)}=
  \left[\widetilde C_{(i+1)}^0
    \xrightarrow{\widetilde\delta_{(i+1)}^0}C_{(i+1)}^1
    \xrightarrow{\delta_{(i+1)}^1}C_{(i+1)}^2\right].
  \label{eq:hat-tilde-complexes}
\end{equation}
The hat complex omits the symmetry generators of $\cC_{(i)}$;
the tilde complex contains only the newly created generators (i.e. the ones at the smallest scale).
Their degree-one and degree-two spaces and differentials agree
with those of the respective full codes.

The tilde complex is assembled from local pieces
$\cC_G,\cC_B,\cC_C$ as in Section~\ref{sec:spectral-sequence-review},
with the level $(i+1)$ suppressed in the notation. Write
\begin{equation}
  H_{\mathrm{loc}}(\widetilde{\cC})
    =(\operatorname{Tot}(E_1),d)
\end{equation}
for its local cohomology equipped with the transferred differential
$d$, including any higher terms, rather than the zero differential. The point of introducing $ \widetilde{\cC}_{(i+1)}$ was that its differential will be exactly the local differential $d$ from the earlier construction. 

As we will see, for each replacement $\cR$, we first construct
\begin{equation}
  \widetilde{\cF}_{\cR,(i)}=J_{\cR,(i)}\iota_{\cR,(i)}:
  \widehat{\cC}_{(i)}\longrightarrow
  H_{\mathrm{loc}}(\widetilde{\cC}_{(i+1)})
  \xrightarrow{\simeq}\widetilde{\cC}_{(i+1)}
\end{equation}
and show that it induces an isomorphism on $H^1$.
The comparison $J_{\cR,(i)}$ need not be canonical.
We then restore the inherited symmetry generators to obtain
$\cF_{\cR,(i)}:\cC_{(i)}\to\cC_{(i+1)}$, which also induces an
isomorphism on $H^1$.

\subsubsection{$\cR_{\mathrm{Rep}}$ in code embedding formalism}
\label{sec:rep-local-formalism}

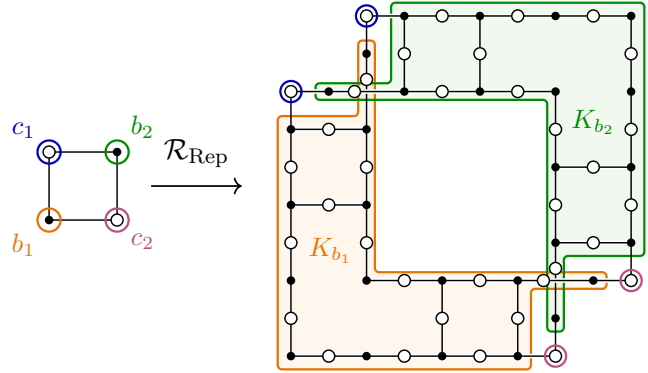
\begin{figure}[htbp]
  \centering
  \begin{tikzpicture}[x=1cm,y=1cm,
    rep edge/.style={line width=0.55pt},
    rep bit/.style={circle,fill=black,inner sep=0pt,minimum size=3.2pt},
    rep check/.style={circle,draw=black,fill=white,line width=0.55pt,
      inner sep=0pt,minimum size=4.5pt},
    rep outline/.style={line width=0.85pt,rounded corners=1.6pt},
    rep label/.style={font=\small,inner sep=2pt}]
    % Seed labels agree with Figure~\ref{fig:base-code}.
    \begin{scope}[xshift=0.1cm,yshift=1.8cm,x=0.9cm,y=0.9cm]
      \draw[rep edge] (0,0) -- (0,1) -- (1,1) -- (1,0) -- cycle;
      \draw[rep outline,orange!90!black] (0,0) circle[radius=0.17];
      \draw[rep outline,green!55!black] (1,1) circle[radius=0.17];
      \draw[rep outline,blue!75!black] (0,1) circle[radius=0.17];
      \draw[rep outline,magenta!75!black] (1,0) circle[radius=0.17];
      \node[rep bit] at (0,0) {};
      \node[rep bit] at (1,1) {};
      \node[rep check] at (0,1) {};
      \node[rep check] at (1,0) {};
      \node[rep label,below left=3pt,text=orange!90!black] at (0,0) {$b_1$};
      \node[rep label,above right=3pt,text=green!55!black] at (1,1) {$b_2$};
      \node[rep label,above left=3pt,text=blue!75!black] at (0,1) {$c_1$};
      \node[rep label,below right=3pt,text=magenta!75!black] at (1,0) {$c_2$};
    \end{scope}
    \draw[->,line width=0.65pt] (1.45,2.25)
      -- node[above=3pt] {$\cR_{\mathrm{Rep}}$} (2.65,2.25);

    \begin{scope}[xshift=3.3cm,x=0.5cm,y=0.5cm]
      % Symmetric contour arms separate the two vertices at each crossing.
      \path[rep outline,draw=orange!90!black,fill=orange,fill opacity=0.07]
        (-0.35,-0.35) -- (6.35,-0.35) -- (6.35,1.78)
        -- (8.35,1.78) -- (8.35,2.22) -- (2.22,2.22)
        -- (2.22,8.35)
        -- (1.78,8.35) -- (1.78,6.35) -- (-0.35,6.35) -- cycle;
      \path[rep outline,draw=green!55!black,fill=green!55!black,fill opacity=0.06]
        (9.35,9.35) -- (2.65,9.35) -- (2.65,7.22)
        -- (0.65,7.22) -- (0.65,6.78) -- (6.78,6.78)
        -- (6.78,0.65)
        -- (7.22,0.65) -- (7.22,2.65) -- (9.35,2.65) -- cycle;

      % Vertical rails and rungs.
      \draw[rep edge]
        (0,0) -- (0,7) (7,0) -- (7,7)
        (2,2) -- (2,9) (9,2) -- (9,9)
        (3,7) -- (3,9) (5,7) -- (5,9)
        (4,0) -- (4,2) (6,0) -- (6,2);
      % White underpasses indicate crossings without identifying vertices.
      \draw[rep edge,preaction={draw=white,line width=1.7pt}]
        (0,0) -- (7,0) (0,7) -- (7,7)
        (2,2) -- (9,2) (2,9) -- (9,9)
        (0,4) -- (2,4) (0,6) -- (2,6)
        (7,3) -- (9,3) (7,5) -- (9,5);

      \foreach \point in {
        (0,0),(7,7),(0,2),(0,4),(0,6),(1,7),(3,7),(5,7),
        (7,1),(7,3),(7,5),(2,0),(4,0),(6,0),
        (2,2),(9,9),(2,4),(2,6),(2,8),(3,9),(5,9),(7,9),
        (9,3),(9,5),(9,7),(4,2),(6,2),(8,2)}
        \node[rep bit] at \point {};
      \foreach \point in {
        (0,7),(7,0),(0,1),(0,3),(0,5),(1.68,7),(4,7),(6,7),
        (7,2.32),(7,4),(7,6),(1,0),(3,0),(5,0),
        (2,9),(9,2),(2,3),(2,5),(2,7.32),(4,9),(6,9),(8,9),
        (9,4),(9,6),(9,8),(3,2),(5,2),(6.68,2),
        (1,4),(1,6),(3,8),(5,8),(8,3),(8,5),(4,1),(6,1)}
        \node[rep check] at \point {};

      \foreach \point in {(0,7),(2,9)}
        \draw[rep outline,blue!75!black] \point circle[radius=0.28];
      \foreach \point in {(7,0),(9,2)}
        \draw[rep outline,magenta!75!black] \point circle[radius=0.28];
      \node[rep label,text=orange!90!black] at (1.05,2.8) {$K_{b_1}$};
      \node[rep label,text=green!55!black] at (8,6.2) {$K_{b_2}$};
    \end{scope}
  \end{tikzpicture}
  \caption{Local pieces of the Rep replacement of the seed code, with
    $\ell=7$. Orange and green outlines enclose $K_{b_1}$ and $K_{b_2}$;
    the blue and magenta pairs mark the two copied-check pieces.}
  \label{fig:rep-piece-decomposition}
\end{figure}

Apply the Rep rule with odd $\ell\ge7$ to $T_{(i)}$, assuming no
isolated vertices.
For each check
$c\in\Lambda_{C,(i)}$, the corresponding local complex $\cC_c: 0 \to 0 \to \ff_2^2$ consists of its two checks from the two copies in $\cC_{(i+1)}$.
When we remove these checks from the Tanner graph,
the remaining vertices form a disjoint union of connected pieces $K_b$, one for each bit $b\in\Lambda_{B,(i)}$.
We view the bits as vertices and the checks as edges for the local graph $K_b$.
Then, they form the local complexes $\cC_B: 0 \to \ff_2^{V(K_b)} \to \ff_2^{E(K_b)}$.
In particular,
\begin{equation}
  \cC_B=\left[0\longrightarrow\ff_2^{V(K_b)}
    \xrightarrow{\delta_b}\ff_2^{E(K_b)}\right],
  \qquad
  \cC_c=\left[0\longrightarrow0\longrightarrow\ff_2^2\right].
  \label{eq:rep-local-pieces}
\end{equation}
Figure~\ref{fig:rep-piece-decomposition} illustrates the decomposition
for the seed code.

% Rep artwork at ell=7, using the same coordinates and crossing offsets
% as fig-code-iterations.tex. Colors identify local pieces, not rung age.

Set $\cC_G=0$, $\cC_B=\bigoplus_{b\in\Lambda_{B,(i)}} \cC_B$, and
$\cC_C=\bigoplus_{c\in\Lambda_{C,(i)}} \cC_c$.
Let $\kappa_{BC}^1:C_B^1\to C_C^2$ sum the level-$(i+1)$ bits adjacent
to each copy of a check $c\in\Lambda_{C,(i)}$.
The assembled complex $\widetilde{\cC}_{(i+1)}$ is

% Layout adapted from refs/memory-paper/paper.tex, graded replacement diagrams.
% All local pieces are at level (i+1).
\begin{equation}
 \begin{tikzpicture}[baseline]
 \matrix (a) [
   matrix of math nodes, nodes in empty cells,
   nodes={minimum size=25pt},
   row sep=2em, column sep=2em,
   text height=1.25ex, text depth=0.25ex
 ] {
   {} & {} & \textcolor{blue}{0}
     & \textcolor{blue}{0} & \textcolor{blue}{0} \\
   {} & 0 & C_B^1 & C_B^2 & {} \\
   \textcolor{red}{0}
     & \textcolor{red}{0}
     & \textcolor{red}{C_C^2} & {} & {} \\
 };
 \path[->,blue,font=\scriptsize]
   (a-1-3) edge node[above]{$\delta_G^0$} (a-1-4)
   (a-1-4) edge node[above]{$\delta_G^1$} (a-1-5);
 \path[->,font=\scriptsize]
   (a-2-2) edge node[above]{$\delta_B^0$} (a-2-3)
   (a-2-3) edge node[above]{$\delta_B^1$} (a-2-4);
 \path[->,red,font=\scriptsize]
   (a-3-1) edge node[above]{$\delta_C^0$} (a-3-2)
   (a-3-2) edge node[above]{$\delta_C^1$} (a-3-3);
 \path[->,font=\scriptsize]
   (a-1-3) edge node[right]{$\kappa_{GB}^0$} (a-2-3)
   (a-1-4) edge node[right]{$\kappa_{GB}^1$} (a-2-4)
   (a-2-2) edge node[right]{$\kappa_{BC}^0$} (a-3-2)
   (a-2-3) edge node[right]{$\kappa_{BC}^1$} (a-3-3);
 \end{tikzpicture}
 \label{eq:rep-local-complexes}
\end{equation}
Explicitly,
\begin{equation}
  \widetilde{\cC}_{(i+1)}:
  0\longrightarrow C_B^1
    \longrightarrow C_B^2\oplus C_C^2.
\end{equation}
No symmetry generators are introduced.

\subsubsection{Properties of the code $\cC_{(i+1)} = \cR_{\mathrm{Rep}}(\cC_{(i)})$}
\label{sec:rep-comparison}

We now analyze the properties of the code after $\cR_{\mathrm{Rep}}$ is applied.
The key result is a cochain map
$\widetilde{\cF}_{\cR_{\mathrm{Rep}},(i)}:\widehat{\cC}_{(i)}\to\widetilde{\cC}_{(i+1)}$
that induces an isomorphism on $H^1$.
Restoring the inherited symmetry generators extends this map to a
cochain map
$\cF_{\cR_{\mathrm{Rep}},(i)}:\cC_{(i)}\to\cC_{(i+1)}$
that also induces an isomorphism on $H^1$, as shown in
Corollary~\ref{cor:restore-inherited-generators}.

\begin{lemma}[Rep local cohomology]
\label{lem:rep-local-complexes}
For each $b\in\Lambda_{B,(i)}$ and $c\in\Lambda_{C,(i)}$, the local
complexes in \eqref{eq:rep-local-pieces} satisfy
\begin{equation}
  H^1(\cC_B,\delta_b)\cong\ff_2,
  \qquad
  H^2(\cC_c,\delta_c)\cong\ff_2^2.
  \label{eq:rep-local-cohomology-identifications}
\end{equation}
The group $H^1(\cC_B,\delta_b)$ is generated by the class
$[\mathbf1_b]$ of the all-ones configuration on $V(K_b)$.
\end{lemma}
\begin{proof}
It is clear that $K_b$ is connected.
For $s\in C_B^1$ and any edge
$e=\{u,v\}$ of $K_b$, we have $(\delta_b s)_e=s_u+s_v$.
Thus $\ker\delta_b$ consists exactly of the constant configurations.
Since $C_B^0=0$, it follows that
$H^1(\cC_B,\delta_b)\cong\ff_2$.

Finally, $\cC_c$ is concentrated in degree two, with
$C_c^2=\ff_2^2$, so $H^2(\cC_c,\delta_c)\cong\ff_2^2$.
\end{proof}
Taking direct sums gives the following local-cohomology spaces.

% Layout adapted from refs/memory-paper/paper.tex, local-cohomology diagrams.
% Wider central cells accommodate our parenthesized level indices.
\begin{equation}
 \begin{adjustbox}{max width=\linewidth}
 \begin{tikzpicture}[baseline]
 \def\localcohomfont{\fontsize{10}{9.5}\selectfont}
 \def\localarrowfont{\fontsize{7}{9.5}\selectfont}
 \matrix (a) [
   matrix of math nodes, nodes in empty cells,
   row sep=2em, column sep=3.2em,
   nodes={minimum size=25pt,align=center,text width=2cm,font=\localcohomfont}
 ] {
   {} & {}
     & |[text width=3.4cm]| \textcolor{blue}{
         0}
     & \textcolor{blue}{0} & \textcolor{blue}{0} \\
   {} & 0
     & |[text width=3.4cm]| {H^1(\cC_B,\delta_B)\cong\ff_2^{\Lambda_{B,(i)}}}
     & {H^2(\cC_B,\delta_B)} & {} \\
   \textcolor{red}{0}
     & \textcolor{red}{0}
     & |[text width=3.7cm]| \textcolor{red}{H^2(\cC_C,\delta_C)\cong(\ff_2^2)^{\Lambda_{C,(i)}}}
     & {} & {} \\
 };
 \path[->,font=\localarrowfont]
   (a-1-3.south) edge node[right]{$H(\kappa_{GB}^0)$} (a-2-3.north)
   (a-1-4.south) edge node[right]{$H(\kappa_{GB}^1)$} (a-2-4.north)
   (a-2-2.south) edge node[left]{$H(\kappa_{BC}^0)$} (a-3-2.north)
   (a-2-3.south) edge node[left]{$H(\kappa_{BC}^1)$} (a-3-3.north);
 \path[->,dashed]
   ([xshift=-4pt]a-1-3.south) edge ([xshift=4pt]a-3-2.north)
   ([xshift=-4pt]a-1-4.south) edge ([xshift=4pt]a-3-3.north);
 \end{tikzpicture}
 \end{adjustbox}
 \label{eq:rep-local-cohomology}
\end{equation}

\begin{lemma}[Rep induced complex]
\label{lem:rep-induced-complex}
Identifying $H^1(\cC_B,\delta_B)$ with $C_{(i)}^1$ and
$H^2(\cC_C,\delta_C)$ with $C_{(i)}^2\oplus C_{(i)}^2$, we obtain
\begin{equation}
  H_{\mathrm{loc}}(\widetilde{\cC}_{(i+1)})=
  \left[0\longrightarrow C_{(i)}^1
    \xrightarrow{s\mapsto(\delta_{(i)}^1s,\delta_{(i)}^1s,0)}
    (C_{(i)}^2\oplus C_{(i)}^2)\oplus H^2(\cC_B,\delta_B)\right].
  \label{eq:rep-induced-complex}
\end{equation}
Furthermore, there exists a cochain-homotopy equivalence
$J_{\mathrm{Rep},(i)}:
H_{\mathrm{loc}}(\widetilde{\cC}_{(i+1)})
\to\widetilde{\cC}_{(i+1)}$
with
\[
  J_{\mathrm{Rep},(i)}^1(s)=\sum_{b\in\Lambda_{B,(i)}}s_b\mathbf1_b,
  \qquad s\in C_{(i)}^1.
\]
\end{lemma}
\begin{proof}
Represent $s\in C_{(i)}^1$ by the configuration that is constant
with value $s_b$ on each $K_b$. Its internal checks vanish, while
each copy of a check $c\in\Lambda_{C,(i)}$ has syndrome
$\sum_{b\sim c}s_b=(\delta_{(i)}^1s)_c$. Thus the induced
differential $H(\kappa_{BC}^1)$ is $s \mapsto (\delta_{(i)}^1s,\delta_{(i)}^1s)$.

On the first page $E_1^{p,q}$, only a few entries are nonzero,
  so all higher differentials vanish.
The comparison construction in Section~\ref{sec:spectral-sequence-review}
therefore gives the stated cochain-homotopy equivalence with
the displayed degree-one map.
\end{proof}

\begin{lemma}[Rep comparison map]
\label{lem:rep-comparison-map}
Fix a comparison $J_{\mathrm{Rep},(i)}$ as above. The maps
\begin{equation}
  \iota_{\mathrm{Rep},(i)}^1(s)=s,
  \qquad
  \iota_{\mathrm{Rep},(i)}^2(\sigma)=(\sigma,\sigma,0)
\end{equation}
define a cochain inclusion into \eqref{eq:rep-induced-complex}, and
\begin{equation}
  \widetilde{\cF}_{\cR_{\mathrm{Rep}},(i)}
  =J_{\mathrm{Rep},(i)}\iota_{\mathrm{Rep},(i)}:
  \widehat{\cC}_{(i)}\lhook\joinrel\longrightarrow
  H_{\mathrm{loc}}(\widetilde{\cC}_{(i+1)})
  \xrightarrow{\simeq}\widetilde{\cC}_{(i+1)}
  \label{eq:rep-comparison-map}
\end{equation}
induces an isomorphism on $H^1$.
\end{lemma}
\begin{proof}
The differential in \eqref{eq:rep-induced-complex} sends
$\iota_{\mathrm{Rep},(i)}^1(s)$ to
$(\delta_{(i)}^1s,\delta_{(i)}^1s,0)
=\iota_{\mathrm{Rep},(i)}^2\delta_{(i)}^1s$.
Thus $\iota_{\mathrm{Rep},(i)}$ is a cochain map, and its components
are injective.
Hence it induces an $H^1$ isomorphism, as does its composition
with the cochain-homotopy equivalence $J_{\mathrm{Rep},(i)}$.
\end{proof}

Restore the inherited symmetry generators by setting
\begin{equation}
  C_{(i+1)}^0=C_{(i)}^0,
  \qquad
  \delta_{(i+1)}^0
    =\widetilde{\cF}_{\cR_{\mathrm{Rep}},(i)}^1\delta_{(i)}^0,
  \label{eq:rep-restored-generators}
\end{equation}
and retaining the degree-one and degree-two spaces and differential
of $\widetilde{\cC}_{(i+1)}$. Define
\begin{equation}
  \cF_{\cR_{\mathrm{Rep}},(i)}^0=\mathrm{id},
  \qquad
  \cF_{\cR_{\mathrm{Rep}},(i)}^m
    =\widetilde{\cF}_{\cR_{\mathrm{Rep}},(i)}^m
    \quad(m=1,2).
\end{equation}

\begin{cor}[Rep full-code comparison]
\label{cor:rep-full-code-comparison}
The restored complex satisfies $\delta_{(i+1)}^1\delta_{(i+1)}^0=0$,
and the map
\begin{equation}
  \cF_{\cR_{\mathrm{Rep}},(i)}:
    \cC_{(i)}\longrightarrow\cC_{(i+1)}
  \label{eq:rep-full-code-comparison}
\end{equation}
is a cochain map inducing $H^1(\cC_{(i)})\cong H^1(\cC_{(i+1)})\cong\ff_2$.

In particular,
  if $\alpha_{(i)}$ represents the nontrivial class in
  $H^1(\cC_{(i)})$, then
  \begin{equation}
    \alpha_{(i+1)}
    =\cF_{\cR_{\mathrm{Rep}},(i)}^1\alpha_{(i)}
  \end{equation}
  also represents the nontrivial class in $H^1(\cC_{(i+1)})$.
\end{cor}

\begin{proof}
The cochain identities follow directly from
\eqref{eq:rep-restored-generators} and those of the reduced map.
By Lemma~\ref{lem:rep-comparison-map}, $\widetilde{\cF}_{\cR_{\mathrm{Rep}},(i)}^1$ is an isomorphism $\ker\delta^1_{(i)}\to\ker\delta^1_{(i+1)}$. By \eqref{eq:rep-restored-generators}, this isomorphism maps $\operatorname{im}\delta^0_{(i)}$ to $\operatorname{im}\delta^0_{(i+1)}$.
So it induces an isomorphism $H^1(\cC_{(i)})\to H^1(\cC_{(i+1)})$, which transports the logical class.
\end{proof}

\subsubsection{$\cR_{\mathrm{Par}}$ in code embedding formalism}
\label{sec:par-local-formalism}

Apply the Par rule with odd $\ell\ge7$ to $T_{(i)}$, assuming no
isolated vertices.
For each bit $b\in\Lambda_{B,(i)}$, the corresponding local complex $\cC_B: 0\to\ff_2^2\to0$ consists of its two copied bits in $\cC_{(i+1)}$.
When we remove these bits from the Tanner graph,
  the remaining vertices form a disjoint union of connected pieces
  $K_c$, one for each check $c\in\Lambda_{C,(i)}$.
We view the checks as vertices and the bits as edges for the local graph $K_c$.
This defines the linear map $\delta_c^1: \ff_2^{E(K_c)}\to\ff_2^{V(K_c)}$,
  which sends an edge to the sum of its endpoints.

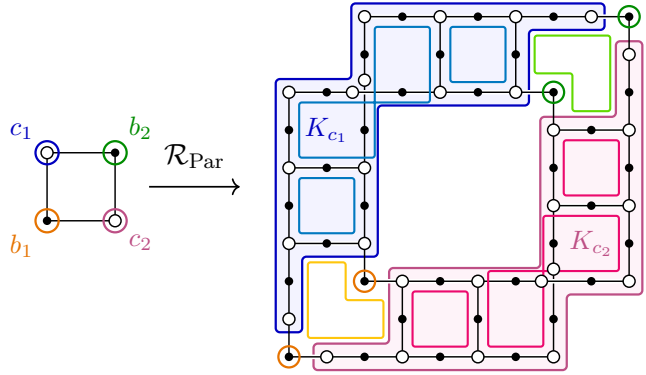
\begin{figure}[htbp]
  \centering
  \begin{tikzpicture}[x=1cm,y=1cm,
    par edge/.style={line width=0.55pt},
    par bit/.style={circle,fill=black,inner sep=0pt,minimum size=3.2pt},
    par check/.style={circle,draw=black,fill=white,line width=0.55pt,
      inner sep=0pt,minimum size=4.5pt},
    par outline/.style={line width=0.85pt,rounded corners=1.6pt},
    par label/.style={font=\small,inner sep=2pt},
    par cycle/.style={line width=0.8pt,rounded corners=1pt}]
    \begin{scope}[xshift=0.1cm,yshift=1.8cm,x=0.9cm,y=0.9cm]
      \draw[par edge] (0,0) -- (0,1) -- (1,1) -- (1,0) -- cycle;
      \draw[par outline,orange!90!black] (0,0) circle[radius=0.17];
      \draw[par outline,green!55!black] (1,1) circle[radius=0.17];
      \draw[par outline,blue!75!black] (0,1) circle[radius=0.17];
      \draw[par outline,magenta!75!black] (1,0) circle[radius=0.17];
      \node[par bit] at (0,0) {};
      \node[par bit] at (1,1) {};
      \node[par check] at (0,1) {};
      \node[par check] at (1,0) {};
      \node[par label,below left=3pt,text=orange!90!black] at (0,0) {$b_1$};
      \node[par label,above right=3pt,text=green!55!black] at (1,1) {$b_2$};
      \node[par label,above left=3pt,text=blue!75!black] at (0,1) {$c_1$};
      \node[par label,below right=3pt,text=magenta!75!black] at (1,0) {$c_2$};
    \end{scope}
    \draw[->,line width=0.65pt] (1.45,2.25)
      -- node[above=3pt] {$\cR_{\mathrm{Par}}$} (2.65,2.25);

    \begin{scope}[xshift=3.3cm,x=0.5cm,y=0.5cm]
      % Each contour includes both check vertices at its rail crossing.
      \path[par outline,draw=blue!75!black,fill=blue,fill opacity=0.05]
        (-0.35,0.65) -- (0.35,0.65) -- (0.35,2.65)
        -- (2.35,2.65) -- (2.35,6.65) -- (6.35,6.65)
        -- (6.35,8.65) -- (8.35,8.65) -- (8.35,9.35)
        -- (1.65,9.35) -- (1.65,7.35) -- (-0.35,7.35) -- cycle;
      \path[par outline,draw=magenta!75!black,fill=magenta,fill opacity=0.05]
        (9.35,8.35) -- (8.65,8.35) -- (8.65,6.35)
        -- (6.65,6.35) -- (6.65,2.35) -- (2.65,2.35)
        -- (2.65,0.35) -- (0.65,0.35) -- (0.65,-0.35)
        -- (7.35,-0.35) -- (7.35,1.65) -- (9.35,1.65) -- cycle;

      \draw[par edge]
        (0,0) -- (0,7) (7,0) -- (7,7)
        (2,2) -- (2,9) (9,2) -- (9,9)
        (4,7) -- (4,9) (6,7) -- (6,9)
        (3,0) -- (3,2) (5,0) -- (5,2);
      \draw[par edge,preaction={draw=white,line width=1.7pt}]
        (0,0) -- (7,0) (0,7) -- (7,7)
        (2,2) -- (9,2) (2,9) -- (9,9)
        (0,3) -- (2,3) (0,5) -- (2,5)
        (7,4) -- (9,4) (7,6) -- (9,6);

      % Cycle basis of each K_c, inset by 0.27 from its edges. The loop
      % through the rail crossing is one cycle; K_{c_2} is K_{c_1} rotated
      % by 180 degrees about (4.5,4.5).
      \begin{scope}[par cycle,draw=cyan!55!blue]
        \draw (0.27,3.27) rectangle (1.73,4.73);
        \draw (4.27,7.27) rectangle (5.73,8.73);
        \draw (0.27,5.27) -- (0.27,6.73) -- (3.73,6.73) -- (3.73,8.73)
          -- (2.27,8.73) -- (2.27,5.27) -- cycle;
      \end{scope}
      \begin{scope}[par cycle,draw=magenta!70!red]
        \draw (7.27,4.27) rectangle (8.73,5.73);
        \draw (3.27,0.27) rectangle (4.73,1.73);
        \draw (8.73,3.73) -- (8.73,2.27) -- (5.27,2.27) -- (5.27,0.27)
          -- (6.73,0.27) -- (6.73,3.73) -- cycle;
      \end{scope}
      % Junction generators of C_G, inset by 0.5 to clear the region
      % outlines: both copies of b plus the shortest path in each K_c.
      \draw[par cycle,draw=yellow!55!orange] (0.5,0.5) -- (0.5,2.5)
        -- (1.5,2.5) -- (1.5,1.5) -- (2.5,1.5) -- (2.5,0.5) -- cycle;
      \draw[par cycle,draw=lime!60!green!85!black] (8.5,8.5) -- (8.5,6.5)
        -- (7.5,6.5) -- (7.5,7.5) -- (6.5,7.5) -- (6.5,8.5) -- cycle;

      \foreach \point in {
        (0,0),(7,7),(0,2),(0,4),(0,6),(1,7),(3,7),(5,7),
        (7,1),(7,3),(7,5),(2,0),(4,0),(6,0),
        (2,2),(9,9),(2,4),(2,6),(2,8),(3,9),(5,9),(7,9),
        (9,3),(9,5),(9,7),(4,2),(6,2),(8,2),
        (1,3),(1,5),(4,8),(6,8),(8,4),(8,6),(3,1),(5,1)}
        \node[par bit] at \point {};
      \foreach \point in {
        (0,7),(7,0),(0,1),(0,3),(0,5),(1.68,7),(4,7),(6,7),
        (7,2.32),(7,4),(7,6),(1,0),(3,0),(5,0),
        (2,9),(9,2),(2,3),(2,5),(2,7.32),(4,9),(6,9),(8,9),
        (9,4),(9,6),(9,8),(3,2),(5,2),(6.68,2)}
        \node[par check] at \point {};

      \foreach \point in {(0,0),(2,2)}
        \draw[par outline,orange!90!black] \point circle[radius=0.28];
      \foreach \point in {(7,7),(9,9)}
        \draw[par outline,green!55!black] \point circle[radius=0.28];
      \node[par label,text=blue!75!black] at (1,6) {$K_{c_1}$};
      \node[par label,text=magenta!75!black] at (8,3) {$K_{c_2}$};
    \end{scope}
  \end{tikzpicture}
  \caption{Local pieces of the Par replacement of the seed code, with
    $\ell=7$. Blue and magenta outlines enclose the check regions
    $K_{c_1}$ and $K_{c_2}$; the orange and green pairs mark the two
    copied-bit pieces. The light-blue and pink loops are the chosen
    cycle bases of $K_{c_1}$ and $K_{c_2}$, i.e.\ the generators of
    $C_{c_1}^0$ and $C_{c_2}^0$; the loop passing through a rail crossing
    is a single cycle. The yellow and light-green loops are the
    junction generators in $C_G^0$ for $b_1$ and $b_2$: each flips both
    copies of the bit and a path in each adjacent $K_c$.}
  \label{fig:par-piece-decomposition}
\end{figure}

We then choose a basis for the kernel of $\delta_c^1$,
  which is the cycle space of $K_c$, i.e. sums of cycles of $K_c$.
A natural basis for this kernel contains
  the bounded-size cycles within the ladders, formed by neighboring rungs,
  together with cycles around the ladder junctions.
Let $\Lambda^c_{G}$ denote the set of these cycles,
  which is a subset of the new symmetry generators $\Lambda_{G,(i+1)}$.
Let $\delta_c^0$ map each generator to the bit configuration supported on the corresponding cycle.
The local complexes are
\begin{equation}
  \cC_B=\left[0\longrightarrow\ff_2^2\longrightarrow0\right],
  \qquad
  \cC_c=\left[\ff_2^{\Lambda^c_{G}}\xrightarrow{\delta_c^0}
    \ff_2^{E(K_c)}\xrightarrow{\delta_c^1}
    \ff_2^{V(K_c)}\right],
  \label{eq:par-local-pieces}
\end{equation}
Figure~\ref{fig:par-piece-decomposition} illustrates the decomposition
for the seed code.

% Par artwork at ell=7, with the same seed and drawing-only crossing
% offsets as fig-rep-piece-decomposition.tex. Colors identify local pieces.

Set $\cC_B=\bigoplus_{b\in\Lambda_{B,(i)}}\cC_B$ and
$\cC_C=\bigoplus_{c\in\Lambda_{C,(i)}}\cC_c$.
In addition to the symmetry generators indexed by the sets $\Lambda_G^c$,
  introduce one new symmetry generator for each $b\in\Lambda_{B,(i)}$.
Let $\Lambda_{G}^B$ denote the set of these new generators, with $\Lambda_{G}^B \cong \Lambda_{B,(i)}$.
For the base code,
  these are the yellow and light-green loops in Figure~\ref{fig:par-piece-decomposition}.
The generator associated with $b$
  flips the two copies of $b$
  and, for each adjacent check $c\sim b$,
  the bits along a path in $K_c$ joining the two copies of $b$.
These generators again have support of bounded size.

These generators form complex $\cC_G$
\begin{equation}
  \cC_G=\left[\ff_2^{\Lambda_G^B}\longrightarrow0\longrightarrow0\right].
\end{equation}
Their action on the new bits has two components.
The map $\kappa_{GB}^0$ sends the generator indexed by $b$ to the two copied bits,
while $\kappa_{GC}^0$ sends it to the sum
of the paths in the regions $K_c$ with $c\sim b$.

The assembled complex $\widetilde{\cC}_{(i+1)}$ is

% Layout adapted from refs/memory-paper/paper.tex, graded replacement diagrams.
% All local pieces are at level (i+1).
\begin{equation}
 \begin{tikzpicture}[baseline]
 \matrix (a) [
   matrix of math nodes, nodes in empty cells,
   nodes={minimum size=25pt},
   row sep=2em, column sep=2em,
   text height=1.25ex, text depth=0.25ex
 ] {
   {} & {} & \textcolor{blue}{C_G^0}
     & \textcolor{blue}{0} & \textcolor{blue}{0} \\
   {} & 0 & C_B^1 & 0 & {} \\
   \textcolor{red}{C_C^0}
     & \textcolor{red}{C_C^1}
     & \textcolor{red}{C_C^2} & {} & {} \\
 };
 \path[->,blue,font=\scriptsize]
   (a-1-3) edge node[above]{$\delta_G^0$} (a-1-4)
   (a-1-4) edge node[above]{$\delta_G^1$} (a-1-5);
 \path[->,font=\scriptsize]
   (a-2-2) edge node[above]{$\delta_B^0$} (a-2-3)
   (a-2-3) edge node[above]{$\delta_B^1$} (a-2-4);
 \path[->,red,font=\scriptsize]
   (a-3-1) edge node[above]{$\delta_C^0$} (a-3-2)
   (a-3-2) edge node[above]{$\delta_C^1$} (a-3-3);
 \path[->,font=\scriptsize]
   (a-1-3) edge node[right]{$\kappa_{GB}^0$} (a-2-3)
   (a-1-4) edge node[right]{$\kappa_{GB}^1$} (a-2-4)
   (a-2-2) edge node[right]{$\kappa_{BC}^0$} (a-3-2)
   (a-2-3) edge node[right]{$\kappa_{BC}^1$} (a-3-3);
 \path[->,dashed,font=\scriptsize]
   (a-1-3.west) edge[out=210,in=140,looseness=1.25]
     node[left,pos=0.5]{$\kappa_{GC}^0$} (a-3-2.north);
 \end{tikzpicture}
 \label{eq:par-local-complexes}
\end{equation}
Explicitly,
\begin{equation}
  \widetilde{\cC}_{(i+1)}:
  C_G^0\oplus C_C^0
  \longrightarrow C_B^1\oplus C_C^1
  \longrightarrow C_C^2.
  \label{eq:par-total-complex}
\end{equation}
In particular,
\begin{equation}
  C_G^0\oplus C_C^0 \cong \ff_2^{\Lambda_{G}^B} \oplus \ff_2^{\bigsqcup_{c\in\Lambda_{C,(i)}} \Lambda_{G}^c}.
\end{equation}
Thus, the symmetry generators in the degree-zero space are indexed by
\begin{equation}
  \Lambda_G^B
  \sqcup
  \bigsqcup_{c\in\Lambda_{C,(i)}}\Lambda_G^c.
\end{equation}

It is straightforward to check that these maps form a cochain complex, i.e. that $\delta^2=0$.

\subsubsection{Properties of the code $\cC_{(i+1)} = \cR_{\mathrm{Par}}(\cC_{(i)})$}
\label{sec:par-comparison}

We now analyze the properties of the code after $\cR_{\mathrm{Par}}$ is applied.
The key result is a cochain map
$\widetilde{\cF}_{\cR_{\mathrm{Par}},(i)}:\widehat{\cC}_{(i)}\to\widetilde{\cC}_{(i+1)}$
that induces an isomorphism on $H^1$.
Restoring the inherited symmetry generators extends this map to a
cochain map
$\cF_{\cR_{\mathrm{Par}},(i)}:\cC_{(i)}\to\cC_{(i+1)}$
that also induces an isomorphism on $H^1$, as shown in
Corollary~\ref{cor:restore-inherited-generators}.

\begin{lemma}[Par local cohomology]
\label{lem:par-local-complexes}
For each $b\in\Lambda_{B,(i)}$ and $c\in\Lambda_{C,(i)}$, the local
complexes in \eqref{eq:par-local-pieces} satisfy
\begin{equation}
  H^1(\cC_B,\delta_b)\cong\ff_2^2,
  \qquad H^0(\cC_c,\delta_c)=H^1(\cC_c,\delta_c)=0,
  \qquad H^2(\cC_c,\delta_c)\cong\ff_2.
  \label{eq:par-check-cohomology}
\end{equation}
The last identification is induced by $\varepsilon_c$.
\end{lemma}
\begin{proof}
$H^1(\cC_B,\delta_b)\cong\ff_2^2$ is immediate,
  since $\cC_B$ is concentrated in degree one.

By definition, $\delta_c^0$ maps the standard basis of $C_c^0$
  bijectively onto the chosen basis of $\ker\delta_c^1$.
Hence
$\delta_c^0$ is injective and
  $\operatorname{im}\delta_c^0=\ker\delta_c^1$.
Therefore,
  $H^0(\cC_c,\delta_c)=0,
  H^1(\cC_c,\delta_c)=0$.

It is clear that $(\delta_c^1)^T$
  is the incidence matrix of $K_c$.
Since $K_c$ is connected,
  the kernel of $(\delta_c^1)^T$ is spanned by the all-ones configuration on $V(K_c)$.
Thus, $H^2(\cC_c,\delta_c)\cong\ff_2$.
\end{proof}

Taking direct sums gives the following local-cohomology spaces.

% Layout adapted from refs/memory-paper/paper.tex, local-cohomology diagrams.
\begin{equation}
 \begin{adjustbox}{max width=\linewidth}
 \begin{tikzpicture}[baseline]
 \def\localcohomfont{\fontsize{10}{9.5}\selectfont}
 \def\localarrowfont{\fontsize{7}{9.5}\selectfont}
 \matrix (a) [
   matrix of math nodes, nodes in empty cells,
   row sep=2.4em, column sep=3.2em,
   nodes={minimum size=25pt,align=center,text width=1.8cm,font=\localcohomfont}
 ] {
   {} & {}
     & |[text width=4.2cm]| \textcolor{blue}{H^0(\cC_G,\delta_G)\cong\ff_2^{\Lambda_{B,(i)}}}
     & \textcolor{blue}{0} & \textcolor{blue}{0} \\
   {} & 0
     & |[text width=4.2cm]| {H^1(\cC_B,\delta_B)\cong(\ff_2^2)^{\Lambda_{B,(i)}}}
     & {0} & {} \\
   \textcolor{red}{0}
     & \textcolor{red}{0}
     & |[text width=4.2cm]| \textcolor{red}{H^2(\cC_C,\delta_C)\cong\ff_2^{\Lambda_{C,(i)}}}
     & {} & {} \\
 };
 \path[->,font=\localarrowfont]
   (a-1-3.south) edge node[right]{$H(\kappa_{GB}^0)$} (a-2-3.north)
   (a-1-4.south) edge node[right]{$H(\kappa_{GB}^1)$} (a-2-4.north)
   (a-2-2.south) edge node[left]{$H(\kappa_{BC}^0)$} (a-3-2.north)
   (a-2-3.south) edge node[left]{$H(\kappa_{BC}^1)$} (a-3-3.north);
 \path[->,dashed]
   ([xshift=-4pt]a-1-3.south) edge ([xshift=4pt]a-3-2.north)
   ([xshift=-4pt]a-1-4.south) edge ([xshift=4pt]a-3-3.north);
 \end{tikzpicture}
 \end{adjustbox}
 \label{eq:par-local-cohomology}
\end{equation}

\begin{lemma}[Par induced complex]
\label{lem:par-induced-complex}
Identifying $H^1(\cC_B,\delta_B)$ with $C_{(i)}^1\oplus C_{(i)}^1$
via the two copied bits, and $H^2(\cC_C,\delta_C)$ with $C_{(i)}^2$, we obtain
\begin{equation}
  H_{\mathrm{loc}}(\widetilde{\cC}_{(i+1)})=
  \left[C_{(i)}^1\xrightarrow{\Delta}
    C_{(i)}^1\oplus C_{(i)}^1
    \xrightarrow{\delta_{(i)}^1\Sigma}C_{(i)}^2\right],
  \label{eq:par-induced-complex}
\end{equation}
where $\Delta:\ C_{(i)}^1\to C_{(i)}^1\oplus C_{(i)}^1$ is the diagonal map and $\Sigma:\ C_{(i)}^1\oplus C_{(i)}^1\to C_{(i)}^1$ is the sum map.

Furthermore, there exists a cochain-homotopy equivalence
$J_{\mathrm{Par},(i)}:H_{\mathrm{loc}}(\widetilde{\cC}_{(i+1)})
\to\widetilde{\cC}_{(i+1)}$.
\end{lemma}
\begin{proof}
On the first page $E^{p,q}_1$, only a few entries are nonzero, so all higher differentials vanish. The construction in Section~\ref{sec:spectral-sequence-review} therefore gives the stated cochain-homotopy equivalence.
\end{proof}

\begin{lemma}[Par comparison map]
\label{lem:par-comparison-map}
Fix $J_{\mathrm{Par},(i)}$ as in Lemma~\ref{lem:par-induced-complex}.
The cochain inclusion
\begin{equation}
  \iota_{\mathrm{Par},(i)}^1(s)=(s,0),
  \qquad \iota_{\mathrm{Par},(i)}^2(\sigma)=\sigma
\end{equation}
maps $\widehat{\cC}_{(i)}: 0\longrightarrow C_{(i)}^1
    \xrightarrow{\delta_{(i)}^1} C_{(i)}^2$ into \eqref{eq:par-induced-complex}, and
\begin{equation}
  \widetilde{\cF}_{\cR_{\mathrm{Par}},(i)}
  =J_{\mathrm{Par},(i)}\iota_{\mathrm{Par},(i)}:
  \widehat{\cC}_{(i)}\lhook\joinrel\longrightarrow
  H_{\mathrm{loc}}(\widetilde{\cC}_{(i+1)})
  \xrightarrow{\simeq}\widetilde{\cC}_{(i+1)}
  \label{eq:par-comparison-map}
\end{equation}
is a cochain map inducing an isomorphism on $H^1$.
\end{lemma}
\begin{proof}
Since $\Sigma(s,0)=s$, the maps commute with the differentials.
A degree-one cocycle $(x,y)$ satisfies
$\delta_{(i)}^1(x+y)=0$. Modulo $\operatorname{im}\Delta$, it has
the unique representative $(x+y,0)$. Thus
$\iota_{\mathrm{Par},(i)}$ induces an $H^1$ isomorphism, as does
its composition with $J_{\mathrm{Par},(i)}$.
\end{proof}

We now fix the comparison maps used throughout. 
For a bit, check, or generator $v$, write $\mathbf e_v$ for its indicator vector. 
For Rep, Lemma~\ref{lem:rep-induced-complex} and $\iota^2_{\mathrm{Rep},(i)}$ give

\begin{equation}
    \widetilde{\cF}^1_{\cR_{\mathrm{Rep}},(i)}(\mathbf e_b)=\mathbf1_b,\qquad \widetilde{\cF}^2_{\cR_{\mathrm{Rep}},(i)}(\mathbf e_c)=\mathbf e_{c^0}+\mathbf e_{c^1},
\end{equation}
where $\mathbf1_b$ is the all-ones configuration on $V(K_b)$
and $c^0, c^1$ are the copies of $c$.

For Par, we have
\begin{equation}
   \widetilde{\cF}^1_{\cR_{\mathrm{Par}},(i)}(\mathbf e_b)=\mathbf1_b^0,\qquad \widetilde{\cF}^2_{\cR_{\mathrm{Par}},(i)}(\mathbf e_c)=\mathbf e_{c^0}, 
\end{equation}
where $\mathbf1_b^0$ is the indicator of $b^0$ together with the internal bits on the rail from $b^0$ to $c^0$ in each ladder replacing $(b,c)$, $c\sim b$.

We now restore the inherited symmetry generators in addition to the new ones.
For either of the replacements $\cR$, define
\begin{equation}
  \begin{gathered}
  C_{(i+1)}^0=\widetilde C_{(i+1)}^0\oplus C_{(i)}^0,
  \qquad C_{(i+1)}^m=\widetilde C_{(i+1)}^m\quad(m=1,2),\\
  \delta_{(i+1)}^0(g,g')
    =\widetilde\delta_{(i+1)}^0 g
       +\widetilde{\cF}_{\cR,(i)}^1\delta_{(i)}^0 g',
  \qquad \delta_{(i+1)}^1=\widetilde\delta_{(i+1)}^1.
  \end{gathered}
  \label{eq:restore-inherited-generators}
\end{equation}
For Par, the symmetry generators in the full code $\cC_{(i+1)}$
  are indexed by
\begin{equation}
  \Lambda_{G,(i+1)}
  =
  \left(\Lambda_G^B
  \sqcup
  \bigsqcup_{c\in\Lambda_{C,(i)}}\Lambda_G^c\right)
  \sqcup
  \Lambda_{G,(i)}.
\end{equation}
For Rep, one instead has
  $\Lambda_{G,(i+1)}=\Lambda_{G,(i)}$.
The inherited symmetry generators in $C_{(i)}^0$
  are mapped to the new bits by passing through $\widetilde{\cF}_{\cR,(i)}^1$.
For Par, $\widetilde C_{(i+1)}^0$
  contains the generators from the local cycles.
For Rep, it is zero.
Define
\begin{equation}
  \cF_{\cR,(i)}^0(g)=(0,g),
  \qquad
  \cF_{\cR,(i)}^m=\widetilde{\cF}_{\cR,(i)}^m
  \quad(m=1,2).
\end{equation}

\begin{cor}[Full-code comparison]
\label{cor:restore-inherited-generators}
For either replacement, $\cC_{(i+1)}$ is a cochain complex. The map
\begin{equation}
  \cF_{\cR,(i)}:\cC_{(i)}\longrightarrow\cC_{(i+1)}
  \label{eq:full-code-comparison}
\end{equation}
is a cochain map whose induced map
\begin{equation}
  \cF_{\cR,(i)}:
  H^1(\cC_{(i)})\longrightarrow H^1(\cC_{(i+1)})
\end{equation}
is an isomorphism. In particular, if $\alpha_{(i)}$ represents a
nontrivial class in $H^1(\cC_{(i)})$, then
\begin{equation}
  \alpha_{(i+1)}=\cF_{\cR,(i)}^1\alpha_{(i)}
\end{equation}
also represents a nontrivial class in $H^1(\cC_{(i+1)})$.
\end{cor}
\begin{proof}
The cochain identities follow directly from
\eqref{eq:restore-inherited-generators} and those of the reduced map.
By Lemmas~\ref{lem:rep-comparison-map} and~\ref{lem:par-comparison-map},
the reduced map identifies $\ker\delta_{(i)}^1$ with
$H^1(\widetilde{\cC}_{(i+1)})$, which already quotients out the
new symmetry images.
By \eqref{eq:restore-inherited-generators}, the inherited generators add exactly $\widetilde{\cF}^1_{\cR,(i)}(\operatorname{im}\delta^0_{(i)})$ to the coboundaries. So $H^1(\cC_{(i+1)})$ is the quotient of $H^1(\widetilde{\cC}_{(i+1)})$ by the image of $\operatorname{im}\delta^0_{(i)}$ under this identification, just as $H^1(\cC_{(i)})$ is the quotient of $\ker\delta^1_{(i)}$ by $\operatorname{im}\delta^0_{(i)}$.
Thus the map induces the stated $H^1$ isomorphism. In particular, it
sends every nontrivial class to a nontrivial class, which proves the
claim about $\alpha_{(i+1)}$.
\end{proof}

In fact, for Par iteration,
  we will include additional symmetry generators in $C_{(i+1)}^0$
  that are linear combinations of the generators in $\widetilde C_{(i+1)}^0 \oplus C_{(i)}^0$,
  which will be convenient for the memory time analysis.
  (See Definition~\ref{def:crossing-gauge-generators}.)
This does not change the result of Corollary~\ref{cor:restore-inherited-generators},
  since the new generators are linear combinations of the old ones
  and thus do not change the image of $\delta_{(i+1)}^0$.

\subsection{Main statements}
\label{sec:ladder-geometry}

In this section, we collect the main statements about the code family $\{\cC_{(i)}\}_{i\ge0}$.
We first recall the construction.
Fix an odd integer $\ell\ge7$.
Starting from the two-bit, two-check
base code $\cC_{(0)}$ in~\eqref{eq:ladder-base-code},
the code family $\{\cC_{(i)}\}_{i\ge0}$ is constructed by
alternating between applying Rep and Par replacements:
\begin{equation}
  \cC_{(2k+1)}=\cR_{\mathrm{Rep}}(\cC_{(2k)}),
  \qquad
  \cC_{(2k+2)}=\cR_{\mathrm{Par}}(\cC_{(2k+1)}).
\end{equation}

\begin{theorem}[Local realization and one encoded bit]
\label{thm:ladder-code-structure}
For every $i\ge0$:
\begin{enumerate}
  \item The locality bounds \eqref{eq:ladder-locality-definition} hold with
    $L=1$ and $M=2$, independently of $i$:
    \begin{equation*}
      \begin{gathered}
      |I_{(i)}(b)-I_{(i)}(c)|\le 1\quad \forall\, (b,c)\in T_{(i)}(1),\\
      |\{v\in T_{(i)}(0):I_{(i)}(v)\in \textnormal{Ball}_{1/2}(x)\}|\le 2\quad \forall x\in\RR^2.
      \end{gathered}
    \end{equation*}
  \item The code encodes one classical bit:
    \begin{equation}
      H^1(\cC_{(i)})\cong\ff_2,
      \qquad [\alpha_{(i)}]\ne0,
      \qquad \alpha_{(0)}=(1,1),
      \qquad
      \alpha_{(i+1)}=\cF_{\cR,(i)}^1\alpha_{(i)}.
      \label{eq:iterated-logical-class}
    \end{equation}
    Here $\cR=\cR_{\mathrm{Rep}}$ for even $i$ and
    $\cR=\cR_{\mathrm{Par}}$ for odd $i$.
\end{enumerate}
\end{theorem}

The first part of the theorem is a geometric statement about the Tanner graph $T_{(i)}$,
  which relies on the following lemma.
\begin{lemma}[Geometry of replacement]
\label{lem:ladder-geometry}
Suppose the Tanner graph $T_{(i)}$ has the following properties:
\begin{enumerate}
  \item vertex degrees two or three,
  \item an integral realization by unit horizontal or vertical edges,
  \item no two distinct edges occupying the same lattice segment.
\end{enumerate}
Either replacement with odd $\ell\ge7$ preserves these properties.
\end{lemma}
\begin{proof}
  The vertex degree bounds follow from the construction,
    since the new vertices have degree two or three (by observing the ladder structure),
    and the old vertices retain their degrees.

  Again by construction, the new vertices are placed at lattice points,
    and the new edges are horizontal or vertical segments of length one.

  The third property follows from the observation of the residue classes modulo $\ell$.
  During replacement,
    the first copy of the rails lies on the lattice lines with coordinates congruent to $0$ modulo $\ell$,
    while the second copy lies on the lattice lines with coordinates congruent to $2$ modulo $\ell$.
  For both replacement types,
    the rungs lie on the lattice lines with coordinates congruent to $3, 4, ..., \ell-1$ modulo $\ell$.
  This implies there is no overlap between the three families.

  It remains to check that edges within each family do not overlap.
  Within either copy, the rails do not overlap because they are obtained by scaling and subdividing the old edges.

  Within one ladder, the rungs have distinct attachment coordinates and hence do not overlap.
  Between the ladders,
    we see that the rungs of one ladder lie in a rectangle that is disjoint from the rectangle containing the rungs of any other ladder.
  For an old horizontal edge from $(a,b)$ to $(a+1,b)$, all its rungs lie in
  \[
    (\ell a+2,\ell a+\ell)\times[\ell b,\ell b+2],
  \]
  while for an old vertical edge from $(a,b)$ to $(a,b+1)$, all its rungs lie in
  \[
    [\ell a,\ell a+2]\times(\ell b+2,\ell b+\ell).
  \]
  It is straightforward to check that these rectangles are disjoint.
\end{proof}

\begin{proof}[Proof of Theorem~\ref{thm:ladder-code-structure}]
  $|I_{(i)}(b)-I_{(i)}(c)|\le 1$ follows from the construction,
    where there are only unit incidence edges.

  By induction from the base code using Lemma~\ref{lem:ladder-geometry},
    every vertex has degree at least two, and no lattice edge is used twice.
  Since only four lattice edges meet at each lattice point,
    at most two vertices can occupy the same point.
  An open ball of radius $1/2$ contains at most one lattice point
    and therefore at most two vertices.

  The second part follows by iterating Corollary~\ref{cor:restore-inherited-generators}.
\end{proof}

\subsection{Structure of the gauge group and Wilson operators}
\label{sec:gauge-hierarchy}

From now on, we will say that a \emph{full iteration} of the construction consists of a Rep + Par step taking $\cC_{(i)} \rightarrow \cC_{(i+2)}$. Thus, the construction $\cC_{(2i)}$ will be said to reside at level $i$ of the hierarchy.  There will be $n$ total levels in the hierarchy.

For the cochain map introduced in Subsec.~\ref{sec:ladder-replacements}, write
\begin{align}
  \cF_{(i)}
  &=\cF_{\cR_{\mathrm{Par}},(2i+1)}\circ\cF_{\cR_{\mathrm{Rep}},(2i)},\\
  \cF_{i\to n}
  &=\cF_{(n-1)}\circ\cdots\circ\cF_{(i)}.
\end{align}
Let $G_0^{(i)}$ be the standard basis of $\widetilde C_{(2i)}^0$, labeling the generators created in the Par step at level $i$. At the final level, define
\begin{equation}
  \cG_0^{(i)}=\{\cF_{i\to n}^0g:g\in G_0^{(i)}\},
  \qquad
  \cG^{(i)}=\operatorname{span}_{\ff_2}\cG_0^{(i)}.
  \label{eq:gauge-gen-created}
\end{equation}

We will also define the following change of basis for the gauge generators at level $i$. It is used to capture the geometric locality of the code upon embedding, which will be needed later. In particular, if two wires intersect at some level $k$ of the construction, upon the subsequent Par replacement, two newly introduced gauge generators (which correspond to the smallest faces/plaquettes in the construction) will spatially overlap; however, their graph distance can be much larger than their spatial distance. The following definition adds the sums of such gauge generators to the basis.  The gauge group itself remains unchanged.
\begin{definition}[Basis of gauge generators capturing locality]
\label{def:crossing-gauge-generators}
At hierarchy level $i$, let $\Omega_i$ consist of the unordered pairs $\{g,h\}\subseteq\mathcal G_0^{(i)}$ with $g\ne h$ such that the faces $f_g$ and $f_h$, whose boundaries correspond to the supports of $g$ and $h$, obey the property that their images in $\mathbb{R}^2$ overlap.

Based on these pairs, define a new basis of generators at level $i$ as well as their representatives at level $n$:
\begin{align}
 \overline{G}_0^{(i)}&=G_0^{(i)}\cup\{g+h:\{g,h\}\in \Omega_i\},\\
 \overline{\cG}_0^{(i)}&=\{\cF_{i\to n}^0g:g\in\overline{G}_0^{(i)}\}.
  \label{eq:crossing-gauge-generators}
\end{align}
\end{definition}
Moving forward, this will be our default hierarchical basis for the gauge generators, and we will drop the bar in the notation.

To describe Wilson operators, we use the chain complex that is dual to $\cC_{(j)}$:
\begin{equation}
  C_{k,(j)}=(C_{(j)}^k)^*,
  \qquad
  \partial_{k,(j)}=(\delta_{(j)}^{k-1})^T,
  \qquad
  B_{1,(j)}=\operatorname{im}\partial_{2,(j)}.
\end{equation}
For a bit $b\in\Lambda_{B,(j)}$, in an abuse of notation, let $b\in C_{1,(j)}$ also denote the vector whose support coincides with the bit. A chain $z=\sum_b z_b b$ specifies the Pauli-$Z$ operator
\begin{equation}
  Z^z=\prod_{b\in\Lambda_{B,(j)}}Z_b^{z_b}.
\end{equation}
We therefore identify chains with the supports of Pauli-$Z$ operators. In particular, the chains in $B_{1,(j)}$ are the supports of products of checks.

Transposing the cochain map gives a map from chains at hierarchy level $i+1$ to chains at level $i$:
\begin{equation}
  p_i=(\cF_{(i)}^1)^T:
  C_{1,(2i+2)}\longrightarrow C_{1,(2i)}.
  \label{eq:local-homology-projection}
\end{equation}

For later use, we also define
\begin{equation}
    P_i=p_i p_{i+1}\cdots p_{n-1},
    \qquad P_n=\mathrm{id}. \label{eq:Pi}
\end{equation}

We also define a lift in the opposite direction. Start with a bit labeled by $b\in\Lambda_{B,(2i)}$. Recall that each replacement duplicates the subdivided graph and shifts one copy by $(2,2)$. After the Rep step, choose the bit corresponding to $b$ in the unshifted copy and denote its label by $\widehat b$. In the subsequent Par step, this bit appears in two copies, labelled by $\widehat b^0$ and $\widehat b^1$. Define the linear map
\begin{equation}
  \cL_{(i),1}:C_{1,(2i)}\longrightarrow C_{1,(2i+2)},
  \qquad
  \cL_{(i),1}(b)=\widehat b^0+\widehat b^1.
\end{equation}
Thus the lift replaces $Z_b$ by $Z_{\widehat b^0}Z_{\widehat b^1}$: it selects one bit through Rep and then takes both copies of that bit through Par. This product commutes with the new Par gauge generators, since each flips both selected bits or neither.

The lift satisfies
\begin{equation}
  p_i\cL_{(i),1}=\mathrm{id},
  \qquad
  \cL_{(i),1}(B_{1,(2i)})\subseteq B_{1,(2i+2)}.
\end{equation}
To see the second property, in Rep the internal repetition checks allow a bit in a check support to be replaced by the chosen bit in the same connected piece $K_b$. In Par, summing all checks in a connected piece $K_c$ cancels the internal bits and leaves the two copies of each adjacent input bit. Thus a product of coarse checks lifts to a product of fine checks.

Write
\begin{equation}
  \cL_{i\to n,1}=\cL_{(n-1),1}\cdots\cL_{(i),1}.
\end{equation}
For each bit $b\in\Lambda_{B,(2i-2)}$ at hierarchy level $i-1$, define its partial Wilson support at the final level by
\begin{equation}
  W_b^{(i)}=\cL_{i-1\to n,1}(b).
\end{equation}
Each Rep + Par replacement step sends a one-bit support to a two-bit support. Applying $n-i+1$ Rep + Par steps (full iterations) therefore gives
\begin{equation}
  |W_b^{(i)}|=2^{n-i+1}.
  \label{eq:partial-wilson-weight}
\end{equation}
Either bit of the base code gives a global Wilson operator $Z^{W_b^{(1)}}$.

Finally, let $\pi_i:C_{0,(2n)}\to(\cG^{(i)})^*$ be restriction to the subspace $\cG^{(i)}$ generated by level-$i$ gauge generators. Define
\begin{align}
  \partial_{1,(2n)}^{(i)}&=\pi_i\partial_{1,(2n)}
\end{align}
The condition $\partial_{1,(2n)}^{(i)}z=0$ means that $Z^z$ commutes with every level-$i$ gauge transformation. We define the space of partial level-$i$ Wilson operators and gauge-invariant operators:
\begin{align}
  \cW_i=\operatorname{span}_{\ff_2}\{W_b^{(i)}:b\in\Lambda_{B,(2i-2)}\}, \qquad
  K_i=\bigcap_{s \geq i}\ker\partial_{1,(2n)}^{(s)}.
\end{align}
Thus $K_i$ consists of the supports of Pauli-$Z$ operators commuting with all gauge transformations at levels $i$ and above. Additionally, we have
\begin{equation}
     K_i=B_{1,(2n)}+\cW_i.
\end{equation}

\begin{lemma}
\label{lemma:weights}
For $1\leq i\leq n$, for any $z\in K_i\setminus B_{1,(2n)}$, it holds
\begin{equation}
  |z|\geq2^{n-i+1}
\end{equation}
In particular, the minimum weight of a nontrivial global Wilson operator is $2^n$.
\end{lemma}
\begin{proof}
We prove the bound by induction on $n\geq i$, keeping $i$ fixed. For the base case $n=i$, every spin belongs to the support of a gauge generator at level $n$. Thus, every weight-one Pauli-$Z$ operator anticommutes with some level-$n$ generator, so the weight of any operator in $K_n$ must be $|z|\geq2$.

Now let $n>i$ and assume the bound holds for the code $\cC_{(2n-2)}$; perform the full Rep + Par iteration from $\cC_{(2n-2)}$ to $\cC_{(2n)}$.

First, given $z\in K_i\setminus B_{1,(2n)}$, we construct its projection to $C_{1,(2n-2)}$:
\begin{equation}
  w=p_{n-1}z=(\cF_{(n-1)}^1)^Tz.
\end{equation}
Since $z\in K_i$, it commutes with all generators introduced in the final iteration as well. The lift $\cL_{(n-1)}w \in C_{1, (2n)}$ (we drop the degree index in $\cL_{(n-1),1}$ for conciseness) also commutes with the new generators and has the same projection $w$, since $p_{n-1}\cL_{(n-1),1}=\mathrm{id}$. Therefore
\begin{equation} \label{eq:boundary-relation}
  z+\cL_{(n-1)}w\in B_{1,(2n)}.
\end{equation}
Recall that $z \in K_i$, $w$ cannot be a boundary either. Moreover, $w$ commutes with the generators at levels $i,\ldots,n-1$.

Next, consider two representatives of the class of $w$ using the two Par rails. For each bit $x\in\Lambda_{B,(2n-1)}$ present after the Rep step, let $A_x^a$ consist of its copied bit $x^a$ and the subdivision bits on rail $a = 0,1$ of every incidence edge adjacent to $x$, excluding rung bits. For $b\in\Lambda_{B,(2n-2)}$, define
\begin{equation}
  (w^a)_b=(p_{n-1}^az)_b
  =\sum_{x\in K_b(0)}\sum_{u\in A_x^a}z_u
\end{equation}
Thus each projection takes parity along one Par rail and then over the entire Rep piece $K_b$ associated with the coarse bit $b$. The sets summed over are disjoint, so
\begin{equation}
  |z|\geq |w^0|+|w^1|.
\end{equation}
Both projections $p^0_{n-1}$ and $p^1_{n-1}$ preserve boundaries; additionally, $p_{n-1}^a\cL_{(n-1),1}=\mathrm{id}$. Applying them to \eqref{eq:boundary-relation} gives
\begin{equation}
  w^a+w\in B_{1,(2n-2)},\qquad a=0,1.
\end{equation}
Thus neither $w^0$ nor $w^1$ is a boundary, and are in $K_i\setminus B_{1,(2n-2)}$ for the code $\cC_{(2n-2)}$.

Applying the induction hypothesis with the same $i$ gives
\begin{equation}
  |z|\geq |w^0|+|w^1|
  \geq2\cdot2^{(n-1)-i+1}
  =2^{n-i+1}.
\end{equation}
This completes the induction.
\end{proof}

Finally, $K_1=\ker\partial_{1,(2n)}$ consists of all fully gauge-invariant chains, including check boundaries and global Wilson supports. Write
\begin{equation}
  K_i=K_1\sqcup\widehat\cW_i,
  \qquad \widehat\cW_i=K_i\setminus K_1.
  \label{eq:partial-centralizer}
\end{equation}
The chains in $\widehat\cW_i$ are partially but not fully gauge invariant; each has weight at least $2^{n-i+1}$.

\section{Effective perturbation strength} \label{sec:perturbation-strength}

\subsection{Hierarchical gauge averaging}

The proof of perturbative stability for the Wegner gauge theory is simple because of the finite degree property: each spin is in the support of a bounded number of symmetry generators, and each symmetry generator acts on a bounded number of spins (see Ref.~\cite{ryan-wegner}).  In the 2D code, gauge group generators can have unbounded support since independent generators are introduced at all scales. This significantly complicates the analysis, since existing cluster expansion methods cannot be directly used.  Instead, we iteratively averaging over generators appearing at each level of the construction, thus creating a sequence of effective potentials.  We show that the norm of these effective potentials can be carefully controlled, which allows us to eventually apply a Peierls argument. 

We now introduce the hierarchy of gauge generators.  Recall that we say that a \emph{full iteration} of the construction consists of a Rep + Par step corresponding to $\cC_{(i)} \rightarrow \cC_{(i+2)}$. Thus, the construction $\cC_{(2i)}$ will be said to reside at level $i$ of the hierarchy.  So there are a total of $n$ levels of the hierarchy.

Define the effective gauge-averaged interaction as:
\begin{equation}
    e^{-\beta V_{\rm eff}}  = \bE_G e^{- \beta V}   = \bE_{1}   \bE_{2}...  \bE_{n}   e^{- \beta V}  .
\end{equation}
where gauge-averaging over the gauge generators at level $i$ is defined to be the procedure
\begin{equation}
    \bE_{i} f = \frac{1}{|\cG^{(i)}|}\sum_{g \in \cG^{(i)}} X^{\partial^T_{1,n} g} f X^{\partial^T_{1,n} g}.
\end{equation}

We will now introduce the hierarchical effective interactions. First, define $V^{(n+1)}_{\rm eff} = V$.
Consider the first step of hierarchical gauge averaging
\begin{equation}
    e^{-\beta V_{\rm eff}}
  = \bE_{1}   ...  \bE_{n-1}  \left ( \bE_{n}   e^{- \beta V} \right )
    = \bE_{1}   ...   \bE_{n-1}  e^{- \beta V_{\rm eff}^{(n)}}  .
\end{equation}
Where we introduced
\begin{equation}
    e^{-\beta V_{\rm eff}^{(n)}}  = \bE_{n}  e^{- \beta V}.
\end{equation}

The intermediate effective interaction $\beta V_{\rm eff}^{(n)}$ is gauge-invariant under gauge transformations generated by $\cG_0^{(n)}$; the supports of all such operators are in $K_n$.

We split the effective interaction into:
\begin{equation}
    V_{\rm eff}^{(n)} = \bE_G  V_{\rm eff}^{(n)} + (1 - \bE_G)V_{\rm eff}^{(n)}   =  \cH_{\rm eff}^{(n)} + \cU^{(n)}
\end{equation}
where the operators in $\cH_{\rm eff}^{(n) }\equiv \bE_G  V_{\rm eff}^{(n)} $ are fully gauge-invariant, and their supports are thus in $K_1$. The terms in $\cU^{(n)} \equiv (1 - \bE_G)V_{\rm eff}^{(n)}$ are only gauge-invariant at level $n$, and their supports are in $\widehat \cW_n  \equiv K_n \setminus K_1$.

Thus, we write
\begin{equation}
    e^{-\beta V_{\rm eff}}
    =
    e^{-\beta \cH_{\rm eff}^{(n)}} \bE_{1}   ...  \bE_{n-2} \left (   \bE_{n-1}  e^{- \beta \cU^{(n)}} \right )   = e^{-\beta \cH_{\rm eff}^{(n)}} \bE_{1}   ...  \bE_{n-2}  e^{- \beta V_{\rm eff}^{(n-1)}}   .
\end{equation}
Continuing this recursively, we define
\begin{equation} \label{eq:hierarchical-HU-def}
    e^{-\beta V_{\rm eff}^{(i)}}  = e^{-\beta \cH_{\rm eff}^{(i)} - \beta \cU^{(i)}} =  \bE_i  e^{- \beta \cU^{(i+1)}},
\end{equation}
where we similarly split
\begin{equation}
    V_{\rm eff}^{(i)} = \bE_G  V_{\rm eff}^{(i)} + (1 - \bE_G)V_{\rm eff}^{(i)}   =  \cH_{\rm eff}^{(i)} + \cU^{(i)}
\end{equation}
where the supports of the operators entering $\cH_{\rm eff}^{(i)} \equiv \bE_G  V_{\rm eff}^{(i)} $ are in $K_1$ and the ones in $\cU^{(i)} \equiv  (1 - \bE_G)V_{\rm eff}^{(i)} $ are supported in $\widehat \cW_i \equiv K_i \setminus K_1$.

Finally, collecting everything together, the full effective interaction is
\begin{equation} \label{eq:full-interaction}
     V_{\rm eff} = \sum_{i = 1}^n \cH_{\rm eff}^{(i)}
\end{equation}
where, for each $i$, $\cH_{\rm eff}^{(i)}$ are  fully gauge invariant; they are products of terms in the unperturbed Hamiltonian $\cH_0$ as well as global Wilson operators.

We will also need the following statement:

\begin{cor} \label{cor:weights}
    The weight of each Pauli-$Z$ monomial in $\cU^{(i)}$ is lower-bounded by $2^{n-i+1}$.
\end{cor}
\begin{proof}
    Straightforwardly follows from Lemma ~\ref{lemma:weights}. 
\end{proof}

An analogous statement, however, is not true for $\cH^{(i)}_{\rm eff}$. To see this, consider its definition $\cH^{(i)}_{\rm eff} = \bE_G V_{\rm eff}^{(i)} = \bE_G \log \left ( \bE_i  e^{- \beta \cU^{(i+1)}} \right )$. While the supports of the operators in $\cU^{(i+1)}$ are bounded from below by the claim above, the logarithm in the definition of $\cH_{\rm eff}$ needs to be expanded in a series, which will multiply high-weight terms, resulting in terms of much smaller weight that are still gauge-invariant.

\subsection{Improved gauge-averaging theorem}
\label{sec:gauge-averaging-thm}

Below, we prove a substantially stronger version of the gauge-averaging theorem of
Ref.~\cite{ryan-wegner}.  This strengthening is needed to provide the necessary bounds on the norm of Hamiltonian terms.

Let $(G,G_0)$ be a finite abelian local symmetry with group $G$ and generator set $G_0$, acting on a finite set
of spins $E$ and let $Q$ be
its bipartite gauge-interaction graph (see Ref.~\cite{ryan-wegner}), namely
\begin{equation}
 Q=(E\sqcup G_0,\cE),
 \qquad
 (e,g)\in\cE\quad\Longleftrightarrow\quad e\in\mathrm{supp} \,  g.
\end{equation}
For each nonempty $X\subseteq E$, choose
$H(X)\subseteq E$ of minimum cardinality subject to $X\subseteq H(X)$
and $H(X)\cup G_0(H(X))$ being connected in $Q$, and set
\begin{equation}
    l(X)=|H(X)|,
    \qquad
    D=\max_{x\in E}|G_0(\{x\})|.
\end{equation}
For an interaction $V=\sum_{X\subseteq E}V_X$, define its norm as
\begin{equation}
    |V|_\eta
    =\sup_{g\in G_0}
      \sum_{X:\,g\in G_0(X)}
      \|V_X\|_\infty e^{\eta l(X)}.
    \label{eq:GA-norm}
\end{equation}

The effective interaction is defined as:
\begin{equation}
    e^{-\beta V_{\rm eff}}=\bE_G e^{- \beta V}.
\end{equation}

\begin{theorem}[Modified gauge-averaging theorem]
\label{lemma:gauge-averaging}
Fix $0<m<\eta$.  If $D$ is bounded by a constant and
\begin{equation}
    \beta|V|_\eta
    <\min\left\{\log 2,\frac{\eta-m}{4D}\right\},
    \label{eq:GA-smallness}
\end{equation}
then the cluster expansion applied to $e^{-\beta V_{\rm eff}}=\bE_G e^{-\beta V}$ defines an absolutely convergent and
exponentially local $G$-invariant interaction $V_{\rm eff}$ such that
\begin{equation}
    |V_{\rm eff}|_m\leq 2|V|_\eta.
    \label{eq:GA-conclusion}
\end{equation}
\end{theorem}

We provide the proof idea, leaving the details to Appendix~\ref{app:gauge-averaging}.
Denoting $f_X=e^{-\beta V_X}-1$, write the gauge-averaged Boltzmann weight as $e^{-\beta V_{\rm eff}}=\bE_G\prod_X(1+f_X)$.
Let $G_0(X)$ denote the generators whose supports meet $X$. We link two terms when they share such a generator, and call a connected group of terms a polymer. Each polymer $\gamma$ has weight $\bE_G\prod_{X\in\gamma}f_X$. Since the commuting generators can be averaged independently, unlinked polymers have factorized averages, and the expansion sums products of these weights over families of unlinked polymers.

Taking the logarithm to obtain $V_{\rm eff}$ cancels disconnected contributions and gives a cluster expansion, which is a weighted sum over connected families of polymers, linked through their constituent terms. 
We show that the Koteck\'y--Preiss convergence criterion~\cite{kotecky1986cluster} holds if the interaction strength is sufficiently small; by this criterion, the cluster expansion converges absolutely. 
Informally the Koteck\'y--Preiss convergence criterion is obtained by observing that cluster contributions can be bounded by sums over trees whose edges join incompatible polymers, which can be bounded in a self-consistent way. 
Given this, the bounds by Fern\'andez and Procacci~\cite{fernandez2007cluster} control the total absolute weight of all clusters containing a specified polymer. Summing these bounds over polymers meeting a fixed gauge generator then helps us bound the interaction strength $|V_{\rm eff}|_m$.

The strengthened theorem allows an arbitrarily small loss of decay: for every $0<m<\eta$, the bound $|V_{\rm eff}|_m\le2|V|_\eta$ holds when $\beta|V|_\eta$ is small compared to $\eta-m$ without any restriction on the latter. In comparison, the estimate in Ref.~\cite{ryan-wegner} requires that $\eta-m$ exceeds an $O(1)$ constant. This difference is crucial, as the gauge-averaging theorem is applied at each level of the hierarchy, and each application decreases the decay rate from $m_{i+1}$ to $m_i < m_{i+1}$, and the total decrease $\sum_i(m_{i+1}-m_i)$ must remain bounded uniformly in $n$ for the resulting effective interaction to be local.
The improvement comes from the way the sum over polymers ``pinned'' at interaction region $X$ (i.e., the sum is over only polymers containing $X$) is bounded in the cluster expansion. This sum can be bounded inductively rather than through a uniform bound; removing a term $X$ from a polymer leaves smaller polymers, each attached to $X$ through a shared gauge generator, which can then be treated recursively. This estimate 
The full proof is deferred to Appendix~\ref{app:gauge-averaging}.

\subsection{Hierarchical interaction graphs}
\label{sec:hierarchical-interaction-graphs}

\subsubsection{Block-spins $E_i$}
Recall that hierarchy level $i$ refers to the code $\cC_{(2i)}$, after $i$ full Rep+Par iterations. Additionally, recall that at level $2i$, the latest replacement was of the Par type. The finest checks are degree-3 and degree-2.  We will define a set of regions $E_i$ (which will be used as interaction supports) in the code $\cC_{(2i)}$, which is a set of subsets of $\Lambda_{B,(2i)}$. They are obtained from $\Lambda_{B,(2i)}$ by grouping all bits connected via degree-2 checks into one region. 

More formally, at level $i$, let $\sim_2$ be the equivalence relation on $\Lambda_{B,(2i)}$ generated by $b\sim_2 b'$ whenever $\partial_{2,(2i)}c=b+b'$ for a degree-two check $c$. Define
\begin{equation}
    E_i=\Lambda_{B,(2i)}/\sim_2.
\end{equation}
It is easy to see from the replacement definition that $|e| \leq 3$ for any $e \in E_i$.  We will call elements $e \in E_i$ \emph{blocks} or \emph{block-spins} at level $i$ 

$E_i$ has the following nice property. Since $\delta_{(2i)}^1\delta_{(2i)}^0=0$, every coboundary of the gauge operator $s=\delta_{(2i)}^0g$, for any  $e\in E_i$ satisfies
\begin{equation}
    s|_e\in\{0,\mathbf1_e\},
\end{equation}
Thus each gauge transformation flips an entire location in $E_i$ or leaves it unchanged. We thus can define  $\mathrm{supp} \,  g$ to be the set of block-spins in $\delta^0_{(2i)} g$:
\begin{equation}
    \mathrm{supp}\,g
    :=\left\{e\in E_i:\left.(\delta^0_{(2i)}g)\right|_e\ne0\right\},
    \label{eq:gauge-generator-support}
\end{equation}
which does not lose any information about the gauge generator support on actual spins.

\subsubsection{Hierarchical interaction graphs}
Recall that, by Definition~\ref{def:crossing-gauge-generators},  $G_0^{(i)}$ is now the modified generating set that includes sums of generators whose embedded faces overlap geometrically. Using this set ensures that the gauge-interaction graph also connects locations that are close geometrically.

We define the gauge-interaction graph
\begin{equation}
 Q_i=(E_i\sqcup G_0^{(i)},\cE_i),
 \qquad
 (e,g)\in\cE_i\quad\Longleftrightarrow\quad e\in\mathrm{supp} \,  g.
\end{equation}

For $X\subseteq E_i$, let $G_0^{(i)}(X)$ be the generators meeting $X$. A connected hull of a nonempty $X$ is a set $H\supseteq X$ for which $H\cup G_0^{(i)}(H)$ is connected in $Q_i$. Choose the smallest hull $H_i(X)$ and define
\begin{equation}
 \|X\|_i=|G_0^{(i)}(X)|,
 \qquad
 l_i(X)=|H_i(X)|.
 \label{eq:GA-length}
\end{equation}

We now define the \emph{coarse shadow} $R:E_{i+1}\to E_i$. Consider the full iteration $\cC_{(2i)}\to\cC_{(2i+1)}\to\cC_{(2i+2)}$, and let $e'\in E_{i+1}$ be a block-spin at level $i+1$. 
First, we define a related map $R_b = R_{b, \rm Rep}\circ R_{b, \rm Par}$ on bits. For each bit $u\in e'$, the Par and Rep parts of the map are defined to have the following action:
\begin{itemize}
    \item For each intermediate bit $b'\in\Lambda_{B,(2i+1)}$, define $K_{b'}^{\mathrm T}$ to be the region obtained by replacing the star of $b'$ excluding the endpoint checks. Denote $V(K_{b'}^{\mathrm T})$ to be its bit set; these sets partition $\Lambda_{B,(2i+2)}$. The map $R_{b, \rm Par}$ sends each $u\in V(K_{b'}^{\mathrm T})$ to $b'$. Note that $K_{b'}^T$ is unrelated to $\cC_{b'}$. 
    \item The map $R_{b, \rm Rep}$ sends each bit $b'\in V(K_b)$ to the coarse bit $b\in\Lambda_{B,(2i)}$.
\end{itemize}

We then define $R$ to satisfy $R(e')=e$ when the bits in $e'$ are mapped into $e$ under $R_b$. Equivalently,
\begin{equation}
  R(e')=e
  \quad\Longleftrightarrow\quad
  e'\subseteq\bigcup_{b\in e}\;\bigcup_{b'\in V(K_b)}V(K_{b'}^{\mathrm T}).
\end{equation}
We note that $R$ is a well-defined map from $E_{i+1}$ to $E_i$; indeed, any two bits joined by a degree-two check are mapped to coarse bits that either coincide or are also joined by a coarse degree-two check.  Thus, they must end up in the same $e \in E_i$. For $X\subseteq E_{i+1}$, define
\begin{equation}
  R(X)=\{R(e'):e'\in X\}.
  \label{eq:full-coarse-shadow}
\end{equation}

\begin{lemma}[Properties of the gauge-interaction graphs]
\label{lemma:properties-gauge-interaction-graph}
For odd $\ell\ge7$, and for all $1\le i\le n$:
\begin{enumerate}
\item $Q_i$ is a connected bounded-degree graph. Every gauge generator has degree at most $16$, and every block-spin $e \in E_i$ has uniformly bounded degree
\begin{equation}
 D_i=\max_{e\in E_i}|G_0^{(i)}(e)|\le D,
 \label{eq:spin-degree-bound}
\end{equation}
with $D$ an $O(1)$ constant independent of $i,n,$ and $\ell$. 

\item For $i<n$, the support of any generator $g\in G_0^{(i)}$, lifted to level $i+1$, overlaps with the support of at most
\begin{equation}
 \left|G_0^{(i+1)}\bigl(R^{-1}(\mathrm{supp} \,  g)\bigr)\right|
 \le A_\ell,\qquad A_\ell=C\ell^2,
 \label{eq:generator-cover}
\end{equation}
gauge generators at the level $i+1$. Here, $C$ is an $O(1)$ constant and $R^{-1}(Y)$ denotes all fine blocks whose coarse shadow is in $Y$. 

\item For $i<n$, every nonempty $X \subseteq E_{i+1}$ satisfies
\begin{equation}
 l_i(R(X))\le l_{i+1}(X).
 \label{eq:pivot-hull-nonexpansion}
\end{equation}
\end{enumerate}
\end{lemma}
This lemma follows straightforwardly from directly inspecting the gauge-interaction graph and the supports of the finest-scale gauge generators. We provide its full proof in Appendix~\ref{app:Qi}.

\subsubsection{Hierarchical norms}
\label{sec:hierarchical-norms}

An interaction decomposition on $E_{i+1}$ can be regrouped at level $i$ by summing terms with the same coarse shadow:
\begin{equation}
    V=\sum_{Y\subseteq E_i}V_Y,
    \qquad
    V_Y=\sum_{\substack{X\subseteq E_{i+1}:\\R(X)=Y}}V_X.
    \label{eq:interaction-regrouping}
\end{equation}
The level-$i$ norm is a local norm with respect to the gauge-interaction graph $Q_i$ which we define to be
\begin{equation}
    |V|_{\eta,i}
    =\sup_{g\in G_0^{(i)}}
      \sum_{\substack{X\subseteq E_i:\\g\in G_0^{(i)}(X)}}
      \|V_X\|_\infty e^{\eta l_i(X)}.
    \label{eq:def-level-i-norm}
\end{equation}

\subsubsection{Geometric locality}

The interaction graph $Q_n$ was introduced in a way so that the graph distance is close to geometric distance in $\zz_2$. Moreover, graphs $Q_i$ are self-similar, and at level $i$, the graph is geometrically local with respect to an appropriately coarse-grained lattice. Similarly, we will define a local interaction norm which will be close to the hierarchical interaction norm defined through the interaction graph.  We state and provide proofs of these claims in this subsection.

Choose any one spin in each block $e$ and denote its lattice position by $x_i(e)\in\mathbb Z^2$. Define the lattice metric by
\begin{equation}
 d_i(e,f)
 =\max\left\{\|x_i(e)-x_i(f)\|_1,\,\mathbf1_{e\ne f}\right\}.
 \label{eq:lattice-metric}
\end{equation}
where the $\mathbf1_{e\ne f}$ part is responsible for adding nonzero distance for distinct overlapping spins.

\begin{lemma}[Locality of the $Q_i$ graphs]
\label{lemma:uniform-metric-comparison}
There exist constants $0<c_1\le c_2<\infty$, independent of $i,n,$ and $\ell\ge7$, such that
\begin{equation}
 c_1d^{Q_i}(e,f)\le d_i(e,f)\le c_2d^{Q_i}(e,f),
 \qquad e,f\in E_i,
 \label{eq:uniform-metric-comparison}
\end{equation}
where $d^{Q_i}$ is the graph distance in $Q_i$.
\end{lemma}
The proof of this is provided in Appendix~\ref{app:local-hierarchical-norms}.

Similarly, the graph-decay norm for the interaction strength that we introduced in Eqn.~\ref{eq:def-level-i-norm} is related to an analogous geometric-decay norm.  For nonempty $X\subseteq E_n$, define
\begin{equation}
    l_{\mathrm{loc}}(X)
    =|H_{\zz^2}(X)|
    \label{eq:geometric-tree-length}
\end{equation}
where $H_{\zz^2}$ is the smallest connected hull of $X$ on $\zz^2$.
The geometrically local norm is
\begin{equation}
    |V|_{\eta',\mathrm{loc}}
    =\sup_{e\in E_n}
      \sum_{\substack{X\subseteq E_n:\\e\in X}}
      \|V_X\|_\infty e^{\eta'l_{\mathrm{loc}}(X)}.
    \label{eq:def-geometric-local-norm}
\end{equation}

\begin{lemma}[Comparison of local and hierarchical norms]
\label{lemma:local-hierarchical-norms}
There is an $O(1)$ constant $\kappa \ge1$, independent of $n$ and $\ell\ge7$, such that for every $\eta>0$  we have 
\begin{equation}
    \frac1{16}|V|_{\eta,n}
    \le |V|_{\kappa \eta,\mathrm{loc}}
    \le |V|_{\kappa^2\eta,n}.
    \label{eq:local-hierarchical-comparison}
\end{equation}
\end{lemma}

The proof of this is also provided in Appendix~\ref{app:local-hierarchical-norms}.

\subsection{Why gauge-averaging theorem can be applied in the induction step}

To obtain our main result, we will perform gauge-averaging iteratively, resulting in the following equation for the effective potential:
\begin{equation}
     e^{-\beta V_{\rm eff}^{(i)}} = e^{-\beta \cH_{\rm eff}^{(i)} - \beta \cU^{(i)}} =  \bE_i e^{- \beta \cU^{(i+1)}}.
\end{equation}
Here, we show that the conditions of the gauge-averaging theorem (more specifically, its version in Sec.~\ref{sec:gauge-averaging-thm}) apply to the procedure $\bE_i$, which is simply one step of the hierarchical gauge averaging procedure. The theorem takes as input (i) a set of spins, (ii) a finite abelian gauge group with a generating set, whose action on the spins is used to define the symmetry graph $Q$, (iii) an interaction written as a sum of terms supported on subsets of the spins, and (iv) the bounded-degree and smallness assumptions. We discuss these here; the ``spins'' are the block-spins $E_i$, the gauge group is $\cG^{(i)}$ with generators $G_0^{(i)}$, the symmetry graph is $Q_i$, the interaction is $\cU^{(i+1)}$ regrouped at level $i$, and the norm in the theorem becomes $|\cU^{(i+1)}|_{\eta,i}$.

For $e\in E_i$, define its set of descendants at level $n$ by
\begin{equation}
D_{i\rightarrow n}(e)
=
 \left \{e'\in E_n : R^{\,n-i}(e')=e\right\}.
\end{equation}
This implies
\begin{equation}
    E_n = \bigsqcup_{e \in E_i} D_{i\rightarrow n}(e).
\end{equation}
We also define  $D_{i\to n}(X) = \bigcup_{e\in X}D_{i\to n}(e)$.

The physical Hilbert space has the exact tensor-product
decomposition
\begin{equation}
\mathscr H_n
=\bigotimes_{e\in E_i}\mathscr H_e^{(i)},
\qquad
\mathscr H_e^{(i)}
=\bigotimes_{f\in D_{i\to n}(e)}
  \bigotimes_{b\in f}\cc^2.
\label{eq:block-hilbert-space}
\end{equation}
To apply Theorem~\ref{lemma:gauge-averaging} at level $i$, we
will treat the $\mathscr H_e^{(i)}$ as the local Hilbert spaces of its
local ``spins''.  This only regroups tensor factors and removes no
physical spins. 
At level $i+1$, the interaction $\cU^{(i+1)}$ is a sum over subsets of $E_{i+1}$.  Regrouping its terms  and using $D_{i+1\to n}(X)\subseteq D_{i\to n}(R(X))$, we obtain
\begin{equation} \label{eq:induction-input-support}
    \cU^{(i+1)} = \sum_{X \subseteq E_{i}}  \cU^{(i+1)}_{X},  \qquad
    \operatorname{supp}\bigl(\cU_X^{(i+1)}\bigr)\subseteq D_{i\rightarrow n}(X).
\end{equation}

We now confirm that the symmetry graph for these new degrees of freedom can be chosen to be $Q_i$. For $g \in G_0^{(i)}$ and $ g' = \cF^0_{i\to n}g \in\cG_0^{(i)}$ , we have 
\begin{equation}
\operatorname{supp}(\partial^T_{1,(2n)}g')
\cap D_{i\to n}(e)\neq\varnothing
\quad\Longleftrightarrow\quad
e\in\operatorname{supp}g.
\label{eq:block-gauge-support}
\end{equation}
Thus the symmetry graph of $\cG^{(i)}$ acting on the block-spins is $Q_i$. Colloquially, if at least one spin in $D_{i \rightarrow n}(e)$ participates in the gauge transformation $g' \in \cG_0^{(i)}$, then $e$ and $g$ are connected in the symmetry graph.

Finally, $|\cU^{(i+1)}|_{\eta,i}$ is the suitable norm for the interaction. Since the symmetry graph is
$Q_i$, the generators overlapping $Y\subseteq E_i$ are
exactly $G_0^{(i)}(Y)$, and the connected-hull size appearing in the
gauge-averaging theorem is $l_i(Y)$. The tensor-factor regrouping in
\eqref{eq:block-hilbert-space} does not change the operator norm.
Therefore, the interaction norm appearing in the gauge-averaging
theorem should be
\begin{equation}
    \sup_{g\in G_0^{(i)}}
    \sum_{\substack{Y\subseteq E_i:\\
                    g\in G_0^{(i)}(Y)}}
     \left \|\cU_Y^{(i+1)}\right\|_\infty
    e^{\eta l_i(Y)}
    =
     \left |\cU^{(i+1)}\right|_{\eta,i}.
    \label{eq:gauge-averaging-norm-identification}
\end{equation}
It remains to check the assumptions of the theorem. The group $\cG^{(i)}$ is finite and abelian. 
The degree bound $D_i\le D$  was shown Lemma~\ref{lemma:properties-gauge-interaction-graph}(1), and it is independent of $i$ and $n$. Consequently, whenever condition in Eq.~\eqref{eq:GA-smallness}, Theorem~\ref{lemma:gauge-averaging} applies at level $i$ with $\bE_G=\bE_i$, $G=\cG^{(i)}$, $Q=Q_i$, $V=\cU^{(i+1)}$ and $|V|_\eta=|\cU^{(i+1)}|_{\eta,i}$.

\subsection{Main result}
\label{sec:main-result}

We now show the main result of the section, Theorem~\ref{thm:full-induction}, which bounds the strength of the effective interaction obtained from hierarchical gauge averaging. Recall that after $i$-th step, we obtain $V_{\rm eff}^{(i)} = \cH_{\rm eff}^{(i)} + \cU^{(i)}$  and the full effective interaction can be written as $V_{\rm eff} = \sum_{i = 1}^n \cH_{\rm eff}^{(i)}$. 
Each application of the gauge-averaging theorem (Theorem~\ref{lemma:gauge-averaging}) can at best give a bound on the effective interaction at the next level where the norm has constant $m_i < m_{i+1}$. Thus, the hierarchical gauge-averaging will lead to a sequence of bounds with parameters $\{ m_i\}$ where $m_i$ decreases as $i$ decreases. We will show that this sequence is bounded from below and above by a positive $O(1)$ constant.

More precisely, to perform a step of gauge-averaging, $|\cU^{(i+1)}|_{m_{i+1},i}$ has to be smaller than $(m_{i+1} - m_{i})/(4D)$. We will show that the decay constants and interaction strengths behave as
\begin{equation}
    m_{i+1} - m_{i} \lesssim \frac{1}{c^{n-i}}, \quad |\cH_{\rm eff}^{(i)}|_{m_i,i} \lesssim \frac{|V|_{\eta, \rm loc}}{c^{n-i}}, \quad |\cU^{(i)}|_{m_i,i} \lesssim \frac{|V|_{\eta, \rm loc}}{c^{n-i}}.
\end{equation}
respectively, for some $c > 1$. These bounds are used in Sec.~\ref{sec:peierls} to show perturbative stability of our classical code when  $\beta |V|_{\eta, \rm loc}$ is sufficiently small.

Since we will use different hierarchical norms at different levels, we will first need a result comparing such norms:
\begin{lemma}[Interaction strength change upon coarse-graining]
\label{lemma:size-i}
Assuming $m>0$, for $1 \leq i \leq n-1$ and interaction $V$, one has
\begin{equation}
    |V|_{m,i}\leq A_\ell |V|_{m,i+1}.
    \label{eq:corrected-level-comparison}
\end{equation}
where $A_\ell = C \ell^2$ from Lemma~\ref{lemma:properties-gauge-interaction-graph} with an absolute constant $C > 0$.
\end{lemma}

\begin{proof}

Recall that for $Y \subseteq E_i$,
$$V_Y = \sum_{\substack{X\subseteq E_{i+1}: \\ R(X) = Y}} V_X.$$
Then, we can write
\begin{align*}
    \sum_{\substack{Y\subseteq E_i:
    \\ g\in G_0^{(i)}(Y)}} \|V_Y\|_\infty e^{ml_i(Y)}
    &\leq
    \sum_{\substack{Y\subseteq E_i:
    \\ g\in G_0^{(i)}(Y)}} \sum_{\substack{X\subseteq E_{i+1}:
    \\ R(X) = Y}} \| V_X \|_\infty e^{ml_i(R(X))}
    \\
    &\leq \sum_{\substack{Y\subseteq E_i:
    \\ g\in G_0^{(i)}(Y)}} \sum_{\substack{X\subseteq E_{i+1}:
    \\ R(X) = Y}} \| V_X\|_\infty e^{ml_{i+1}(X)}
\end{align*}
where the second inequality uses $l_i(R(X))\leq l_{i+1}(X)$ by Lemma~\ref{lemma:properties-gauge-interaction-graph}(3).

For $g\in G_0^{(i)}$, define its level-$(i+1)$ cover as
\begin{equation}
    \Gamma_{i+1}(g)
    =G_0^{(i+1)}\bigl(R^{-1}(\mathrm{supp}\,g)\bigr).
\end{equation}
By part (2) of Lemma~\ref{lemma:properties-gauge-interaction-graph}, $|\Gamma_{i+1}(g)|\leq A_\ell$. For $g\in G_0^{(i)}(R(X))$, choose $e'\in X$ such that $R(e')\in\mathrm{supp}\,g$. Since $Q_{i+1}$ is connected, $e'$ belongs to the support of some $g'\in G_0^{(i+1)}$. This generator belongs to both $\Gamma_{i+1}(g)$ and $G_0^{(i+1)}(X)$.

 Then, we can replace the summation over $Y \subseteq E_i:
 g\in G_0^{(i)}(Y)$ with the summation over  $g' \in \Gamma_{i+1}(g)$, continuing the sequence of inequalities:
\begin{align}
    \sum_{g'\in\Gamma_{i+1}(g)} \sum_{\substack{X\subseteq E_{i+1}:
    \\  g'\in G_0^{(i+1)}(X)}} \| V_X\|_\infty e^{ml_{i+1}(X)}
    \leq
      |\Gamma_{i+1}(g)|
      |V|_{m,i+1}
    \leq
      A_\ell
      |V|_{m,i+1}.
\end{align}

Taking the supremum over $g \in G_0^{(i)}$ proves the statement of the Lemma.
\end{proof}

\vspace{8 pt}

We can now outline the rest of the logic of this section. Recall that
\begin{equation}
    \cH_{\rm eff}^{(i)} = \bE_G V_{\rm eff}^{(i)}, \quad \cU^{(i)} = (1-\bE_G) V_{\rm eff}^{(i)},
\end{equation}
where $\bE_G$ denotes gauge averaging over the entire gauge group. 
Since $\bE_i$ is an average of unitary conjugations, it follows that
\begin{equation}
    \| \bE_G \cO\|_\infty \leq \|\cO \|_\infty, \qquad  \| (1- \bE_G)\cO\|_\infty \leq 2 \|\cO \|_\infty.
\end{equation}
Additionally, because conjugation by $X^h$ does not change the support of individual terms in $\cO = \sum_X \cO_X$, we also have
\begin{equation}
    | \bE_G \cO|_{m,i} \leq |\cO |_{m,i}, \qquad  | (1-\bE_G) \cO|_{m,i} \leq 2 |\cO |_{m,i}. \label{eq:norm-PiK1}
\end{equation}

For the main result of this section, we would like to bound the strength of $\cU^{(i)}$ and $\cH_{\rm eff}^{(i)}$. The latter is straightforwardly determined by the strength of the previous term $\cU^{(i+1)}$. For $\cU^{(i)}$, we will prove a recursive statement, relating $|\cU^{(i)}|_{m_i,i}$ to $|\cU^{(i+1)}|_{m_{i+1},i+1}$, and use this to establish that the strength decreases exponentially in $n-i$.

Unfortunately, one \emph{cannot} show exponential suppression of $|\cU^{(i)}|_{m_i,i}$ by naively applying the gauge-averaging theorem, Lemma~\ref{lemma:size-i} and the considerations so far.  Assuming that the required smallness condition holds, we would obtain
\begin{equation}
    |V_{\rm eff}^{(i)}|_{m_i,i}
    \leq 2|\cU^{(i+1)}|_{m_{i+1},i}
    \leq 2A_\ell|\cU^{(i+1)}|_{m_{i+1},i+1},
\end{equation}
where $m_{i+1} > m_i$ and $A_\ell > 1$. Then, by Eq.~\eqref{eq:norm-PiK1}, we obtain $$|\cH_{\rm eff}^{(i)}|_{m_i,i} \leq 2A_\ell|\cU^{(i+1)}|_{m_{i+1},i+1}$$ which is good enough.  However, we would also have $$|\cU^{(i)}|_{m_i,i} \leq 4A_\ell|\cU^{(i+1)}|_{m_{i+1},i+1},$$ which gives a growing, instead of decreasing, bound!

Thus, we need to consider more carefully how the terms in  $\cU^{(i)}$ are generated from $\cU^{(i+1)}$. Write
\begin{equation}
V_{\rm eff}^{(i)} = -\frac{1}{\beta} \log (\bE_i e^{-\beta \cU^{(i+1)}})
= \mathbb E_i \cU^{(i+1)} -\frac{\beta}{2} \left[
\mathbb E_i \left ((\cU^{(i+1)})^2\right ) -(\mathbb E_i \cU^{(i+1)})^2 \right]
+...
\end{equation}
Note that $\bE_i \cU^{(i+1)}$ only contributes to $\cU^{(i)}$ and not to $\cH^{(i)}_{\rm eff}$, since $\bE_G\bE_i=\bE_G$ and $\bE_G\cU^{(i+1)}=0$.  We thus have
\begin{equation} \label{eq:cumulant}
\cU^{(i)} = (1 - \bE_G)V_{\rm eff}^{(i)}
= \mathbb E_i \cU^{(i+1)} - (1 - \bE_G) \left ( \frac{\beta}{2} \left[
\mathbb E_i \left ((\cU^{(i+1)})^2\right ) -(\mathbb E_i \cU^{(i+1)})^2 \right]
+...\right )
\end{equation}
As a result, $\cU^{(i)}$ can be written as a sum of a linear part $\bE_i \cU^{(i+1)}$, and a remainder term that starts at quadratic order in $\cU^{(i+1)}$.  We will provide bounds on each of these parts separately.

Let us first bound the strength of the linear part of the interaction, $\mathbb E_i \cU^{(i+1)}$. Any nontrivial term in $\cU^{(i+1)}$ must contain a Pauli monomial whose support is in $K_i \setminus K_1$ in order to survive the next step of gauge-averaging. The following shows that the size of such supports must decrease upon coarse-graining:
\begin{lemma} [Norm decrease lemma] \label{lemma:norm-decrease}
    Assume odd $\ell \geq 7$. For $1 \leq i \leq n-1$, consider $z \in K_i \setminus K_1$ and let $X \subseteq E_{i+1}$ contain the coarse-grained support of $z$, i.e.:
    \begin{equation}
        R^{n - (i+1)}({\rm supp} \, z) \subseteq X.
    \end{equation}
    Then
    \begin{equation}
    l_{i+1}(X)-l_i(R(X))\geq1.
    \label{eq:largesupport}
\end{equation}
\end{lemma}
\begin{proof}
By Lemma~\ref{lemma:properties-gauge-interaction-graph},
\begin{equation}
    l_i(R(X))
    \leq |R(H_{i+1}(X))|
    \leq |H_{i+1}(X)|
    =l_{i+1}(X).
\end{equation}
Suppose that $l_i(R(X)) = l_{i+1}(X)$. Then, from the expression above, $R$ must be injective on $H_{i+1}(X)$.

Consider the full Rep+Par replacement of the code  $\cC_{(2i)} \mapsto \cC_{(2i+2)}$ from hierarchy level $i$ to $i+1$. 
For an edge $(b,c)\in T_{(2i)}(1)$, let $e\in E_i$ be the block-spin containing $b$. Every fine block $e'\in E_{i+1}$ containing a spin produced by the full Rep+Par replacement of $(b,c)$ satisfies $R(e')=e_b$, where $e_b\in E_i$ is the block-spin containing $b$. Thus, coarse-graining under $R$ collapses all the fine block-spins along the replacement of $(b,c)$ to the single coarse block-spin $e_b$ containing $b$.

Since $R$ is injective on $H_{i+1}(X)$, the hull $H_{i+1}(X)$ can contain at most one fine block-spin inside the replacement of any edge $(b,c)\in T_{(2i)}(1)$.  Any two distinct block-spins $e',f'\in H_{i+1}(X)$ connected through a common gauge generator in $Q_{i+1}$ satisfy $R(e')\ne R(f')$.
Inspecting the overlapping faces used to form the added generators $g+h$ within one Rep+Par replacement shows that fine level-$i+1$ block-spins $e',f'\in E_{i+1}$ with $R(e')\ne R(f')$ can be connected in $Q_{i+1}$ only through a junction or the geometric crossing of the coarse Tanner graph $T_{(2i)}$ (recall that the gauge generators whose associated faces overlap upon embedding contribute new $g+h$-type gauge generators according go the definition~\ref{def:crossing-gauge-generators}).
This, together with connectedness, confines the hull to a single coarse junction or crossing, one in $T_{(2i)}$.

A coarse junction in $T_{(2i)}$ can be a bit or a check. A bit will enter exactly one block-spin $e \in E_i$, while checks have degree at most 3, so a check corresponds to at most 3 block-spins, from its neighboring bits. At a coarse crossing in $T_{(2i)}$, the two overlapping bits or checks have degree two by Lemma~\ref{lem:ladder-geometry}; each therefore enters only one block-spin. Therefore the connected hull $H_{i+1}(X)$ restricted to a junction or a crossing can coarse-grain under $R$ into at most 3 block-spins, i.e..
\begin{equation} \label{eq:3}
    |H_{i+1}(X)|
    =|R(H_{i+1}(X))|
    \leq 3.
\end{equation}

On the other hand, by assumption of the lemma, the projected chain $P_{i+1}z\in C_{1,(2i+2)}$ must commute with the generators at levels $i$ and $i+1$. Applying eq.~\eqref{eq:boundary-relation} iteratively shows that it is not a boundary, since $z\notin B_{1,n}$.
We can thus use Lemma~\ref{lemma:weights} with $i$ and $n=i+1$, giving
\begin{equation}
    4
    \leq \min_{u\in B_{1,(2i+2)}}|P_{i+1}z+u|.
\end{equation}
There exists $u_0 \in B_{1,(2i+2)}$ such that:
\begin{equation}
    \min_{u\in B_{1,(2i+2)}}|P_{i+1}z+u| \leq |P_{i+1}z+u_0| \leq |X|.
\end{equation}
Indeed, the bits in $P_{i+1}z$ are supported in the same block-spins as $X$, by definition. However, $|P_{i+1}z|$ can still be larger than $|X|$ if in some block-spins it is supported on multiple bits. Recall that a block-spin at level $i$ is defined by uniting the spins connected by degree-2 checks into one block-spin; choosing $u_0$ to be the boundary of degree-2 checks only, one can reduce the support of $P_{i+1}z$ to at most one bit per block-spin, which shows inequality below.

Combining the last two equations together, we obtain $|X| \geq 4$. However, at the same time, by \eqref{eq:3}, $|X|  \leq |H_{i+1}(X)| \leq 3$, giving a contradiction. Therefore, $l_{i+1}(X)-l_i(R(X))\geq1$.
\end{proof}

\begin{lemma}[Bounding $\cU^{(i)}$, linear part] \label{lemma:linear}
For $1 \leq i \leq n-1$ and $0 < m_i < m_{i+1}$,
    \begin{equation}
        |\bE_i \cU^{(i+1)}|_{m_{i},i} \leq e^{-m_{i}} A_\ell|\cU^{(i+1)}|_{m_{i+1},i+1}  \label{eq:expsuppression}
    \end{equation}
\end{lemma}
\begin{proof}
    Consider a term in $\cU^{(i+1)}$ supported on $X \subseteq E_{i+1}$. If any Pauli monomial inside this term survives the average, the conditions of Lemma~\ref{lemma:norm-decrease} are satisfied, and thus, $l_i(R(X)) \leq l_{i+1}(X) - 1$. Because $m_{i+1} > m_i$,
    \begin{equation}
        e^{m_i l_i(R(X))} \leq e^{-m_i} e^{m_{i+1}l_{i+1}(X)},
    \end{equation}
    combining this inequality with the rest of the proof of Lemma~\ref{lemma:size-i}, and using that $\bE_i$ does not increase the operator norm, the desired statement follows.
\end{proof}

Next, we focus on bounding the remainder term in $\cU^{(i)}$, which starts at the quadratic order.  For this, we will utilize the smallness of $\cU^{(i+1)}$ along with the holomorphicity of the cluster expansion.

\begin{lemma}[Bounding $\cU^{(i)}$, ``the rest''] \label{lemma:the-rest}
In addition to the assumptions in the previous lemma, suppose that
\begin{equation} \label{eq:diff-condition}
    0 < \frac{m_{i+1} - m_i}{4D} \leq \log 2.
\end{equation}
If the interaction $\cU^{(i+1)}$ satisfies
\begin{equation}    \label{eq:condition-lemma-rest}
    A_\ell\beta|\cU^{(i+1)}|_{m_{i+1},i+1}
    \leq\frac{m_{i+1}-m_i}{16D}.
\end{equation}
Then
\begin{equation}
    \beta|\cU^{(i)}-\bE_i\cU^{(i+1)}|_{m_i,i}
    \leq\frac{32D}{m_{i+1}-m_i}
    \left(A_\ell\beta|\cU^{(i+1)}|_{m_{i+1},i+1}\right)^2.
\end{equation}
\end{lemma}
\begin{proof}
    By Lemma~\ref{lemma:size-i} and the assumption of this lemma,
    \begin{equation}
        \beta |\cU^{(i+1)}|_{m_{i+1},i} \leq A_\ell \beta |\cU^{(i+1)}|_{m_{i+1},i+1} < \frac{m_{i+1}-m_i}{4D}.
    \end{equation}
    Together with~\eqref{eq:diff-condition}, this verifies both smallness requirements in~\eqref{eq:GA-smallness}. Therefore, the gauge-averaging theorem~\ref{lemma:gauge-averaging} applies at level $i$ with $\eta=m_{i+1}$ and $m=m_i$, and the corresponding cluster expansion converges absolutely.

    We will bound $\beta |\cU^{(i)} - \bE_i \cU^{(i+1)}|_{m_{i},i} $ which we wrote earlier through cumulant expansion in eq.~\eqref{eq:cumulant}. We will rewrite it in the form
    \begin{align}
    \cU^{(i)} - \mathbb E_i \cU^{(i+1)}
    = ( 1- \bE_G) \left ( V_{\rm eff}^{(i)}  -  \mathbb E_i \cU^{(i+1)}  \right)
    \end{align}
    where the term $ V_{\rm eff}^{(i)}  -  \mathbb E_i \cU^{(i+1)}$ is at least second-order in $\cU^{(i+1)}$.
    Using eq.~\eqref{eq:norm-PiK1}, we obtain
    \begin{equation} \label{eq:factoroftwo}
    |\cU^{(i)} - \mathbb E_i \cU^{(i+1)}|_{m_i,i} \leq 2 \left  |V_{\rm eff}^{(i)}  -  \mathbb E_i \cU^{(i+1)} \right |_{m_i,i},
    \end{equation}
    so we will focus on bounding the strength of $V_{\rm eff}^{(i)}  -  \mathbb E_i \cU^{(i+1)}$.

    Let us introduce an auxiliary perturbation $\lambda \cU^{(i+1)}$. We introduce an effective $\lambda$-dependent perturbation $F(\lambda)$:
    \begin{equation}
        e^{- \beta F(\lambda)} = \bE_i e^{- \beta \lambda \cU^{(i+1)}}, \quad F(0) = 0.
    \end{equation}
    As long as
    \begin{equation} \label{eq:lambda-disk}
         0 \leq |\lambda| < \quad \frac{m_{i+1}- m_i}{4 D \beta |\cU^{(i+1)}|_{m_{i+1},i}}
    \end{equation}
    by \eqref{eq:condition-lemma-rest} and \eqref{eq:diff-condition}, the gauge-averaging theorem applies at level $i$ with $\eta = m_{i+1}$ and $m = m_i$, to $V = \lambda \cU^{(i+1)}$ and $\bE = \bE_i$. The cluster expansion for $F(\lambda)$ converges absolutely and
    \begin{equation}\label{eq:F-strength}
        |F(\lambda)|_{m_i,i} \leq 2 |\lambda| |\cU^{(i+1)}|_{m_{i+1},i}.
    \end{equation}

    Additionally, by \cite{kotecky1986cluster,fernandez2007cluster}, in the disk \eqref{eq:lambda-disk}, $F(\lambda)$ is a holomorphic function in the $|\cdot|_{m_i,i}$ norm. Note that by \eqref{eq:condition-lemma-rest}, the radius of the disk $\frac{m_{i+1}- m_i}{4 D \beta |\cU^{(i+1)}|_{m_{i+1},i}} \geq 4$, so $\lambda = 1$ is well within its radius.  By
    \begin{equation}
        \left.\left ( \frac{d}{d\lambda}\bE_i e^{-\lambda\beta \cU^{(i+1)}} \right) \right|_{\lambda=0}= -\beta\bE_i \cU^{(i+1)}
    \end{equation}
    we have
    \begin{equation}
         \quad F'(0) = \bE_i \cU^{(i+1)}.
    \end{equation}
    Thus, we define a function $h(\lambda)$ which satisfies:
    \begin{equation} \label{eq:square}
        h(\lambda) = F(\lambda) - \lambda \bE_i \cU^{(i+1)} = F(\lambda) - F'(0)\lambda = O(\lambda^2)
    \end{equation}

    With this,
    \begin{equation} \label{eq:anotherfactoroftwo}
    |\cU^{(i)} - \mathbb E_i \cU^{(i+1)}|_{m_i,i} \leq 2 \left|  h(1) \right |_{m_i,i}.
    \end{equation}
    From \eqref{eq:square}, $\frac{h(\lambda)}{\lambda^2}$ is also holomorphic inside the same disk as $F(\lambda)$, and also with the $|\cdot|_{m_i,i}$ norm. Pick any radius $1 < r < \frac{m_{i+1}- m_i}{4 D \beta |\cU^{(i+1)}|_{m_{i+1},i}}$ By the maximum-modulus principle that uses the $|\cdot|_{m_i,i}$ norm,
    \begin{align}
        \left | h(1) \right|_{m_i,i} &\leq \sup _{|\lambda| = r} \left | \frac{h(\lambda)}{\lambda^2} \right|_{m_i,i}
        \\
        &\leq \frac{1}{r^2} \sup _{|\lambda| = r} \left ( \left| F(\lambda)\right|_{m_i,i} + r\left|  \bE_i \cU^{(i+1)} \right|_{m_i,i} \right) \\
        &\leq \frac{1}{r^2} \left (  2 r |\cU^{(i+1)}|_{m_{i+1},i} + r\left|  \bE_i \cU^{(i+1)} \right|_{m_i,i} \right)
        \\
        &\leq \frac{1}{r^2} \left (  2 r |\cU^{(i+1)}|_{m_{i+1},i} + r|\cU^{(i+1)}|_{m_{i+1},i}  \right) \\
        &\leq \frac{3 |\cU^{(i+1)}|_{m_{i+1},i}}{r}
    \end{align}
    where in the second inequality, we used the definition of $h(\lambda)$, in the third inequality, we used \eqref{eq:F-strength}, and in the fourth inequality, we used \eqref{eq:norm-PiK1} and that $m_{i+1} > m_i$ and the norm is monotone in the size parameter. Letting $r$ approach $ \frac{m_{i+1}- m_i}{4 D \beta |\cU^{(i+1)}|_{m_{i+1},i}}$, one obtains
    \begin{equation}
         \left | h(1) \right|_{m_i,i}  \leq \frac{12 D \beta }{m_{i+1}- m_i}(|\cU^{(i+1)}|_{m_{i+1},i})^2 \leq  \frac{12 D \beta }{m_{i+1}- m_i}(A_\ell |\cU^{(i+1)}|_{m_{i+1},i+1})^2.
    \end{equation}
    Using this in \eqref{eq:anotherfactoroftwo} completes the proof of the Lemma.
\end{proof}

Together, the last two lemmas can be combined into a full induction step.

\begin{definition}[Choice of constants] \label{def:parameters}
     Fix an odd $\ell \geq 7$ and $c>1$, independently of $n$. Choose an $n$-independent constant $m_1$ such that
    \begin{align}  \label{eq:parameter-initial-bounds}
        A_\ell e^{-m_1} \leq \frac{1}{2c},
    \end{align}
    and, for $1\leq i\leq n-1$, define $m_{i+1}$ recursively via
    \begin{equation} \label{eq:sum-m}
        m_{i+1} =  m_i + \frac{D}{c^{n - (i+1)}}.
    \end{equation}
    where $D$ is the uniform spin-degree bound in~\eqref{eq:spin-degree-bound}.
    Let $\kappa\geq1$ denote the constant in Lemma~\ref{lemma:local-hierarchical-norms} and choose $\eta>m_n \kappa$ such that
    \begin{equation} \label{eq:eta-condition}
        0 < \frac{\eta/\kappa - m_n}{4D} \leq \log 2.
    \end{equation}
    Finally, set $\eta=\kappa\eta'$.
\end{definition}
Such a choice always exists.  For example, after fixing $c>1$, take $m_1=\log(2cA_\ell)$.  Then \eqref{eq:parameter-initial-bounds} holds, and \eqref{eq:sum-m} defines an increasing sequence $m_1,\ldots,m_n$.  After this sequence has been fixed, take $\eta'=m_n+D$ and $\eta=\kappa\eta'$.      Then $(\eta/\kappa-m_n)/(4D)=1/4$, so all the requirements in Definition~\ref{def:parameters} are satisfied simultaneously.

The choices in Definition~\ref{def:parameters} give the following properties that we will need:

\begin{enumerate}
\item The condition \eqref{eq:diff-condition} is automatically satisfied for $1\leq i\leq n-1$:
\begin{align}
    0<\frac{m_{i+1}-m_i}{4D}
    \overset{\eqref{eq:sum-m}}{=} \frac{1}{4c^{n-i-1}}
    \leq \frac{1}{4}
    < \log 2.
    \label{eq:parameter-increment-smallness}
\end{align}

\item The factor in Lemma~\ref{lemma:linear} satisfies
\begin{equation}
    A_\ell e^{-m_i} \leq A_\ell e^{-m_1} \overset{\eqref{eq:parameter-initial-bounds}}{\leq} \frac1{2c}.
    \label{eq:parameter-linear-bound}
\end{equation}

\item  The smallness condition needed for applying the gauge averaging theorem at level $i$, assuming \eqref{eq:Ui-hierarchical-induction} as the induction hypothesis and condition \eqref{eq:V-hierarchical-condition}
\begin{align}
      A_\ell\beta|\cU^{(i+1)}|_{m_{i+1},i+1}
\overset{\eqref{eq:Ui-hierarchical-induction}}{\leq}
\frac{4A_\ell\beta|V|_{\eta',n}}{c^{n-i-1}}
\overset{\eqref{eq:V-hierarchical-condition}}{\leq}
\frac{m_{i+1}-m_i}{16D}.
    \label{eq:parameter-input-smallness}
\end{align}

\item Assuming \eqref{eq:V-hierarchical-condition} and the induction hypothesis \eqref{eq:Ui-hierarchical-induction} at level $i+1$, the factor in Lemma~\ref{lemma:the-rest} satisfies
\begin{align}
    \frac{32DA_\ell^2}{m_{i+1}-m_i}
    \beta|\cU^{(i+1)}|_{m_{i+1},i+1}
    &\leq
    \frac{32DA_\ell^2}{m_{i+1}-m_i}
    \frac{4\beta|V|_{\eta',n}}{c^{n-i-1}}
    \nonumber\\
    &\overset{\eqref{eq:sum-m}}{=}
    128A_\ell^2\beta|V|_{\eta',n}
    \overset{\eqref{eq:V-hierarchical-condition}}{\leq}
    \frac{1}{2c}.
    \label{eq:parameter-quadratic-bound}
\end{align}

\item For all $1 \leq i \leq n$, we uniformly and independent of $n$ have $m_i>0$:
\begin{equation}
    m_i\geq m_1\geq\log(2cA_\ell)>0.
    \label{eq:uniform-positive-m-bound}
\end{equation}
as well as  $m_n<\eta'\leq\eta=\Theta(1)$, since
\begin{equation}
    \eta=\kappa\eta'
    \leq\kappa\left(m_1+\frac{Dc}{c-1}+4D\log2\right).
\end{equation}
\end{enumerate}

\begin{theorem}[Induction step]
\label{thm:full-induction}
Choose any $c > 1$ and suppose that all other parameters are chosen according to Def.~\ref{def:parameters}.  If the perturbation satisfies
        \begin{equation} \label{eq:V-condition}
            \beta|V|_{\eta,\mathrm{loc}}
            <\min\left\{
                \frac{\eta/\kappa-m_n}{64D},
                \frac{1}{4096cA_\ell^2}
            \right\}.
        \end{equation}
    Then, for $ 1 \leq i \leq n$
        \begin{equation}
         |\cU^{(i)}|_{m_{i},i} \leq \frac{64|V|_{\eta,\mathrm{loc}}}{c^{n-i}}.
         \label{eq:Ui-induction}
         \end{equation}
        For arbitrarily chosen $n$-independent constant $c>1$.
\end{theorem}
\begin{proof}
By Lemma~\ref{lemma:local-hierarchical-norms} with $\eta=\kappa\eta'$, and the assumption~\eqref{eq:V-condition} of the theorem, the geometric strength bound implies the following bound on the finest-level interaction norm:
\begin{equation}
    \beta|V|_{\eta',n}
    \leq16\beta|V|_{\eta,\mathrm{loc}}
    <\min\left\{
        \frac{\eta'-m_n}{4D},
        \frac{1}{256cA_\ell^2}
    \right\}.
    \label{eq:V-hierarchical-condition}
\end{equation}
Thus, in the rest of the proof, we will focus on the bound
\begin{equation}
    |\cU^{(i)}|_{m_i,i}
    \leq\frac{4|V|_{\eta',n}}{c^{n-i}}
    \label{eq:Ui-hierarchical-induction}
\end{equation}
from which the theorem statement follows.

Let us first check the base case of the induction at level $n$. By \eqref{eq:eta-condition} and \eqref{eq:V-condition}, the conditions of the gauge-averaging theorem are satisfied, and thus,
\begin{equation}
    |V_{\rm eff}^{(n)}|_{m_n,n} \leq 2 |V|_{\eta',n}
\end{equation}
Using \eqref{eq:norm-PiK1}, we then obtain
\begin{equation}
    |\cU^{(n)}|_{m_n,n} \leq 2  |V_{\rm eff}^{(n)}|_{m_n,n} \leq 4 |V|_{\eta',n} ,
\end{equation}
The last inequality uses the earlier bound on
$|V_{\rm eff}^{(n)}|_{m_n,n}$. This proves the base case.

The induction assumption~\eqref{eq:Ui-hierarchical-induction} at level $i+1$ together with Lemma~\ref{lemma:size-i} give
 \begin{equation}
          |\cU^{(i+1)}|_{m_{i+1},i} \leq \frac{4 A _\ell |V|_{\eta',n}}{c^{n-(i+1)}},
  \end{equation}
and by \eqref{eq:parameter-input-smallness}, the conditions of Lemma~\ref{lemma:the-rest} hold. Writing
\begin{align}
    |\cU^{(i)}|_{m_{i},i} \leq  |\bE_i \cU^{(i+1)}|_{m_{i},i} +  |\cU^{(i)} - \bE_i \cU^{(i+1)}|_{m_{i},i},
\end{align}
we apply Lemma~\ref{lemma:linear} and Lemma~\ref{lemma:the-rest} to obtain
\begin{align}
    \beta |\cU^{(i)}|_{m_{i},i} &\leq e^{-m_{i}} A_\ell\beta |\cU^{(i+1)}|_{m_{i+1},i+1}   + \frac{32 D}{m_{i+1}- m_i} \left ( A_\ell \beta|\cU^{(i+1)}|_{m_{i+1},i+1} \right )^2
    \\
    &\leq \left ( e^{-m_{i}} A_\ell  + \frac{32 D}{m_{i+1}- m_i}  A_\ell^2 \beta|\cU^{(i+1)}|_{m_{i+1},i+1}\right ) \beta |\cU^{(i+1)}|_{m_{i+1},i+1}   .
\end{align}
Applying the induction assumption, we find that each of the terms in round brackets is at most $1/(2c) $ by eqs.~\eqref{eq:parameter-linear-bound} and \eqref{eq:parameter-quadratic-bound}, and thus,
\begin{align}
    |\cU^{(i)}|_{m_{i},i} \leq \frac{1}{c} |\cU^{(i+1)}|_{m_{i+1},i+1} \leq \frac{4 |V|_{\eta',n}}{c^{n-i}}.
\end{align}
Since $|V|_{\eta',n}\leq16|V|_{\eta,\mathrm{loc}}$, this proves~\eqref{eq:Ui-induction}.

\end{proof}

\begin{cor}[Bound on $\cH_{\rm eff}$] \label{cor:bound-Heff}
Assume that the conditions of Theorem \ref{thm:full-induction} hold.
Then
    \begin{equation}
        |\cH^{(i)}_{\rm eff}|_{m_{i},i} \leq 2 A_\ell |\cU^{(i+1)}|_{m_{i+1},i+1}, \quad 1 \leq i \leq n-1.
    \end{equation}
and
\begin{equation}
    |\cH_{\rm eff}^{(n)}|_{m_n,n} \leq 32|V|_{\eta,\mathrm{loc}}.
\end{equation}

    Using Theorem~\ref{thm:full-induction}, it also follows that
    \begin{equation}
        |\cH^{(i)}_{\rm eff}|_{m_{i},i} \leq \frac{128A_\ell|V|_{\eta,\mathrm{loc}}}{c^{n-(i+1)}}, \quad 1 \leq i \leq n-1.
    \end{equation}
\end{cor}
\begin{proof}
By Lemma~\ref{lemma:size-i}, we have $|\cU^{(i+1)}|_{m_{i+1},i} \leq A_\ell|\cU^{(i+1)}|_{m_{i+1},i+1}$. We then apply the modified gauge-averaging theorem in Thm.~\ref{lemma:gauge-averaging} using $\cU^{(i+1)}$ regrouped at level $i$.  The conditions of the gauge-averaging theorem with $\eta = m_{i+1}$ and $m = m_i$ are satisfied because by the proof of Theorem~\ref{thm:full-induction}.
Thus, by Theorem~\ref{lemma:gauge-averaging}
\begin{equation}
    |V^{(i)}_{\rm eff}|_{m_{i},i} \leq 2|\cU^{(i+1)}|_{m_{i+1},i}  \leq  2 A_\ell |\cU^{(i+1)}|_{m_{i+1},i+1}
\end{equation}

At level $n$, the base-case estimate in the proof of Theorem~\ref{thm:full-induction} and~\eqref{eq:norm-PiK1} give
\begin{equation}
    |\cH_{\rm eff}^{(n)}|_{m_n,n}
    \leq|V_{\rm eff}^{(n)}|_{m_n,n}
    \leq2|V|_{\eta',n}
    \leq32|V|_{\eta,\mathrm{loc}}.
\end{equation}
\end{proof}

\section{Peierls argument}\label{sec:peierls}

\subsection{Main idea}
\label{subsec:peierls-main-idea}

In this section, we argue that the memory lifetime of the 2D code is exponentially long in a power of the number of spins.  Recall that the model of dynamics is a reversible (detailed-balance) Markov chain whose stationary distribution $\pi$ is the Gibbs distribution at inverse temperature $\beta$.  As discussed in the introduction, we identify a subset $A \subseteq \Omega$ corresponding to configurations in a certain logical class.  We then define
\begin{equation}
    \phi(A) = \frac{\pi(\partial A)}{\pi(A)},
\end{equation}
such that $1/\phi(A)$ bounds the escape time from a distribution on $A$, which allows us to bound the memory lifetime.  This is argued in Lemma~\ref{lem:conductancetomemlife} (see similar statements for quantum codes in Refs.~\cite{alicki2010thermal, chesi2010thermodynamic, balasubramanian2026passive}).  

To determine the states in a logical class, we define a decoder, which is a map from an error syndrome to the correction it applies.  Suppose we want to prove stability of a particular encoded bit to arbitrary perturbations.  We say that a spin configuration $s$ whose corresponding syndrome is $\sigma$ is \emph{unstable} if there exists $s' = s \oplus e$ and corresponding $\sigma'$ where $e$ is a single spin flip such that $s + {\rm Dec}(\sigma)$ and  $s' + {\rm Dec}(\sigma')$ are in different equivalence classes corresponding to different logical values of that encoded bit. We denote the set of unstable configurations by $\unstable=A_{\rm unstable}$.

Then, divide $\Omega = A_0 \oplus A_1 \oplus A_{{\rm unstable}}$ where $A_b$ consists of the stable configurations decoding to logical value $b$.  Promote $\Omega$ to a graph where $\omega, \omega' \in \Omega$ are connected by an edge if $\omega$ and $\omega'$ are related by a single spin flip.  It follows that any path from $A_0$ to $A_1$ must pass through $A_{{\rm unstable}}$.  We bound the total probability mass of $A_{{\rm unstable}}$ by noting that for any $S \in A_{{\rm unstable}}$, we may write $\sigma = \tau \oplus \sigma'$ where $\tau$ is a \emph{witness} which must have large weight.  A Peierls estimate then proves that the probability of having a large witness weight is exponentially small in a power of the number of spins.  Since $\partial A_0 \subseteq A_{\rm unstable}$ and $\partial A_1 \subseteq A_{\rm unstable}$, together with a uniform lower bound on $\pi(A_0)$ and $\pi(A_1)$, this implies that both $\pi(\partial A_{0,1})/\pi(A_{0,1})$ are exponentially small.  See Ref.~\cite{balasubramanian2026passive, alicki2010thermal, chesi2010thermodynamic} for a similar chain of reasoning which was used to establish slow mixing bounds in quantum stabilizer codes.

Below, we provide details of the outline sketched above. Gauge averaging preserves the Gibbs probabilities of the gauge-invariant sets. The key place where the gauge-averaged perturbation enters is in the Peierls estimate, where we must argue that if the syndrome configuration contains a witness $\tau$ (whose contribution to the bare energy is $|\tau|$), the effect of the perturbation is to change the energy difference upon removing $\tau_n$ by at most $\epsilon|\tau_n|$, where $\epsilon$ is small.

\subsection{Decoder}
\label{subsec:decoder}

We construct a hierarchical RG-like decoder. Note that the Hamiltonian in this section is the perturbed Hamiltonian $\cH_0 + V$; however, one can define the decoder solely based on the checks entering $\cH_0$ and the associated code structure. 

Starting from the physical syndrome $\sigma_{2n}$, the decoder applies local corrections on the Rep pieces $K_b$ and the Par pieces $K_c$ from Secs.~\ref{sec:rep-local-formalism} and~\ref{sec:par-local-formalism}, respectively. These pieces are associated with the local cochain complexes $\cC_B$ in Rep and $\cC_c$ in Par. Each step produces a residual syndrome that is the image of a syndrome of the preceding code under the comparison map.

Level $i$ in this subsection will count individual Rep and Par replacements, Rep for odd $i$ and Par for even $i$. Given a spin configuration $s_i \in C_{(i)}^1$ and its syndrome $\sigma_i=\delta_{(i)}^1s_i$, each step a correction $f_i$ and a coarse syndrome $\sigma_{i-1}$ satisfying
\begin{equation}
  \sigma_i+\delta_{(i)}^1f_i= \cF_{(i-1)}^2 \sigma_{i-1},
  \label{eq:decoder-step}
\end{equation}
where $\cF_{(i-1)}^\bullet: \cC^\bullet_{(i-1)} \rightarrow \cC^\bullet_{(i)}$ is the comparison map introduced in Sec.~\ref{sec:ladder-replacements}, and we drop the subscript for Rep or Par since it can be deduced from the level. Note that the correction determined by the decoder must satisfy the condition that $\sigma_i+\delta_{(i)}^1f_i$ is in the image of $\cF_{(i-1)}^2$. We define the operation of the decoder below, and we also trace the change in the total syndrome weight at each step.

\paragraph{Rep step.} Upon Rep replacement, all bits are contained in the union of $K_b$ pieces. In each piece $K_b$, solve 
\begin{equation}
    \delta_b f_{i,b}=\sigma_i|_{K_b}.
\end{equation} 
There are two possible solutions, which differ by the local all-ones configuration $\mathbf1_b$. The decoder chooses the one that minimizes the support on terminal bits, i.e., the bits adjacent to the checks in the neighboring $\cC_c$ subcomplexes. There are a total of $2 \deg(b)$ terminal bits.

Then, the correction at level $i$ is $f_i=\sum_b f_{i,b}$. $s_i+f_i$ is constant on each $K_b$; its common value defines $(s_{i-1})_b$. Thus $s_i+f_i=\cF_{(i-1)}^1 s_{i-1}$, and eq.~\eqref{eq:decoder-step} holds. The residual syndrome consists of two identical copies of $\sigma_{i-1}=\delta_{(i-1)}^1s_{i-1}$, with $\cF_{(i-1)}^2(c)=c^0+c^1$.

To bound the residual syndrome weight, let $t_b\leq \deg(b)$ count how many of the terminal bits are flipped by $f_{i,b}$:
\begin{itemize}
    \item If $f_{i,b}$ takes the same value at $b^0$ and $b^1$, then $|\delta_b f_{i,b}| \geq \min (t_b, 2 \deg(b) - t_b) \geq t_b$, because at least $t_b$ of all the rails have unequal value of $f_{i,b}$ on their endpoints.
    \item If the values at $b^0$ and $b^1$ differ, then $|\delta_b f_{i,b}|\geq \deg(b) \geq t_b$ because these bits are connected to each other through $\deg(b)$ ladders.
\end{itemize}
In both cases, $|\delta_b f_{i,b}| \geq t_b$. 
Writing $\sigma_{i,C}$ for the restriction of $\sigma_i$ to the union of the check subcomplexes $\cC_c$, we obtain
\begin{align}
  2|\sigma_{i-1}|
  &=|\sigma_i+\delta_{(i)}^1f_i|\nonumber\\
  &\leq |\sigma_{i,C}|+\sum_b t_b \nonumber
  \\
  &\leq |\sigma_{i,C}|+\sum_b|\delta_b f_{i,b}|
  =|\sigma_i|.
  \label{eq:decoder-rep-contraction}
\end{align}
where the first equality holds by Eq.~\eqref{eq:decoder-step}.

\paragraph{Par step.} The bits are contained in subcomplexes $\cC_c$ and $\cC_B$. 
 The subcomplexes $K_c$ from Sec.~\ref{sec:par-local-formalism} have checks as vertices and internal bits as edges, with local boundary map $\delta_c^1$. 

In each $K_c$, fix a spanning tree rooted at $c^0$, one of the two copies of the coarse level-$(i-1)$ check $c$. Choose the comparison map so that $\cF_{\cR,(i-1)}^2(c)=c^0$. 
Let $f_{i,c}$ be the sum of the paths on the spanning tree from the violated checks to $c^0$ supported on internal bits, and recall that $H^2(\cC_c) \cong \ff_2$

As a result of the correction, the syndrome  can only be at $c^0$, and its value is the parity of the original syndrome in $K_c$:
\begin{equation}
  (\sigma_{i-1})_c=\sum_{v\in V(K_c)}(\sigma_i)_v,
  \label{eq:decoder-par-step}
\end{equation}
and the local correction satisfies
\begin{equation}
      \delta_c^1f_{i,c}=\sigma_i|_{K_c}+(\sigma_{i-1})_c c^0.
\end{equation}
The total correction in iteration $i$ is $f_i=\sum_c f_{i,c}$. 

Then~\eqref{eq:decoder-step} holds and 
\begin{equation}
|\sigma_{i-1}|\leq|\sigma_i|.
\end{equation}
One can directly verify that the coarse syndrome is that of the configuration
\begin{equation}
  (s_{i-1})_b=(s_i)_{b^0}+(s_i)_{b^1}.
  \label{eq:decoder-par-coarse-spin}
\end{equation}
The corrected configuration $s_i+f_i$ equals $\cF_{\cR,(i-1)}^1s_{i-1}$ up to coboundaries of finest gauge generators at level $i$.

\paragraph{Final correction.} Repeating the above repeatedly, we arrive at the code $\cC_{(1)}$, which is a fixed-size repetition code. Use a minimum-weight correction $f_1$ with $\delta_{(1)}^1f_1=\sigma_1$. This final step corrects every single-bit error. Lifting the corrections back to the physical code and telescoping~\eqref{eq:decoder-step} gives
\begin{align}
  f(\sigma_{2n})
  =\sum_{i=1}^{2n}\cF_{\cR,(2n-1)}^1\cdots\cF_{\cR,(i)}^1 f_i
  \label{eq:decoder-total-correction}
\end{align}
such that $\delta_{(2n)}^1f(\sigma_{2n})=\sigma_{2n}$.
The decoded bit is the class of $s_{2n}+f(\sigma_{2n})$ in the logical-state space $H^1(\cC_{(2n)})$ of Sec.~\ref{sec:logical-states}.

\subsection{Decoding graph and induced subgraph}

We will organize each of the replacement subcomplexes into a structure called a \emph{decoding graph}, defined similarly to Ref.~\cite{balasubramanian2026passive}.  The decoding graph will help us determine how the syndrome weight changes under decoding and help us identify unstable configurations.  For a vertex $v\in T_{(i-1)}(0)$ of the Tanner graph from Sec.~\ref{sec:tanner-embedding}, let $F_i(v)$ contain the fine bits and checks in its local piece associated with $v$. In Rep, $F_i(b)$ is $K_b$, while $F_i(c)=\{c^0,c^1\}$. In Par, $F_i(b)=\{b^0,b^1\}$, while $F_i(c)$ contains the bits and checks of $K_c$. These sets partition $T_{(i)}(0)$. 

\begin{definition}[Decoding graph]
\label{def:decoding-graph}
For $2\leq i\leq2n$, let $G_i$ be a copy of the Tanner graph $T_{(i-1)}$. $G_1$ consists of one root $v^{(0)}$. We will formally redefine $F_i$ at level $i=1$ to be $F_1(v^{(0)})=T_{(1)}(0)$. 

The decoding graph $G_{\mathrm{Dec}}$ consists of $\bigcup_{i=1}^{2n} G_i$, including their Tanner edges, with additional edges between $v\in G_i$ and $w\in G_{i+1}$ whenever $w\in F_i(v)$, for $1\leq i<2n$.
\end{definition}
For fixed $\ell$, the maximum degree $\Delta$ of the decoding graph is independent of $n$. Define $f_{1,v^{(0)}}=f_1$.

\begin{definition}[Subgraph induced by a syndrome]
\label{def:syndrome-subgraph}
For an input syndrome $\sigma_{2n}$, mark $v\in G_i$ if it contains nonzero syndrome 
\begin{equation} \label{eq:sub-1}
    \sigma_i|_{F_i(v)}\ne0
\end{equation}
or if the correction changes the syndrome in it, i.e., for some $w\in G_i$,  
\begin{equation}\label{eq:sub-2}
    (\delta_{(i)}^1f_{i,w})|_{F_i(v)}\ne0.
\end{equation}
Let $G_{\sigma_{2n}}$ be the subgraph of,  $G_{\mathrm{Dec}}$ consisting of the marked nodes and all the edges connecting them. We call $G_{\sigma_{2n}}$ the subgraph induced by syndrome $\sigma_{2n}$. 
\end{definition}

\subsection{Unstable configurations and the witness}

\begin{lemma}
\label{lem:decoder-stability}
If $\sigma_1=0$, the decoded logical class is unchanged by any single spin flip, so the configuration is stable in the sense of Sec.~\ref{subsec:peierls-main-idea}.
\end{lemma}
\begin{proof}
Call $S$ the spin configuration corresponding to $\sigma_1=0$. Denote the associated configuration at level $i$ obtained throughout the decoding process $s_i$. Since $\sigma_1=0$, $s_1$ must be either all-1 or all-0. 

A single spin flip $e$ that takes $S + e = S'$ changes the Rep correction only in the piece containing that spin, so the coarse configurations differ by at most one bit. In Par, formula~\eqref{eq:decoder-par-coarse-spin} sends an internal flip to nothing and a copied-bit flip (i.e. $b^0$ or $b^1$) to one coarse flip. Iterating this from the physical level all the way to level $1$ gives the spin configuration $s_1'$ that differs from $s_1$ by a single spin flip. Since $s_1$ is a uniform configuration, $s_1'$ must be simply a configuration with a single error, and the final minimum-weight correction removes this difference.
\end{proof}

Thus, unstable configurations are a subset of spin configurations whose syndrome $\sigma_1$ is nonempty. Such syndrome configurations contain a \emph{witness}, which we define below:

\begin{definition}[Witness subgraph and witness syndrome, adapted from Ref.~\cite{balasubramanian2026passive}]
Let $K$ be the connected component of $G_{\sigma_{2n}}$ containing the root $v^{(0)}$ if the root is marked, and is otherwise the empty graph.
 Write $K_i=K\cap G_i$ and define $\tau_i$ to be the syndrome in $\sigma_i$ restricted to the regions used to define the witness subgraph at level $i$, i.e.
\begin{equation}
  \tau_i=\sigma_i\big|_{\bigcup_{v\in K_i}F_i(v)}.
\end{equation}

We call $K$ the witness subgraph and $\tau_{2n}$ the witness syndrome.
\end{definition}

The witness subgraph is non-empty only if $\sigma_1$ is nonempty. We show that the level-$n$ witness is associated to an error:

\begin{lemma}
\label{lem:boundary}
Decoding $\tau_{2n}$ reproduces the corrections in $K$. In particular, $\delta_{(2n)}^1f(\tau_{2n})=\tau_{2n}$, so the witness syndrome is a coboundary of an error.
\end{lemma}
\begin{proof}
At level $i$, consider a node $u\in G_i\setminus K_i$ that neighbors some $v\in K_i$. The node $u$ cannot be marked, since it would then belong to $K$. Thus $F_i(u)$ contains no syndrome and the decoder applies no correction there. Also, no correction in $K_i$ changes the syndrome in $F_i(u)$, since this would mark $u$.

Define the correction
\begin{equation}
  f_i^{K_i}=\sum_{v\in K_i}f_{i,v}.
\end{equation}
Then
\begin{equation}
  \tau_i+\delta_{(i)}^1f_i^{K_i}=\cF_{(i-1)}^2\tau_{i-1}.
  \label{eq:witness-restricted-step}
\end{equation}
Indeed, suppose a correction outside $K_i$ contributed to the syndrome in a piece belonging to $K_i$. Its node would have to neighbor $K_i$, contradicting the observation above. Also, if the residual syndrome corresponds to a violated coarse check $c$, the node $c\in G_i$ and the node whose piece contains $c$ at level $i-1$ are both marked and joined by an edge. Thus restricting Eq.~\eqref{eq:decoder-step} to the pieces in $K$ gives Eq.~\eqref{eq:witness-restricted-step}.

Now apply the decoder to $\tau_{2n}$. At a step with input $\tau_i$, the local syndrome agrees with $\sigma_i$ on the pieces in $K_i$ and is zero elsewhere. The decoder therefore applies $f_i^{K_i}$, and Eq.~\eqref{eq:witness-restricted-step} gives the next input $\tau_{i-1}$.  At level $1$, $\tau_1=\sigma_1$, so the final correction is $f_1^{K_i}=f_1$.
Applying the comparison maps as in Eq.~\eqref{eq:decoder-total-correction}, we obtain
\begin{align}
  f(\tau_{2n})
  =\sum_{i=1}^{2n}\cF_{(2n-1)}^1\cdots\cF_{(i)}^1f_i^{K_i},\nonumber
\end{align}
and 
\begin{equation}
      \delta_{(2n)}^1 f(\tau_{2n})=\tau_{2n},
\end{equation}
where the second equality follows by summing Eq.~\eqref{eq:witness-restricted-step} after applying the comparison maps.
\end{proof}

\begin{lemma}[Witness bounds]
\label{lem:witness-bounds}
The witness satisfies
\begin{equation}
  \frac{|K|}{1+\Delta}\leq\sum_{i=1}^{2n}|\tau_i|\leq4|\tau_{2n}|.
\end{equation}
If $K$ is nonempty, then $|\tau_{2n}|\geq2^n$.
\end{lemma}
\begin{proof}
A node $v\in K_i$ either contains a syndrome (condition \eqref{eq:sub-1}) or neighbors a node at the same level that contains a syndrome (condition \eqref{eq:sub-2}). To see this, suppose $F_i(v)$ contains no syndrome. Then condition \eqref{eq:sub-2} must be true, and $v$ was marked because a correction $f_{i,u}$ changes the syndrome in $F_i(v)$. The decoder only applies correction in local pieces containing syndrome, so $F_i(u)$ must contain syndrome, and $u\in K_i$ neighbors $v$. Thus, at least $|K|/(1+\Delta)$ nodes in $K$ contain a syndrome, giving the first inequality.

Next, by Lemma~\ref{lem:boundary}, decoding $\tau_{2n}$ gives the syndromes $\tau_{2n},\ldots,\tau_1$. The bounds proved in Sec.~\ref{subsec:decoder} therefore give, for $2\leq i\leq2n$,
\begin{equation}
  |\tau_{i-1}|\leq
  \begin{cases}
    |\tau_i|/2,&\text{if level $i$ is a Rep step},\\
    |\tau_i|,&\text{if level $i$ is a Par step}.
  \end{cases}
\end{equation}
Combining these bounds gives
\begin{equation}
  \sum_{i=1}^{2n}|\tau_i|
  \leq2|\tau_{2n}|\sum_{k=0}^{\infty}2^{-k}
  =4|\tau_{2n}|.
\end{equation}

Finally, the repetition graph of $\cC_{(1)}$ is such that a nonzero syndrome at level $1$ has weight at least two. If $K$ is nonempty, then $\tau_1=\sigma_1\ne0$, and  we have
\begin{equation}
  |\tau_{2n}|\geq2^{n-1}|\tau_1|\geq2^n.
\end{equation}
\end{proof}

\begin{cor}[Counting possible witness syndrome configurations]
\label{cor:witness-count}
For fixed $\ell$, there is a constant $\Gamma>1$, independent of $n$, such that the number of distinct witness syndromes of weight $m$ is at most $\Gamma^m$ for every $m\geq1$.
\end{cor}
\begin{proof}
The number of witness subgraphs of size $|K|$ is at most $\Delta^{2|K|}$, which comes the standard count of connected subgraphs containing a fixed node and the fact that the decoding graph is bounded-degree, see Refs.~\cite{balasubramanian2026passive,ryan-wegner}.

Lemma~\ref{lem:witness-bounds} gives $|K|\leq4(1+\Delta)m$ when $|\tau_{2n}|=m$. For fixed $\ell$, each piece $F_{2n}(v)$ contains at most $q$ syndrome locations, where $q$ depends on $\ell$ but is independent of $n$. A fixed $K$ supports at most $2^{q|K|}$ syndrome configurations, and
\begin{equation}
  \#\{\text{witness syndromes }\tau_{2n}:|\tau_{2n}|=m\}
  \leq \sum_{|K|=1}^{4(1+\Delta)m}(2^q\Delta^2)^{|K|}
  \leq \Gamma^m,
\end{equation}
where $\Gamma>1$ depends only on $\ell$.
\end{proof}

\subsection{Effect of the perturbation}\label{subsec:pertshift}

In this subsection, the hierarchy level index $i=1,\ldots,n$ counts full Rep+Par iterations, as in Sec.~\ref{sec:gauge-hierarchy}. The code construction levels run up to $2n$, so the physical code is $\cC_{(2n)}$, and its physical witness syndrome is $\tau_{2n}$. We assume throughout that $\ell\geq7$ is odd.

In this section, we use the previous results on the cluster expansion to determine how the energy of the witness configuration is modified by the perturbation.  Recall that, from Sec.~\ref{sec:perturbation-strength}, we can write
\begin{equation}
    V_{\rm eff} = \sum_{i = 1}^n \cH_{\rm eff}^{(i)} = \sum_{i = 1}^n \sum_{X_i \subseteq E_i} \cH_{{\rm eff}, X_i}^{(i)},
\end{equation}
where each part $\cH_{{\rm eff}, X_i}^{(i)}$   acts on physical spins and can contain terms of small weight, as long as they are gauge-invariant. They are defined as:
\begin{equation}
    \cH_{{\rm eff}, X_i}^{(i)} = \sum_{\substack{Y \subseteq E_n: \\ R^{n-i}(Y) = X_i}} \cH_{{\rm eff}, Y}^{(i)}.
\end{equation}

\begin{theorem} \label{thm:perturbation-bound-K}
    Consider the spin configuration $S \in \unstable$. Call its witness syndrome $\tau_{2n} \subseteq \partial^T_{2,(2n)}S$ with associated correction $f(\tau_{2n})$ from Lemma~\ref{lem:boundary}. Then, under the assumptions of Theorem~\ref{thm:full-induction}, there exists a constant $\alpha$ independent of $n$ such that
    \begin{equation}
        |V_{\rm eff} (S) - V_{\rm eff}  (S + f(\tau_{2n})) | \leq \alpha |V|_{\eta, \rm loc} |\tau_{2n}|.
    \end{equation}

\end{theorem}
Before proving the theorem, lus introduce some additional notation. We use the coarse shadow map $R:E_{i+1}\to E_i$ from Sec.~\ref{sec:hierarchical-interaction-graphs}. For a chain $z \in C_{1,(2i+2)}$ at hierarchy level $i+1$, we define its coarse shadow to be
\begin{equation}
    R(z)=R\bigl(\{e\in E_{i+1}:e\cap\operatorname{supp}z\ne\varnothing\}\bigr).
\end{equation}
Thus $R^{n-i}(z)$ denotes the level-$i$ shadow of the blocks whose support overlaps the physical chain $z$.

\begin{definition}[Graph neighborhoods]\label{def:graph-neighborhoods}
For $A\subseteq E_i$ and $s\geq0$, define
\begin{equation}
    \cN_s^{(i)}(A)=\{e\in E_i:d^{Q_i}(e,A)\leq s\}.
\end{equation}
For $b \in C_{2,(2n)}$, let $\cN_0^{(i)}(b)$ be obtained by applying $R^{n-i}$ to the blocks in $E_n$ containing a bit adjacent to $\operatorname{supp}b$. Set
\begin{equation}
    \cN_s^{(i)}(b)
    =\cN_s^{(i)}\bigl(\cN_0^{(i)}(b)\bigr).
\end{equation}

\end{definition}
Since every check has degree at most three, $|\cN_0^{(i)}(\tau_{2n})|\leq3|\tau_{2n}|$.

\vspace{5pt}

\begin{proof}[Proof of Theorem~\ref{thm:perturbation-bound-K}]
Recall that $f(\tau_{2n})$ can be understood as a 1-cochain. We start with the bound
    \begin{align}
         |V_{\rm eff} (S) - V_{\rm eff}  (S + f(\tau_{2n}))|
         \leq
         \sum_{i = 1}^n   \sum_{X_i \subseteq E_i}  \left |\cH_{{\rm eff}, X_i}^{(i)}(S) - \cH_{{\rm eff}, X_i}^{(i)}(S + f(\tau_{2n}))  \right|
    \end{align}
    Next, we note that for $\cH_{{\rm eff}, X_i}^{(i)}(S) - \cH_{{\rm eff}, X_i}^{(i)}(S + f(\tau_{2n})) \neq 0$, the support $X_i$ must satisfy $X_i \in \cX_i(\tau_{2n})$, where
    \begin{equation}\label{eq:condition}
    \cX_i(\tau_{2n})
    =\{X\subseteq E_i:\exists z\in Z_{1,(2n)},\ R^{n-i}(z)\subseteq X,\ \langle z,f(\tau_{2n})\rangle\ne0\}.
    \end{equation}
    Here $Z_{1,(2n)}=\ker\partial_{1,(2n)}=K_1$ is the space of fully gauge-invariant chains from Sec.~\ref{sec:gauge-hierarchy}, and $\langle z,f(\tau_{2n})\rangle$ is the usual chain-cochain pairing.
     Then, we have:
    \begin{align}
         |V_{\rm eff} (S) - V_{\rm eff}  (S + f(\tau_{2n}))|
        &\leq \sum_{i = 1}^n   \sum_{\substack{X_i \subseteq E_i: \\ X_i \in \cX_i(\tau_{2n}) }}  \left |\cH_{{\rm eff}, X_i}^{(i)}(S) - \cH_{{\rm eff}, X_i}^{(i)}(S + f(\tau_{2n}))  \right|
        \\
        &\leq  \sum_{i = 1}^n   \sum_{\substack{X_i \subseteq E_i: \\ X_i \in \cX_i(\tau_{2n}) }}  2 \|\cH_{{\rm eff}, X_i}^{(i)} \|_\infty
    \end{align}
Suppose $z\in Z_{1,(2n)}\setminus B_{1,(2n)}$ and $R^{n-i}(z)\subseteq X_i$. Repeatedly applying the rail projections $p_j^a$ from the proof of Lemma~\ref{lemma:weights} gives a nontrivial cycle at hierarchy level $i$, supported on bits in the blocks $X_i$. Recall the projection $P_i$ from Eq.~\eqref{eq:Pi}. There exists $u_0\in B_{1,(2i)}$ such that $w=P_i z+u_0$ is supported in the blocks $X_i$ and has at most one bit per block. Thus, Lemma~\ref{lemma:weights} applied at level $i$ gives
\begin{equation}
    2^i\leq |w|\leq |X_i|.
\end{equation}
    Using this, we split the inequality above into two parts according to how $|X_i|$ compares to $2^i$:
    \begin{align} \label{eq:2i}
         |V_{\rm eff} (S) - V_{\rm eff}  (S + f(\tau_{2n}))|
        \leq 2 \sum_{i = 1}^n
        \left (
        \sum_{\substack{X_i \subseteq E_i: \\ X_i \in \cX_i(\tau_{2n}), |X_i| < 2^i }}
         \|\cH_{{\rm eff}, X_i}^{(i)} \|_\infty
        +
        \sum_{\substack{X_i \subseteq E_i: \\ X_i \in \cX_i(\tau_{2n}) , |X_i| \geq 2^i}}
         \|\cH_{{\rm eff}, X_i}^{(i)} \|_\infty
        \right )
    \end{align}
\vspace{5pt}
\textbf{Bounding the first term in \eqref{eq:2i}.}
When $|X_i|<2^i$, every $z\in Z_{1,(2n)}$ with $R^{n-i}(z)\subseteq X_i$ lies in $B_{1,(2n)}$. Thus $X_i\in\cX_i(\tau_{2n})$ is equivalent to $X_i\in\cX'_i(\tau_{2n})$, where
   \begin{equation}
       \cX'_i(\tau_{2n}) = \{X \subseteq E_i \big| \ \exists z \in B_{1,(2n)}: \ R^{n-i}(z) \subseteq X, \ \langle z, f(\tau_{2n}) \rangle \neq 0 \},
    \end{equation}
Since $z \in B_{1,(2n)}$, there exists $b \in C_{2,(2n)}: \ z = \partial_{2}b$\footnote{Note that such $b$ is not unique because our code has nontrivial relations.}.
Then, for all such $b$, the following holds
\begin{equation} \label{eq:syndrome}
     \langle z, f(\tau_{2n}) \rangle =  \langle \partial_2 b, f(\tau_{2n}) \rangle =  \langle b, \partial_2^T f(\tau_{2n}) \rangle = \langle b, \tau_{2n} \rangle.
\end{equation}
where we used $\partial_{2,(2n)}^T f(\tau_{2n})=\tau_{2n}$ from Lemma~\ref{lem:boundary}.

The condition $\langle z, f(\tau_{2n}) \rangle \neq 0$ tests whether an operator $Z^z$ that can enter perturbation $\cH^{(i)}_{{\rm eff}, X_i}$ anticommutes with the correction $X^{f(\tau_{2n})}$. In this case, $Z^z$ changes sign under the correction and $\cH^{(i)}_{{\rm eff}, X_i}$ can contribute to the energy difference between $S$ and $S + f(\tau_{2n})$.

Eq.~\eqref{eq:syndrome} translates this operator condition into the syndrome language. For $z \in B_{1,(2n)}$, we have $z = \partial_{2,(2n)} b$ with $b \in C_{2,(2n)}$. Then $Z^z$ is a product of checks in $\cH_0$ indexed by ${\rm supp}(b)$, and it measures the parity of syndrome at those locations.  The equivalent condition  $ \langle b, \tau_{2n} \rangle = \langle z, f(\tau_{2n}) \rangle \neq 0$ tests if $\tau_{2n}$ has an odd overlap with $b$. If it does, the operator $Z^z$ detects $\tau_{2n}$.\footnote{For example, in the 3D Wegner gauge theory, when $z$ is a Wilson loop, $b$ is a plaquette surface with boundary $z$, while $\tau_{2n}$ consists of loops on the dual lattice. The pairing $\langle b,\tau_{2n}\rangle$ counts the intersections of these syndrome loops with $b$ mod 2, or equivalently their total linking number with $z$ mod 2.}

This gives
   \begin{equation}
       \cX'_i(\tau_{2n}) = \{X \subseteq E_i \big| \ \exists b \in C_{2,(2n)}: \ R^{n-i}(\partial_{2,(2n)} b) \subseteq  X, \ \langle b, \tau_{2n} \rangle \neq 0 \}.
    \end{equation}
    %
%%%%%%%%%%%%%%%%%%%%%%%%%%%%%%%%%%%%%%%%%%%%%%%%%%%%%%%%
%%%%%%%%%%%%%%%%%%%%%%%%%%%%%%%%%%%%%%%%%%%%%%%%%%%%%%%%
%%%%%%%%%%%%%%%%%%%%%%%%%%%%%%%%%%%%%%%%%%%%%%%%%%%%%%%%
%%%%%%%%%%%%%%%%%%%%%%%%%%%%%%%%%%%%%%%%%%%%%%%%%%%%%%%%
%%%%%%%%%%%%%%%%%%%%%%%%%%%%%%%%%%%%%%%%%%%%%%%%%%%%%%%%
The nonzero pairing forces the filling $b$ to meet $\tau_{2n}$.We will show that this implies that $X_i$ lies near $\tau_{2n}$. For this, we first need to choose a ``filling'' for $z$, i.e. $b\in C_{2,(2n)}$ such that $\partial_{2,(2n)}b=z$, that stays close to $X_i$ after coarse-graining.

\begin{claim}\label{claim:b-distance}
Let $X_i\subseteq E_i$ be nonempty, and let $z\in B_{1,(2n)}$ satisfy $R^{n-i}(z)\subseteq X_i$. There is a filling $b\in C_{2,(2n)}$, with $\partial_{2,(2n)}b=z$, such that
\begin{equation}
    \operatorname{diam}_{Q_i}
    \left(X_i\cup\cN_0^{(i)}(b)\right)
    \leq c' l_i(X_i),
    \label{eq:distance-b}
\end{equation}
where $c'$ is independent of $i$ and $n$.
\end{claim}

\begin{proof}
The idea of the proof is as follows. Because $z \in B_{1,(2n)}$, there exists some filling $b'\in C_{2,(2n)} $ such that $z = \partial_{2,(2n)} b'$; additionally, the coarse-grained version of $z$ fits in $X_i$. We will further coarse-grain the connected hull of $X_i$ until it fits in a star-like region $Y_{(j)}$ at some coarser level $j < i$. We can show that the level can be chosen such that the refinement $Y_{(i)}$ consequently has diameter bounded by a constant multiple of $l_i(X_i)$.
The refinement of $Y_{(j)}$ to the physical level, which we denote $Y_{(n)}$, contains $z$, and we can construct a filling $b$ of $z$ that fits inside $Y_{(n)}$, and can be extended to the whole code $\cC_{(2n)}$. At level $i$, the coarse-grained filling lies inside $Y_{(i)}$, giving the required diameter bound.

\noindent \textbf{Determining coarse level $j$.}
Fix a sufficiently large constant $q>1$. Starting from level $i$, choose the first positive coarse level $j$ for which $ql_i(X_i)\leq\ell^{2(i-j)}$. Since $j$ is the first qualifying level, the condition fails at level $j+1$, so $\ell^{2(i-j-1)}<ql_i(X_i)$. Thus:
\begin{equation} \label{eq:j-choice}
    ql_i(X_i)\leq\ell^{2(i-j)}
    \leq q\ell^2l_i(X_i).
\end{equation}
If no such positive level exists, then $\ell^{2(i-1)}<ql_i(X_i)$, and we can use the original filling $b'$ in the whole code, and the statement of the claim follows straightforwardly.  After $2i$ steps of replacement, the Tanner graph lies inside $[0,2\ell^{2i}]^2$. Lemma~\ref{lemma:uniform-metric-comparison} then gives
\begin{align}
    \operatorname{diam}_{Q_i}
    \left(X_i\cup\cN_0^{(i)}(b')\right)
    &\leq \operatorname{diam}_{Q_i}E_i
    \leq \frac{4}{c_1}\ell^{2i}
    < \frac{4q\ell^2}{c_1}l_i(X_i).
\end{align}
Thus Eq.~\eqref{eq:distance-b} holds in this case if we choose $c'\geq4q\ell^2/c_1$. Moving forward, we will consider the case $1\leq j<i$.

\noindent \textbf{Specifying region $Y_{(j)}$.} 
Consider the Tanner graph $T_{(2j)}$ of $\cC_{(2j)}$ (this corresponds to the hierarchy level $j$). The star at a vertex $v$ is the set of incidence edges
\begin{equation}
    \operatorname{St}_{(2j)}(v)
    =\{(b,c) \in T_{(2j)}:  \ b\sim c,\ v\in\{b,c\}\}.
\end{equation}
Level $j$ was chosen so that the Tanner edges incident to the bits of $H_i(X_i)$ can be covered by the level-$i$ refinement of one Tanner star, or of a pair of stars at a geometric crossing. Denote this star or the smallest union thereof required for such covering by $\mathcal S$. If only one coarse edge is involved, choose the star centered at its check endpoint. Define
\begin{equation}
    E_j(\mathcal S)
    =\{e\in E_j:\exists(b,c)\in\mathcal S,\ b\in e\}.
\end{equation}
The set $E_j(\mathcal S)$ for a star centered at a bit contains one block-spin; one for a star centred at a check contains at most three. At a crossing, both centres have degree two, so $E_j(\mathcal S)$ contains at most two block-spins. Thus,
\begin{equation}
    R^{i-j}(H_i(X_i))\subseteq E_j(\mathcal S),
\end{equation}
and $|E_j(\mathcal S)|\leq3$.

Define $Y_{(j)}$ to be the Tanner subgraph with bit set
\begin{equation}
    \Lambda_B(Y_{(j)})=\bigcup_{e\in E_j(\mathcal S)}e,
\end{equation}
together with all checks whose full support is contained in this set.

\noindent \textbf{Refinement of region $Y_{(j)}$.} 
Starting from $Y_{(j)}$, define $Y_{(k+1)}$ by applying Rep followed by Par to $Y_{(k)}$, for $j\leq k<n$.Write $Y_{(k)}$ for the resulting region in $\cC_{(2k)}$.
At each step, for every bit $b$ and check $c$ in the region being refined, we replace
\begin{equation}
    \begin{aligned}
        \mathrm{Rep}:&\qquad b\mapsto K_b,
        &&c\mapsto\{c^0,c^1\},\\
        \mathrm{Par}:&\qquad b'\mapsto\{{b'}^0,{b'}^1\},
        &&c'
        \mapsto K_{c'}.
    \end{aligned}
\end{equation}
Here $b^0,b^1$ and $c^0,c^1$ are the copies of $b$ and $c$ in the definition of the refinement. The pieces $K_b$ and $K_c$ were defined in Secs.~\ref{sec:rep-local-formalism} and~\ref{sec:par-local-formalism}.  The definition above preserves the property that every check of $Y_{(k)}$ has its full support contained in $Y_{(k)}$, and the replacement rules preserve this property at every subsequent level.

For $j\leq k\leq n$, define
\begin{equation}
    \cN_0^{(k)}(Y_{(k)})
    =\{e\in E_k:e\cap\Lambda_B(Y_{(k)})\ne\varnothing\}.
\end{equation}
In particular, $\cN_0^{(j)}(Y_{(j)})=E_j(\mathcal S)$. Under the Rep and Par parts of the coarse-shadow map $R$, every bit in $Y_{(k+1)}$ maps back to a bit in $Y_{(k)}$.  Therefore
\begin{equation}
    R\bigl(\cN_0^{(k+1)}(Y_{(k+1)})\bigr)
    \subseteq\cN_0^{(k)}(Y_{(k)}).
    \label{eq:regional-shadow}
\end{equation}

\noindent \textbf{The refinement $Y_{(n)}$ contains $z$.} By definition of $Y_{(i)}$,
\begin{equation}
    \bigcup_{e\in X_i}\ \bigcup_{b\in e}
    \operatorname{St}_{(2i)}(b)\subseteq Y_{(i)}.
\end{equation}
The refinement rules preserve this property. Iterating this property and using $R^{n-i}(z)\subseteq X_i$ gives $ X_i\subseteq\cN_0^{(i)}(Y_{(i)})$ and
\begin{equation}
    \operatorname{supp}z
    \subseteq\bigcup_{\substack{e\in E_n\\R^{n-i}(e)\in X_i}}e
    \subseteq\Lambda_B(Y_{(n)}).
\end{equation}

\noindent \textbf{$\cN_0^{(i)}(Y_{(i)})$ has $Q_i$-graph diameter $O(l_i(X_i))$.}
Iterating Eq.~\eqref{eq:regional-shadow} gives
\begin{equation}
    R^{i-j}\bigl(\cN_0^{(i)}(Y_{(i)})\bigr)
    \subseteq E_j(\mathcal S).
\end{equation}
The block-spins in $E_j(\mathcal S)$ are connected through the finest gauge generators. At a check, this follows because every incident bit is flipped by a generator, and each generator flips an even number of incident bits. At a crossing, the added generators of Definition~\ref{def:crossing-gauge-generators} connect the two block-spins. Since $|E_j(\mathcal S)|\leq3$,
\begin{equation}
    \operatorname{diam}_{Q_j}E_j(\mathcal S)
    \leq2\bigl(|E_j(\mathcal S)|-1\bigr)
    \leq4.
\end{equation}

To pass this bound to level $i$, the replacement rules and Lemma~\ref{lemma:uniform-metric-comparison} give
\begin{equation}
    d^{Q_i}(e,f)
    \leq C\ell^{2(i-j)}
    d^{Q_j}(R^{i-j}e,R^{i-j}f)
    +\frac C 2 \sum_{m=1}^{2(i-j)}\ell^m
    \leq C\ell^{2(i-j)}
    \left(d^{Q_j}(R^{i-j}e,R^{i-j}f)+1\right),
\end{equation}
where $C> 2$ is a sufficiently large constant that depends on $c_1,c_2$ from Lemma~\ref{lemma:uniform-metric-comparison} but is independent of the levels. The factor $\ell^{2(i-j)}$ accounts for distance increase upon Rep+Par refinement and the extra sum bounds the extra distance because we add block-spins as we refine.
Consequently,
\begin{align}
    \operatorname{diam}_{Q_i}\cN_0^{(i)}(Y_{(i)})
    &\leq C\ell^{2(i-j)}
    \left(\operatorname{diam}_{Q_j}E_j(\mathcal S)+1\right)\nonumber\\
    &\leq5C\ell^{2(i-j)}
    \leq5Cq\ell^2l_i(X_i).
    \label{eq:regional-star-diameter}
\end{align}
where in the last inequality, we used eq.~\eqref{eq:j-choice}.

\noindent \textbf{Filling $z$ inside $Y_{(n)}$.} For $j\leq k\leq n$, associate $Y_{(k)}$ with the chain complex $C_\bullet(Y_{(k)})$ consisting of its checks in degree two, its bits in degree one, and all gauge generators acting on those bits in degree zero. With the restricted boundary maps, this is a subcomplex of $C_{\bullet,(2k)}$, and every check in it has its full support in the region.  Since $z$ is supported in $Y_{(n)}$ and is a global boundary, if 
\begin{equation} \label{eq:injectivity}
    \ker\bigl[H_1(Y_{(n)})\longrightarrow H_1(C_{\bullet,(2n)})\bigr]=0.
\end{equation}
holds, then the class of $z$  also vanishes in $H_1(Y_{(n)})$. Thus, we can choose $b\in C_2(Y_{(n)})$ with $\partial_2b=z$. Since every check has its full support in $Y_{(n)}$, the same chain is also a valid filling in the full code with $\partial_{2,(2n)} \iota b=z$, where $\iota$ is the map invcluding $b$ in the full chain complex. By eq.~\eqref{eq:regional-shadow} and $ X_i\subseteq\cN_0^{(i)}(Y_{(i)})$, we thus have
\begin{equation}
    X_i\cup\cN_0^{(i)}(b)
    \subseteq\cN_0^{(i)}(Y_{(i)}),
\end{equation}
and applying Eq.~\eqref{eq:regional-star-diameter} gives
\begin{equation}
    \operatorname{diam}_{Q_i}
    \bigl(X_i\cup\cN_0^{(i)}(b)\bigr)
    \leq5Cq\ell^2l_i(X_i).
\end{equation}
This proves the claim with $c'=q\ell^2\max\{5C,4/c_1\}$.

\end{proof}

\begin{claim}\label{claim:syndrome-neighborhood-localization}
If $X_i\in\cX'_i(\tau_{2n})$, then
\begin{equation}
    X_i\subseteq
    \cN_{c'l_i(X_i)}^{(i)}(\tau_{2n}).
    \label{eq:linear-coarse-localization}
\end{equation}
Moreover, for $s\geq1$,
\begin{equation}
    \left|\cN_s^{(i)}(\tau_{2n})\right|
    \leq c''s^2|\tau_{2n}|,
    \label{eq:boundonN}
\end{equation}
where $c'$ and $c''$ are independent of $i$ and $n$.
\end{claim}

\begin{proof}
Choose $z$ from the definition of $\cX'_i(\tau_{2n})$ and a filling $b$ from Claim~\ref{claim:b-distance}. By Eq.~\eqref{eq:syndrome},
\begin{equation}
    \langle b,\tau_{2n}\rangle
    =\langle z,f(\tau_{2n})\rangle
    =1.
\end{equation}
Thus $\cN_0^{(i)}(b)\cap\cN_0^{(i)}(\tau_{2n})\ne\varnothing$. Claim~\ref{claim:b-distance} therefore gives
\begin{equation}
    \max_{e\in X_i}d^{Q_i}\!\left(e,\cN_0^{(i)}(\tau_{2n})\right)
    \leq\operatorname{diam}_{Q_i}\left(X_i\cup\cN_0^{(i)}(b)\right)
    \leq c' l_i(X_i),
\end{equation}
which proves Eq.~\eqref{eq:linear-coarse-localization}.

By Lemma~\ref{lemma:uniform-metric-comparison} and the bounded-density embedding in Theorem~\ref{thm:ladder-code-structure}, every radius-$s$ ball in $Q_i$ contains at most $2(2 c_2 +1)^2s^2$ blocks-spins. Covering $\cN_s^{(i)}(\tau_{2n})$ by balls centered at $\cN_0^{(i)}(\tau_{2n})$ gives
\begin{equation}
    \left|\cN_s^{(i)}(\tau_{2n})\right|
    \leq 2(2 c_2 +1)^2 s^2\left|\cN_0^{(i)}(\tau_{2n})\right|
    \leq 6(2 c_2 +1)^2 s^2|\tau_{2n}|.
\end{equation}
Choosing $c'' = 6(2 c_2 +1)^2 s^2$ shows Eq.~\eqref{eq:boundonN}.
\end{proof}

%%%%%%%%%%%%%%%%%%%%%%%%%%%%%%%%%%%%%%%%%%%%%%%%%%%%%%%%
%%%%%%%%%%%%%%%%%%%%%%%%%%%%%%%%%%%%%%%%%%%%%%%%%%%%%%%%
%%%%%%%%%%%%%%%%%%%%%%%%%%%%%%%%%%%%%%%%%%%%%%%%%%%%%%%%
%%%%%%%%%%%%%%%%%%%%%%%%%%%%%%%%%%%%%%%%%%%%%%%%%%%%%%%%
%%%%%%%%%%%%%%%%%%%%%%%%%%%%%%%%%%%%%%%%%%%%%%%%%%%%%%%%

We finally apply this to the first term in Eq.~\eqref{eq:2i}. For each $e\in E_i$, choose a generator meeting $e$. Every support containing $e$ meets this generator. Thus
\begin{align}
    \sum_{i=1}^n
    \sum_{\substack{X_i\subseteq E_i:\\
        X_i\in\cX'_i(\tau_{2n}),\ |X_i|<2^i}}
    \|\cH_{{\rm eff},X_i}^{(i)}\|_\infty
    &\leq
    \sum_{i=1}^n\sum_{a_i=1}^{\infty}
    \sum_{\substack{X_i\subseteq E_i:\ l_i(X_i)=a_i\\
        X_i\subseteq\cN_{c'a_i}^{(i)}(\tau_{2n})}}
    \|\cH_{{\rm eff},X_i}^{(i)}\|_\infty
    \nonumber\\
    &\leq
    \sum_{i=1}^n\sum_{a_i=1}^{\infty}
    e^{-m_i a_i}
    \sum_{e\in\cN_{c'a_i}^{(i)}(\tau_{2n})}
    \sum_{\substack{X_i\subseteq E_i:\\
        l_i(X_i)=a_i,\ e\in X_i}}
    \|\cH_{{\rm eff},X_i}^{(i)}\|_\infty
    e^{m_i l_i(X_i)}
    \nonumber\\
    &\leq
    \sum_{i=1}^n\sum_{a_i=1}^{\infty}
    e^{-m_i a_i}
    |\cN_{c'a_i}^{(i)}(\tau_{2n})|
    |\cH_{\rm eff}^{(i)}|_{m_i,i}.
\end{align}

Applying Corollary~\ref{cor:bound-Heff} and Eq.~\eqref{eq:boundonN}, we obtain
\begin{align}
    \sum_{i=1}^{n}\sum_{a_i=1}^{\infty}
    e^{-m_i a_i}c''(c'a_i)^2|\tau_{2n}|
    |\cH_{\rm eff}^{(i)}|_{m_i,i}
    &\leq128A_\ell c''(c')^2|\tau_{2n}||V|_{\eta,\rm loc}
    \sum_{i=-\infty}^{n}\sum_{a_i=1}^{\infty}
    a_i^2e^{-m_i a_i} \frac{1}{c^{n-i-1}}\nonumber\\
    &\leq128A_\ell c''(c')^2|\tau_{2n}||V|_{\eta,\rm loc}
    \frac{c^2}{c-1}\nonumber\\
    &\lesssim |V|_{\eta,\rm loc}|\tau_{2n}|.
\end{align}
Here we used $m_i\geq m_1\geq\log(2cA_\ell)\geq\log4$, so that $\sum_{a=1}^{\infty}a^2e^{-m_i a}\leq\sum_{a=1}^{\infty}a^2\,4^{-a}<1$. We extended the sum to $i=-\infty$, formally defining $m_i=m_1$ for $i\leq1$.

\vspace{12pt}
\textbf{Bounding the second term in \eqref{eq:2i}.}
For a fixed level $i$, every support in the sum satisfies $l_i(X_i)\geq|X_i|\geq2^i$. Inserting the exponential weight from the interaction norm therefore gives
\begin{align}
    \sum_{\substack{X_i\in\cX_i(\tau_{2n}):\\|X_i|\geq2^i}}
    \|\cH_{{\rm eff},X_i}^{(i)}\|_\infty
    &\leq e^{-m_i2^i}
    \sum_{\substack{X_i\in\cX_i(\tau_{2n}):\\|X_i|\geq2^i}}
    \|\cH_{{\rm eff},X_i}^{(i)}\|_\infty e^{m_i l_i(X_i)}
    \nonumber\\
    &\leq e^{-m_i2^i}
    \sum_{e\in E_i}
    \sum_{\substack{X_i\subseteq E_i:\\e\in X_i}}
    \|\cH_{{\rm eff},X_i}^{(i)}\|_\infty e^{m_i l_i(X_i)}
    \nonumber\\
    &\leq |E_i|e^{-m_i2^i}|\cH_{\rm eff}^{(i)}|_{m_i,i}.
\end{align}
Each Rep or Par replacement multiplies the number of Tanner edges by $3\ell-3$, so $|E_i|\leq4(3\ell-3)^{2i}\leq4(3\ell+2)^{2i}$. Therefore
\begin{align}
    |E_i|e^{-m_i2^i}
    &\leq4(3\ell+2)^{2i}e^{-m_i2^i}\nonumber\\
    &\leq4e^{(2i-2^i)\log(3\ell+2)}
    \leq4.
\end{align}
Here we used the choice of parameters in Definition~\ref{def:parameters} and the choice $A_\ell=240D\ell^2$ in Appendix~\ref{app:Qi}, giving $m_i\geq m_1\geq\log(2cA_\ell)\geq\log(3\ell+2)$.

Summing over $i$ and applying corollary~\ref{cor:bound-Heff} we obtain
\begin{equation}\label{eq:large-support-perturbation-bound}
    \sum_{i=1}^n
    \sum_{\substack{X_i\in\cX_i(\tau_{2n}):\\|X_i|\geq2^i}}
    \|\cH_{{\rm eff},X_i}^{(i)}\|_\infty
    \leq4\sum_{i=1}^n|\cH_{\rm eff}^{(i)}|_{m_i,i}
    \lesssim |V|_{\eta, \rm loc},
\end{equation}
independent of $|\tau_{2n}|$.

\vspace{12pt}
\textbf{The full bound.}
Combining the two parts together into \eqref{eq:2i} and using $|\tau_{2n}|\geq1$ gives
\begin{equation}
    |V_{\rm eff}(S)-V_{\rm eff}(S+f(\tau_{2n}))|
    \lesssim |V|_{\eta, \rm loc}(|\tau_{2n}|+1)
    \lesssim |V|_{\eta, \rm loc}|\tau_{2n}|.
\end{equation}
Taking $\alpha$ to be the implicit $i$ and $n$-independent constant proves the theorem.
\end{proof}

\subsection{Peierls argument}

Finally, we use the above results to argue that the probability of an unstable configuration is very small, by combining the bound on the number of possible distinct witnesses with the minimum energy of such a witness.  This uses a standard Peierls argument, combined with the result from the previous section on the effect of the perturbation.

\begin{theorem}\label{thm:mainresult}
Assume the hypotheses of Theorem~\ref{thm:full-induction} and, assume that, in addition to   Eq.~\eqref{eq:V-condition}, the perturbation satisfies   $\beta |V|_{\eta,\mathrm{loc}}< \epsilon$ for $n$-independent constant $\epsilon >0$. The probability that $S\in\unstable$ under the Gibbs measure at inverse temperature $\beta$ is at most $2^{-2^n}$ for $\beta\geq\beta_c$, where $\beta_c<\infty$ is independent of $n$.
\end{theorem}
\begin{proof}
Denote by $\pi$ the Gibbs distribution for physical spin configurations $S$ at inverse temperature $\beta$. Call $\sigma_i(S)$ the level-$i$ syndrome obtained during decoding. Suppose $S$ is a spin configuration for which $\sigma_1(S)\ne0$. Then it contains a witness syndrome $\tau_{2n}\subseteq\sigma_{2n}(S)$. Fix such a witness and write $f=f(\tau_{2n})$ for its correction. By Lemma~\ref{lem:boundary}, $\delta_{(2n)}^1f=\tau_{2n}$.

The event $\sigma_{2n}(S) \supseteq  \tau_{2n} \neq 0$ is gauge-invariant, so gauge averaging preserves its probability and the partition function $\mathcal Z$. The probability that $S\sim\pi$ has a syndrome containing $\tau_{2n}$ is therefore
\begin{align}
\mathbb P_{S\sim\pi}
    \bigl(\tau_{2n}\subseteq\sigma_{2n}(S)\bigr)
&=\frac{1}{\mathcal Z}\sum_{S:\,\tau_{2n}\subseteq\sigma_{2n}(S)}
    e^{-\beta|\sigma_{2n}(S)|-\beta V_{\rm eff}(S)}
    \nonumber\\
&=e^{-\beta|\tau_{2n}|}\frac{1}{\mathcal Z}
    \sum_{S:\,\tau_{2n}\subseteq\sigma_{2n}(S)}
    e^{-\beta|\tau_{2n}+\sigma_{2n}(S)|-\beta V_{\rm eff}(S+f)}
    e^{-\beta(V_{\rm eff}(S)-V_{\rm eff}(S+f))}
    \nonumber\\
&\leq e^{-\beta|\tau_{2n}|}\frac{1}{\mathcal Z}
    \sum_{S:\,\tau_{2n}\subseteq\sigma_{2n}(S)}
    e^{-\beta|\tau_{2n}+\sigma_{2n}(S)|-\beta V_{\rm eff}(S+f)}
    e^{\beta|V_{\rm eff}(S)-V_{\rm eff}(S+f)|}
    \nonumber\\
&\leq e^{-\beta(1-\alpha|V|_{\eta,\mathrm{loc}})|\tau_{2n}|}\frac{1}{\mathcal Z}
    \sum_{S:\,\tau_{2n}\subseteq\sigma_{2n}(S)}
    e^{-\beta|\tau_{2n}+\sigma_{2n}(S)|-\beta V_{\rm eff}(S+f)}
    \nonumber\\
&\leq e^{-\beta(1-\alpha|V|_{\eta,\mathrm{loc}})|\tau_{2n}|}.
\end{align}
The second-to-last inequality uses Theorem~\ref{thm:perturbation-bound-K}. For the last inequality, we use that $\sigma_{2n}(S+f)=\sigma_{2n}(S)+\tau_{2n}$ and the injective map $S \mapsto S + f$, so the sum is bounded by $\mathcal Z$.

By Lemma~\ref{lem:decoder-stability}, every unstable configuration has $\sigma_1(S)\ne0$ and therefore contains a nonempty witness. Applying the union bound over all possible witnesses, the probability of an unstable configuration is therefore bounded by
\begin{align}
\mathbb P_{S\sim\pi}(S\in\unstable)
&\leq\sum_{m=1}^{\infty}
    \sum_{\tau_{2n}:\,|\tau_{2n}|=m}
    \mathbb P_{S\sim\pi}
    \bigl(\tau_{2n}\subseteq\sigma_{2n}(S)\bigr)
    \nonumber\\
&\leq\sum_{m=1}^{\infty}
    \bigl(\#\text{ witness syndromes with }|\tau_{2n}|=m\bigr)
    e^{-\beta(1-\alpha|V|_{\eta,\mathrm{loc}})m}.
\end{align}
By Corollary~\ref{cor:witness-count}, the number of witness syndromes of weight $m$ is bounded by $\Gamma^m$. Lemma~\ref{lem:witness-bounds} gives $|\tau_{2n}|\geq2^n$ for every nonempty witness. Thus, 
\begin{equation}
\mathbb P_{S\sim\pi}(S\in\unstable)
\leq\sum_{m=2^n}^{\infty}
    \Gamma^m e^{-\beta(1-\alpha|V|_{\eta,\mathrm{loc}})m} \leq \sum_{m=2^n}^{\infty} \left ( \Gamma e^{-\beta+\alpha\epsilon}\right)^m.
\end{equation}
For $\beta\geq\beta_c:=\log(4\Gamma)+\alpha\epsilon$, using $\beta|V|_{\eta,\mathrm{loc}}\leq\epsilon$, we have $\Gamma e^{-\beta+\alpha\epsilon} \leq 1/4$ and thus we obtain 
\begin{equation}
\sum_{m=2^n}^{\infty} \left ( \Gamma e^{-\beta+\alpha\epsilon}\right)^m
    \leq \frac43\,4^{-2^n}\leq2^{-2^n},
\end{equation}
which proves the theorem.
\end{proof}

To bound $\phi(A_b)$, we also need a uniform lower bound on the probabilities of the stable sectors $A_0,A_1$ mentioned in Sec.~\ref{subsec:peierls-main-idea}.  This allows us to bound the memory lifetime, by the following Lemma (standard, see Ref.~\cite{levin2026markov}, but a proof is provided for completeness):
\begin{lemma}\label{lem:conductancetomemlife}
For a Markov process on $N$ spins with single spin updates and stationary distribution $\pi$, start with an initial distribution $\pi(\cdot | A)$ where $A \subseteq \Omega$. After evolving for time $t$,
\begin{equation}
\mathbb{P}_{\pi(\cdot | A)}\bigl(\text{leaves }A\text{ by time }t\bigr) \leq t N \frac{\pi(\partial A)}{\pi(A)}
\end{equation}
\end{lemma}
\begin{proof}
Define the flow out of $A$ by
\begin{equation}
Q(A, A^c) = \sum_{x \in A, y \notin A} \pi(x) q(x,y)
\end{equation}
where $q(x,y)$ is the transition rate from $x$ to $y$.  Using detailed balance $\pi(x) q(x,y) = \pi(y) q(y,x)$, a standard calculation shows that $Q(A, A^c) \leq N \pi(\partial A)$ where the factor of $N$ comes from the fact that the number of single spin updates is $N$ and thus $\sum_{x \in A} q(y, x) \leq N$.  Next, denote $X_0$ to be the initial state and $J_t$ to be the number of jumps from $A$ to $A^c$ in the interval $[0,t]$.  Then, we can bound
\begin{equation}
    \mathbb{P}_{\pi(\cdot | A)}\bigl(\text{leaves }A\text{ by time }t\bigr) \leq \frac{\mathbb{P}_{X_0 \sim \pi}\bigl(X_0 \in A, J_t \geq 1\bigr)}{\pi(A)}.
\end{equation}
It is also the case that $\mathbb{P}_{X_0 \sim \pi}\bigl(X_0 \in A, J_t \geq 1\bigr) \leq \mathbb{P}_{X_0 \sim \pi}\bigl(J_t \geq 1\bigr) \leq \mathbb{E}_\pi[J_t]$.  But this is
\begin{equation}
\mathbb{E}_\pi[J_t] = \int_{0}^t Q(A, A^c) ds = t Q(A, A^c) \leq t N \pi(\partial A).
\end{equation}
Combining these expressions completes the proof.
\end{proof}

Thus, bounding $\phi(A_b)$ directly translates to bounding the memory lifetime.  In our situation, $A$ will be either $A_0$ or $A_1$ and in both cases $\pi(\partial A_b) \leq \pi(\unstable)$.  Theorem~\ref{thm:mainresult} bounds the numerator by $\pi(\unstable)\leq2^{-2^n}$. It remains to bound $\pi(A_b)$ away from zero for both encoded values. 
We do this by comparing the Gibbs weights of the two sectors under adding the logical flip $S\mapsto S+\alpha_{(2n)}$ where the support of $\alpha_{(2n)}$ was defined in Sec.~\ref{sec:logical-states}.  Recall that adding $\alpha_{(2n)}$ does not change the syndrome and maps $A_b$ bijectively onto $A_{1-b}$.
Only nontrivial Wilson operators change under this flip, and their level-$i$ supports satisfy $|X_i|\geq2^i$. By the proof of Theorem~\ref{thm:perturbation-bound-K} (see around Eq.~\eqref{eq:2i}), we have 
\begin{align}
    \sup_S\left|V_{\mathrm{eff}}(S+\alpha_{(2n)})-V_{\mathrm{eff}}(S)\right|
    &\leq 2\sum_{i=1}^{n}
    \sum_{\substack{X_i\subseteq E_i\\|X_i|\geq2^i}}
    \|\mathcal H_{\mathrm{eff},X_i}^{(i)}\|_\infty
    \leq 2\sum_{i=1}^{n}
    |E_i|e^{-m_i2^i}
    |\mathcal H_{\mathrm{eff}}^{(i)}|_{m_i,i}
    \nonumber\\
    &\leq 8\sum_{i=1}^{n}
    |\mathcal H_{\mathrm{eff}}^{(i)}|_{m_i,i}
    \leq c'|V|_{\eta,\mathrm{loc}}.
        \label{eq:logical-flip-energy-bound}
\end{align}

The sectors are gauge-invariant, so gauge averaging preserves their mass with respect the the stationary distribution, as discussed in Sec.~\ref{subsec:peierls-main-idea}. We may therefore compute their probabilities using $\mathcal H_0+V_{\mathrm{eff}}$ and the associated partition function $\mathcal Z$. Applying the logical flip and using Eq.~\eqref{eq:logical-flip-energy-bound} gives
\begin{align}
    \pi(A_{1-b})
    &=\frac{1}{\mathcal Z}\sum_{S\in A_b}
    e^{-\beta[\mathcal H_0(S)+V_{\mathrm{eff}}(S+\alpha_{(2n)})]}
  \leq e^{c'\beta|V|_{\eta,\mathrm{loc}}}\pi(A_b).
    \label{eq:sector-weight-comparison}
\end{align}
Using $\pi(A_0)+\pi(A_1)=1-\pi(\unstable)$ gives 
\begin{equation}
    1-\pi(\unstable)\leq(1+e^{c'\beta|V|_{\eta,\mathrm{loc}}})\pi(A_b).
\end{equation}
Therefore
\begin{equation}
    \min_{b=0,1}\pi(A_b)
    \geq
    \frac{1-\pi(\unstable)}
    {1+e^{c'\beta|V|_{\eta,\mathrm{loc}}}}
    \geq
    \frac{3}{4(1+e^{c'\epsilon})} \equiv C^{-1} >0.
    \label{eq:stable-sector-mass}
\end{equation}
where we used $\pi(\unstable)\leq2^{-2^n}\leq1/4$ for $n\geq1$ (by Theorem~\ref{thm:mainresult}), as well as $\beta|V|_{\eta,\mathrm{loc}}\leq\epsilon$.

Together with $\partial A_b\subseteq\unstable$, Eq.~\eqref{eq:stable-sector-mass} upper bounds Eq.~\eqref{eq:conductance} by $C2^{-2^n}$. By Theorem~\ref{thm:mainresult} and Lemma~\ref{lem:conductancetomemlife}, we obtain $t_{\mathrm{mem}}\gtrsim2^{2^n}/N$. By construction (see Sec.~\ref{sec:ladder-replacements}), $N\leq C_0a^n$ for a fixed $a>1$, so $2^n\geq(N/C_0)^{\log 2/\log a} \cong N^\eta$. Thus, the memory lifetime is exponentially long in a power of $N$.

\section{Acknowledgments}

All ideas and proofs in this paper were developed by the authors.  ChatGPT 5.6 and 6 and Claude Opus 5.5 and Fable 5.1 were only used to check for correctness, simplify notation, and generate TikZ figures.  S.B. and M.D. are supported by the Walter Burke Institute for Theoretical Physics at Caltech and by the Institute for Quantum Information and Matter, an NSF Physics Frontiers Center (NSF Grant PHY-2317110).
TCL is supported by the Stanford Leinweber Institute for Theoretical Physics and a Stanford Q-FARM Bloch Postdoctoral Fellowship.

\appendix

\section*{Appendices}

\section{Proof of the improved gauge-averaging theorem}
\label{app:gauge-averaging}
\begin{proof}[Proof of Lemma~\ref{lemma:size-i}]
Some parts of the proof will be close to that in Ref.~\cite{ryan-wegner}; we write a version that does not use the parameter $b$ here and also accounts for the difference in our definition of $l(X)$.

We will use the notation
\begin{align}
\|X\| = |G_0(X)|, \qquad \|\gamma\| = |G_0(\gamma)| \equiv \left|\bigcup_{X\in\gamma}G_0(X)\right|.
\end{align}

For a polymer $\gamma$, write $S(\gamma)=\bigcup_{X\in\gamma}X\subseteq E$ for the union of its supports, and extend the hull and length of Sec.~\ref{sec:gauge-averaging-thm} to polymers by
\begin{equation}
   H(\gamma)=H\bigl(S(\gamma)\bigr),
  \qquad
l(\gamma)=|H(\gamma)|=l\bigl(S(\gamma)\bigr).
  \label{eq:polymer-hull}
\end{equation}

Write $X\sim Y$ if $G_0(X)\cap G_0(Y)\neq\varnothing$. A polymer $\gamma$ is a nonempty set of distinct labels with a connected dependency graph; let $P$ denote the set of polymers. Two polymers are dependent $\gamma'\sim\gamma$ if some label in one is dependent on a label in the other (called incompatible in Ref.~\cite{kotecky1986cluster}). In particular, each polymer is dependent with itself.
We use
\begin{align}
    \Phi(\gamma) = \bE_G\prod_{X\in\gamma}f_X, \qquad f_X = e^{-\beta V_X}-1.
\end{align}

Gauge averages factorize across independent polymers; therefore, the cluster expansion is defined as
\begin{equation}
  V_{{\rm eff},S}
  =
  -\frac{1}{\beta}
  \sum_{n\ge1}\frac{1}{n!}
  \sum_{\substack{\Gamma=(\gamma_1,\ldots,\gamma_n)\in P^n\\S(\Gamma)=S}}
  \phi^T(\Gamma)\prod_{j=1}^n\Phi(\gamma_j),
  \qquad
  S(\Gamma)=\bigcup_{j=1}^n\bigcup_{X\in\gamma_j}X.
\end{equation}
The tuple $\Gamma$ may contain repeated polymers. For $n\ge2$, $\phi^T(\Gamma)$ is the sum of $(-1)^{|F|}$ over connected spanning edge sets $F$ of its incompatibility graph; for $n=1$, set $\phi^T(\gamma_1)=1$. Thus disconnected tuples contribute zero. We write $V_{\rm eff}=\sum_S V_{{\rm eff},S}$.

Define
\begin{equation}
    \|\Phi(\gamma)\|_\infty
    \leq\prod_{X\in\gamma}\|f_X\|_\infty,
    \label{eq:polymer-activity-product}
\end{equation}
and additionally define
\begin{equation}
    q_\eta
    =\sup_{g\in G_0}
      \sum_{X:\,g\in G_0(X)}
      \|f_X\|_\infty e^{\eta l(X)}.
    \label{eq:pinned-activity}
\end{equation}

The following bound is still unchanged compared to  Ref.~\cite{ryan-wegner}:
\begin{equation}
    q_\eta
    \leq e^{ \beta|V|_\eta} - 1 \leq 2\beta|V|_\eta
    <\frac{\eta-m}{2D}.
    \label{eq:pinned-smallness}
\end{equation}
Note that with the new definition of $l(X) = |H(X)|$, we now have
\begin{equation}
    \|X\|=|G_0(X)|
    \leq |G_0(H(X))| \leq Dl(X).
    \label{eq:spin-generator-incidence}
\end{equation}
This replaces Eq.~(46) of Ref.~\cite{ryan-wegner}. We will split the logic of the proof into several parts.

\textbf{The Kotecky-Preiss (KP) criterion.}
The main theorem in Ref.~\cite{kotecky1986cluster} concerns an abstract polymer model with polymer activity $\Phi(\gamma)$.  Set $z(\gamma) = \| \Phi(\gamma)\|_\infty$, and choose some  nonnegative functions $d(\gamma), a(\gamma): \, P \rightarrow [0,\infty)$. A sufficient condition for the absolute convergence of the cluster expansion is
\begin{equation}
    K_\gamma =\sum_{\gamma': \gamma' \sim \gamma} z(\gamma') e^{d(\gamma') + a(\gamma')} \leq a(\gamma),
\end{equation}
for all $\gamma \in P$.  We call this the ``KP criterion''. We choose:
\begin{equation}
   d(\gamma) = m l (\gamma), \qquad a(\gamma) = \frac{\eta-m}{2D}\|\gamma\|.
    \label{eq:polymer-kp-weights}
\end{equation}
Moreover, by \cite{fernandez2007cluster}, when the KP criterion is satisfied, for every polymer $\gamma_0$,
\begin{equation}
    1 + \sum_{n \geq 1}\frac{1}{n!} \sum_{(\gamma_1, .. \gamma_n) \in P^n} |\phi^T(\gamma_0, \gamma_1, ... \gamma_n)|\prod_{j = 1}^n \left (z(\gamma_j) e^{d(\gamma_j)} \right ) \leq e^{a(\gamma_0)}.
    \label{eq:KP-series}
\end{equation}
This will be useful for bounding the strength $|V_{\rm eff}|_m$.

\textbf{Verification of the KP criterion.}
Define the pinned activity
\begin{align}
    K =\sup_{g\in G_0}
      \sum_{\gamma:\,g\in G_0(\gamma)}
      \|\Phi(\gamma)\|_\infty
     e^{d(\gamma)+a(\gamma)}
\end{align}
The pinned activity can be used to bound $K_\gamma$ in the KP criterion (as in Eq.~(42) of Ref.~\cite{ryan-wegner}):
\begin{equation}
    K_\gamma \leq \sum_{g_0 \in G_0(\gamma)} \sum_{\gamma': g_0 \in G_0(\gamma')}\|\Phi(\gamma')\|_\infty e^{d(\gamma')+a(\gamma')}  \leq \|\gamma\| K.
\end{equation}
Thus, if we show the pinned KP criterion obeys
\begin{equation}
K \leq \frac{\eta - m}{2D}, \label{eq:pinned-KP-criterion}
\end{equation}
the KP criterion is then satisfied.  We proceed by analyzing $K$. Note that
\begin{equation}
    d(\gamma)+a(\gamma) = m l(\gamma) + \frac{\eta - m}{2 D} \|\gamma\|
    \leq \frac{\eta + m}{2}l(\gamma) \leq \frac{\eta + m}{2}\sum_{X\in\gamma}l(X). \label{eq:d+a}
\end{equation}
where we used $G_0(\gamma)\subseteq G_0(H(\gamma))$ so the definition of $D$ gives $\|\gamma\| \leq |G_0(H(\gamma))| \leq Dl(\gamma)$, and from the definition~\eqref{eq:polymer-hull}, one has $l(\gamma) \leq\sum_{X\in\gamma}l(X)$.
From \eqref{eq:d+a} and \eqref{eq:polymer-activity-product}, we thus have
\begin{align}
    K \leq \sup_{g\in G_0}
      \sum_{\gamma:\,g\in G_0(\gamma)}
      \prod_{X\in\gamma}
      \|f_X\|_\infty e^{(\eta + m)l(X)/2}.
\end{align}
A polymer $\gamma$ pinned at $g$ contains at least one label $X$ satisfying $g\in G_0(X)$. Therefore, we can write:
\begin{align}
    K \leq \sup_{g\in G_0}
      \sum_{X:\,g\in G_0(X)} \,
      \sum_{\substack{\gamma\in P: \\ X\in\gamma}} \
      \prod_{Y\in\gamma}
      \|f_Y\|_\infty e^{(\eta + m)l(Y)/2}.
    \label{eq:K-fixed-support-reduction}
\end{align}

We now bound the second sum in \eqref{eq:K-fixed-support-reduction}. Define the weight of a polymer ${\rm wt}(\gamma)$ to be $|\gamma|$. For $\ell\geq1$, define
\begin{equation}
    R_X^{(\ell)}
    =\sum_{\substack{\gamma\in P:\,X\in\gamma\\ {\rm wt }(\gamma) \leq \ell}} \
      \prod_{Y\in\gamma}
      \|f_Y\|_\infty e^{(\eta+m)l(Y)/2} \label{eq:RX}
\end{equation}
We will show by induction that
\begin{equation}
    R_X^{(\ell)}
    \leq \|f_X\|_\infty e^{\eta l(X)}.
    \label{eq:truncated-connected-support-bound}
\end{equation}
This is straightforward for $\ell = 1$ because $m \leq \eta$.   Suppose now that Eqn.~\eqref{eq:RX} holds for $R_X^{\ell-1}$.

Consider $R_X^{(\ell)}$.  Removing $X$ from the
dependency graph of any polymer $\gamma$ leaves some number $k \geq 0$ of connected components.  Each component contains some $Y\sim X$ and has weight at most $\ell-1$ since ${\rm wt}(\gamma) \leq \ell$.  Thus, the contribution for such a component is bounded by $R_Y^{(\ell-1)}$. Therefore, we can write:
\begin{align}
    R_X^{(\ell)}
    &= \|f_X\|_\infty e^{(\eta+m)l(X)/2}  \sum_{\substack{\gamma\in P:\,X\in\gamma\\|\gamma|\leq \ell}}
      \prod_{\substack{Y\in\gamma \\ Y \neq X} }
      \|f_Y\|_\infty e^{(\eta+m)l(Y)/2}
      \\
    &\leq \|f_X\|_\infty e^{(\eta + m)l(X)/2}
      \sum_{k\geq0}\frac{1}{k!}
      \left(
        \sum_{Y:\,Y\sim X}R_Y^{(\ell-1)}
      \right)^k
    \\
    &= \|f_X\|_\infty e^{(\eta + m)l(X)/2}
      \exp \left(
        \sum_{Y:\,Y\sim X}R_Y^{(\ell-1)}
      \right).
    \label{eq:connected-support-recursion}
\end{align}
The $1/k!$ factor in the second line compensates for overcounting by $k!$ reorderings of the $k$ components. This replaces equations (53)-(55) of Ref.~\cite{ryan-wegner} while avoiding counting involving $\Delta$.

Applying the induction hypothesis results in
\begin{align}
    R_X^{(\ell)}
   \leq  \|f_X\|_\infty e^{(\eta + m)l(X)/2}
      \exp \left(
        \sum_{Y:\,Y\sim X}\| f_Y\|_\infty e^{\eta l(Y)}
      \right).
\end{align}
The latter quantity can be bounded directly by using the definition of $q_\eta$ and using \eqref{eq:pinned-smallness}:
\begin{align}
    \sum_{Y:\,Y\sim X}
      \|f_Y\|_\infty e^{\eta l(Y)}
    &\leq
      \sum_{g\in G_0(X)}
      \sum_{Y:\,g\in G_0(Y)}
      \|f_Y\|_\infty e^{\eta l(Y)}
    \\
    &\leq q_\eta \|X\|
    \leq q_\eta Dl(X)
    <\frac{\eta-m}{2}l(X).
    \label{eq:support-neighborhood-bound}
\end{align}
This closes the induction, since
\begin{align}
    R_X^{(\ell)}
   \leq  \|f_X\|_\infty e^{(\eta + m)l(X)/2}
      e^{\frac{\eta-m}{2}l(X)} = \|f_X\|_\infty e^{\eta l(X)}.
\end{align}

Going back to the equation for $K$ (Eqn.~\eqref{eq:K-fixed-support-reduction}), we obtain
\begin{align}
    K \leq \sup_{g\in G_0}
      \sum_{X:\,g\in G_0(X)} \,
      \|f_X\|_\infty e^{\eta l(X)} = q_\eta < \frac{\eta - m}{2 D}.
\end{align}
This proves the pinned KP criterion \eqref{eq:pinned-KP-criterion}. The bound $K \leq q_\eta$ that we showed along the way will be useful below.

\textbf{Bound on $|V_{\rm eff}|_m$. }
Using \eqref{eq:KP-series}, we obtain
\begin{equation}
    \beta|V_{\rm eff}|_m \leq K \leq q_\eta
    \leq e^{\beta |V|_\eta}-1 \leq 2\beta|V|_\eta.
\end{equation}
This proves the Lemma.
\end{proof}

\section{Geometry of the gauge-interaction graphs}
\label{app:Qi}

\begin{proof}[Proof of Lemma~\ref{lemma:properties-gauge-interaction-graph}]
Let us first recall some notations. 
By \eqref{eq:gauge-generator-support}, $\mathrm{supp}\,g$ is the support of coboundary of $g\in G_0^{(i)}$ in $E_i$. We denote the coboundary itself as denoted $s=\delta^0_{(2i)}g$.
Denote $$S_g=\bigcup_{e\in\mathrm{supp}\,g}e$$  its physical spin support in $\Lambda_{B,(2i)}$. Then, $|\mathrm{supp}\,g|\le|S_g|$.

\vspace{6 pt}

(1) We first bound the maximum degree of generators. Consider first the unmodified generators in $G_0^{(i)}$, before the modification in Definition~\ref{def:crossing-gauge-generators}. 
These generators are obtained by taking the code $\cC_{2i-1}$ and performing the Par-type replacement. There are three types of generators: inner generators already present in the replacement of $(b,c)$ in Fig.~\ref{fig:segment-replacements}, and there are two junction-type generators: where $(b,c)$ meet at bit $b$ for $c \sim b$, and where $(b,c)$ meet at $c$ for $b \sim c$. We will call these bit-junction and check-junction generators.  For the latter two, the size of the introduced gauge generator depends on the degree of bits $\leq 3$ and degree of checks $\leq 2$ in $\cC_{2i-1}$, respectively. The size bound then comes from these degree bounds together with inspecting Fig.~\ref{fig:segment-replacements}:
\begin{equation*}
 |S_g| \leq 
 \begin{cases}
  4, & g\text{ is inner generator},\\
  2+2\deg(b)\le8, & g\text{ is a bit-junction generator},\\
  4\deg(c) \leq 8, & g\text{ is a check-junction generator}.
 \end{cases}
\end{equation*}
Next, let us take into account the new generators added in Definition~\ref{def:crossing-gauge-generators}. 
Every added generator is the sum of two original generators, so
\begin{equation}
  |S_{g+h}|\leq |S_g|+|S_h|\leq16.
\end{equation}

Next, we show that the block-spins $e \in E_i$ have uniformly bounded degree. Before the modification of the set of gauge generators in Definition~\ref{def:crossing-gauge-generators}, each spin meets only the neighboring inner and junction generators. The degree of junctions is at most 3, and each junction can have at most 2 independent generators acting on the same spin. Thus, before the modification of Definition~\ref{def:crossing-gauge-generators}, the block-spin degree would be at most 3. 

Definition~\ref{def:crossing-gauge-generators} adds the sums of gauge generators for which the faces bounded by their supports overlap. The faces have bounded diameter, and by the density bound in Theorem~\ref{thm:ladder-code-structure}, at most $O(1)$ generators supports overlap; thus, there exists a universal constant $D$ such that  $G_0^{(i)} \leq D$.

The connectedness of $Q_i$ is easy to see from the definitions of the replacement and of $E_i$.

\vspace{6 pt}

(2) Recall that  the coarse shadow maps $\bigcup_{b'\in K_b}K_{b'}^{\mathrm T} \mapsto \{b\}$.
Fix $g\in G_0^{(i)}$ and let $Y=R^{-1}(\mathrm{supp}\,g)$ be the set of block-spins that coarse-grain to it via map $R$. We also have that the lifted support of the generator $g$ is contained in $Y$, i.e. 
\begin{equation*}
 \mathrm{supp}\bigl(\cF_{(i)}^0g\bigr)
 \subseteq R^{-1}(\mathrm{supp}\,g)=Y.
\end{equation*}

By the definition of $R$,
\begin{equation*}
 \bigcup_{e'\in Y}e'
 =\bigcup_{b\in S_g}\;\bigcup_{b'\in K_b}K_{b'}^{\mathrm T}.
\end{equation*}
The Rep and Par replacements in Fig.~\ref{fig:segment-replacements} give, respectively,
\begin{equation*}
 |K_b|=2+\deg(b)(2 + \ell - 3) \le 3\ell,
 \qquad
 |K_{b'}^{\mathrm T}|
=2+\deg(b')\left(\ell-1+\frac{\ell-3}{2}\right)\leq5\ell,
\end{equation*}
where $b\in\Lambda_{B,(2i)}$, $b'\in\Lambda_{B,(2i+1)}$, and both degrees are at most three. Using (1), we obtain
\begin{align*}
 |G_0^{(i+1)}(Y)|
 \le\sum_{e'\in Y}|G_0^{(i+1)}(e')|
 \le D|Y|
 \le |S_g| D |K_b||K^T_{b'}|
 \le C\ell^2.
\end{align*}
where $C = 240 D$.

\vspace{6 pt}

(3) We will show the following sequence:
\begin{equation}
 l_i(R(X)) = |H_i(R(X))| \le |R(H_{i+1}(X))|\le|H_{i+1}(X)|=l_{i+1}(X),
\end{equation}
which contains two inequalities.

To show the first inequality, we use that if $e,f\in\mathrm{supp}\,g$, then either $R(e) = R(f)$ or $R(e)$ and $R(f)$ participate in a shared coarse gauge generator.
Every path connecting locations in $H_{i+1}(X)$ therefore gives a coarse path whose location vertices lie in $R(H_{i+1}(X))$. Since $X\subseteq H_{i+1}(X)$ also gives $R(X)\subseteq R(H_{i+1}(X))$, the set $R(H_{i+1}(X))$ is a connected hull of $R(X)$. Thus, by the minimality of $H_i$, $|H_i(R(X))|
 \le |R(H_{i+1}(X))|$.

The second one follows because $R$ assigns exactly one coarse location to each fine location:
\begin{equation}
    |R(H_{i+1}(X))|
=\bigl|\{R(e):e\in H_{i+1}(X)\}\bigr|
\le |H_{i+1}(X)|.
\end{equation}
\end{proof}

\section{Comparison of geometric locality statements}
\label{app:local-hierarchical-norms}

\begin{proof}[Proof of Lemma~\ref{lemma:uniform-metric-comparison}]
For the bound $d_i(e,f)\le c_2d^{Q_i}(e,f)$, assume taht $d^{Q_i}(e,f)=2m$ and choose a shortest path in $Q_i$ between $e$ and $f$, i.e. $( e=e_0,g_1,e_1,\ldots,g_m,e_m=f)$.
The representative spins of $e_{j-1},e_j\in\mathrm{supp}\,g_j$ belong to $S_{g_j}$. By locality of embedding and the universal finite bound on the size of the support of the gauge generators, ther must exist a universal constant $c_2 \geq 1$ such taht $\operatorname{diam}_\infty S_g \leq c_2$ for every $g \in G_0^{(i)}$. Thus, $ d_i(e_{j-1},e_j) \leq 2c_2$ and 
\begin{align}
 d_i(e,f)
 \le\sum_{j=1}^m d_i(e_{j-1},e_j)
 \le2c_2m=c_2d^{Q_i}(e,f).
\end{align}

For the second bound $c_1d^{Q_i}(e,f) \leq d_i(e,f)$. Join $x_i(e)$ to $x_i(f)$ by a straight line $\sigma$. The rectangular and corner regions left empty by the construction, including the exterior, allow each segment $\sigma_j$ that is not inside the embedded Tanner graph to be replaced by a boundary arc $\gamma_j$ with the same endpoints and
\begin{equation*}
 |\gamma_j|_1\le A|\sigma_j|_1,
\end{equation*}
for some $1 \leq A = O(1)$. The resulting path $\gamma$ lies fully within the embedded Tanner graph and satisfies
\begin{align*}
 |\gamma|_1
 &\le|\sigma|_1+\sum_j\bigl(|\gamma_j|_1-|\sigma_j|_1\bigr)\\
 &\le|\sigma|_1+(A-1)\sum_j|\sigma_j|_1\\
 &\le A|\sigma|_1 \leq  A d_i(e,f).
\end{align*}
Because the path $\gamma$ lies within the Tanner graph, it goes between at most $2(|\gamma|_1/2 + 1)$ bits connected by checks; the extra factor of two is because there might be at most two bits per site upon embedding. However, any pair of consecutive spins must participate in at least one common gauge generator.  Thus, a path going through such bits on $Q_i$ and connecting $e$ and $f$ must be at most
\begin{align*}
 d^{Q_i}(e,f)
 &\leq 4\left(\frac {|\gamma|_1}{2} + 1\right) \leq 6 A d_i(e,f).
\end{align*}
thus, choosing $c_1 = 1/(6A)$, we obtain the statement of the Lemma. 
\end{proof}

\begin{proof}[Proof of Lemma~\ref{lemma:local-hierarchical-norms}]
First, we will show that there exists $\kappa \geq 1$ such that
\begin{equation}
 \kappa^{-1}l_n(X)\le l_{\mathrm{loc}}(X)\le \kappa l_n(X).
 \label{eq:geometric-hull-size-comparison}
\end{equation}
Recall that $l_n(X) = |H_n(X)|$ and $l_{\rm loc}(X) = |H_{\zz^2}(X)|$.

A spanning tree joining the spin locations in $H_n(X)$ through shared generators has $l_n(X)-1$ edges. By Lemma~\ref{lemma:uniform-metric-comparison}, replacing each edge by a lattice path of length at most $2c_2$ gives
\begin{equation*}
 l_{\mathrm{loc}}(X)\le1+2c_2(l_n(X)-1).
\end{equation*}
Next, traverse a spanning tree of the lattice hull $H_{\zz^2}(X)$, which can be done using each edge twice, and list the locations $X=\{e_1,\ldots,e_k\}$ in the order their positions are first visited. The traverse itself is at most $2 l_{\rm loc}$ edges; there might also be at most two bits at the same location, adding an extra distance per site, and thus, $\sum_j d_n(e_j,e_{j+1})\leq 3 l_{\mathrm{loc}}(X)$. Joining these locations by shortest paths in $Q_n$ gives
\begin{equation*}
 l_n(X)
 \le1+\frac12\sum_j d^{Q_n}(e_j,e_{j+1})
 \le1+\frac{1}{2c_1}\sum_j d_n(e_j,e_{j+1})
 \lesssim l_{\mathrm{loc}}(X).
\end{equation*}

Finally, 
\begin{align*}
 |V|_{\eta,n}
 \le\sup_{g\in G_0^{(n)}}\sum_{e\in\mathrm{supp}\,g}\sum_{X\ni e}
 \|V_X\|_\infty e^{\kappa\eta l_{\mathrm{loc}}(X)}
 \le16|V|_{\kappa\eta,\mathrm{loc}}.
\end{align*}
In the opposite direction, fix a block $e$ and choose a generator $g$ acting on it. Every interaction containing $e$ also meets $g$, so
\begin{align*}
 \sum_{X\ni e}\|V_X\|_\infty e^{\kappa\eta l_{\mathrm{loc}}(X)}
 \le\sum_{X:\,g\in G_0^{(n)}(X)}
 \|V_X\|_\infty e^{\kappa^2\eta l_n(X)}
 \le |V|_{\kappa^2\eta,n}.
\end{align*}
Taking the supremum over $e$ completes the proof.
\end{proof}

\bibliographystyle{unsrturl}
\bibliography{main}

\end{document}